\documentclass[tradiabstract]{aa}  

\usepackage{graphicx}
\usepackage{txfonts}
\usepackage{natbib}

\bibpunct{(}{)}{;}{a}{}{,}
\usepackage[colorlinks=true,allcolors=blue]{hyperref}

\newcommand{\HH}{\mbox{H$_2$}}

\newcommand{\lya}{\rm \mbox{Ly-$\alpha$}}
\newcommand{\Lya}{\rm \mbox{Ly-$\alpha$}}

\newcommand{\Hb}{H\,$\beta$}

\newcommand{\CI}{\ion{C}{i}}
\newcommand{\CII}{\ion{C}{ii}}
\newcommand{\CIV}{\ion{C}{iv}}
\newcommand{\ClI}{\ion{Cl}{i}}

\newcommand{\FeII}{\ion{Fe}{ii}}

\newcommand{\HI}{\ion{H}{i}}
\newcommand{\MgI}{\ion{Mg}{i}}
\newcommand{\MgII}{\ion{Mg}{ii}}

\newcommand{\OII}{\ion{O}{ii}}
\newcommand{\OIII}{\ion{O}{iii}}

\newcommand{\SII}{\ion{S}{ii}}
\newcommand{\SiII}{\ion{Si}{ii}}

\newcommand{\ZnII}{\ion{Zn}{ii}}

\newcommand{\kms}{\ensuremath{{\rm km\,s^{-1}}}}
\newcommand{\cmsq}{\ensuremath{{\rm cm}^{-2}}}

\newcommand{\J}{J0015$+$1842}

\newcommand{\iap}{Institut d'Astrophysique de Paris, CNRS-SU, UMR\,7095, 98bis bd Arago, 75014 Paris, France --- \email{noterdaeme@iap.fr}\label{iap}}
\newcommand{\ioffe}{Ioffe Institute, {Polyteknicheskaya 26}, 194021 Saint-Petersburg, Russia \label{ioffe}}
\newcommand{\iucaa}{Inter-University Centre for Astronomy and Astrophysics, Pune University Campus, Ganeshkhind, Pune 411007, India \label{iucaa}}\newcommand{\dawnone}{Cosmic Dawn Center (DAWN), University of Copenhagen, Jagtvej 128, DK-2200, Copenhagen N, Denmark \label{dawnone}}
\newcommand{\dawntwo}{Niels Bohr Institute, University of Copenhagen, Jagtvej 128, DK-2200, Copenhagen N, Denmark \label{dawntwo}}

\newcommand{\ita}{Institute of Theoretical Astrophysics, University of Oslo, PO Box 1029 Blindern, 0315 Oslo, Norway \label{ita}}

\begin{document} 
   
   \title{Down-the-barrel observations of a multi-phase quasar outflow at high redshift 
   }
   \subtitle{VLT/X-shooter spectroscopy of the proximate molecular absorber at $z=2.631$ towards SDSS\,J001514+184212
      \thanks{Based on observations collected at the European Organisation for Astronomical Research in the Southern Hemisphere under ESO programme 103.B-0260(A).}
   }

     \author{
   P. Noterdaeme\inst{\ref{iap}} 
      \and 
   S. Balashev\inst{\ref{ioffe}}
   \and 
   J.-K. Krogager\inst{\ref{iap}} 
   \and
   P. Laursen\inst{\ref{ita},\ref{dawnone}}
   \and
   R. Srianand\inst{\ref{iucaa}}
     \and \\
   N. Gupta\inst{\ref{iucaa}} %
   \and  
   P. Petitjean\inst{\ref{iap}}
   \and 
   J.~P.~U. Fynbo\inst{\ref{dawnone},\ref{dawntwo}}
   }
   \institute{\iap \and \ioffe \and \ita \and \dawnone \and \iucaa \and \dawntwo}
   
   \date{\today}

  \abstract{
  We present ultraviolet to near infrared spectroscopic observations of the quasar SDSS\,J001514$+$184212 and its proximate molecular absorber at $z=2.631$. 
  The [\OIII] emission line of the quasar is composed of a broad (FWHM $\sim1600$~\kms), spatially unresolved component, blueshifted by about 600~\kms\ from a narrow, spatially-resolved component (FWHM $\sim650$~\kms). The wide, blueshifted, unresolved component is consistent with the presence of outflowing gas in the nuclear region. The narrow component can be further decomposed into a blue and a red blob with a velocity width of several hundred~\kms\ each, seen $\sim$5~pkpc on opposite spatial locations from the nuclear continuum emission, indicating outflows on galactic scales. The presence of ionised gas on kpc scales is also seen from a weak \CIV\ emission component, detected in the trough of a saturated \CIV\ absorption that removes the strong nuclear emission from the quasar.
  
  Towards the nuclear emission, we observe absorption 
  lines from atomic species in various ionisation and excitation stages and confirm the presence of strong H$_2$ lines originally detected in the SDSS spectrum. The overall absorption profile is very wide, spread over $\sim$600~\kms, and it roughly matches the velocities of the narrow blue [\OIII] blob. 
  From a detailed investigation of the chemical and physical conditions in the absorbing gas, we infer densities of about $n_\mathrm{H} \sim 10^4$--$10^5$~cm$^{-3}$ in the cold ($T\sim 100$~K) H$_2$-bearing gas, which we find to be located at $\sim$10~kpc distances from the central UV source.  
  We conjecture that we are witnessing different manifestations of a same AGN-driven multi-phase outflow, where approaching gas is intercepted by the line of sight to the nucleus. We corroborate this picture 
  by modelling the scattering of \lya\ photons from the central source through the outflowing gas, reproducing the peculiar \lya\ absorption-emission profile, with a damped \lya\ absorption in which red-peaked, spatially offset, and extended \lya\ emission is seen. 
  Our observations open up a new way to investigate quasar outflows at high redshift and shed light on the complex issue of AGN feedback. 
    }


   \maketitle
%

\section{Introduction}

Feedback from active galactic nuclei (AGN) is an essential element in modern models of galaxy formation and evolution \citep[e.g.][]{Silk1998}. It may quench star formation \citep[e.g.][]{Zubovas2012,Pontzen2017,Terrazas2020}, impact galaxy morphology \citep{Dubois2016}, regulate the growth of the supermassive black holes \citep{Volonteri2016}, and so on. It may also have positive feedback on star formation through compression of the gas \citep[e.g.][]{Zubovas2013,Richings2018}. The most supportive evidence for AGN feedback is the presence of kiloparsec-scale outflows, which result from the propagation of energy and momentum from accretion disc winds to the host galaxy interstellar medium \citep[e.g.][]{Fabian2012,Costa2015}. At low and intermediate redshifts, such outflows have been observed in ionised and molecular phases, with many studies focusing on Type 2 quasars, where the obscuration of the nuclear emission facilitates the observations \citep[e.g.][]{Sun2017}. Outflows have also been observed in unobscured, luminous AGN (Type 1 quasars), but mostly at low redshift, such as the well-known Mrk~231 at $z=0.04$ \citep{Rupke2011, Feruglio2015}. Naturally, observational constraints on hot and cold outflows in high-redshift Type 1 quasars are much harder to obtain, not only because of the dimming of the light, but also because the coarser angular resolution impedes observations close to the bright nucleus. 

In turn, the cosmological distances and high luminosity of the nuclear emission of high-$z$ quasars make them excellent background sources against which gas in any phase can be studied in absorption. This includes not only gas at any place along the line of sight (from the intergalactic medium, giving rise to the Ly-$\alpha$ forest or closer to intervening galaxies, producing damped Ly-$\alpha$ systems (DLAs) in the quasar spectrum), but also gas close to the supermassive black hole, in the quasar host galaxy or in galaxies belonging to the quasar group environment. 
While broad absorption lines have long been associated with gas flows close 
to the central engine \citep[e.g.][]{Weymann1991,Arav2018}, large-scale outflows, when intercepted by the line of sight to the nucleus, should also produce a detectable signature in absorption. 

Absorption spectroscopy has proven to provide very detailed information about the kinematics, chemical composition, and physical conditions in the absorbing gas, in particular when molecules are detected. Indeed, the formation, survival, and excitation of molecules is very sensitive to the prevailing physical conditions, such as abundance of dust, temperature, density, and ambient UV field (e.g. \citealt{Reimers2003,Srianand2005, Cui2005, Guillard2009, Balashev2017} and \citealt{Wakelam2017}, among many others).

Combining information from both emission and absorption line measurements could hence provide us with a fresh view of quasar outflows and feedback at high redshift. We recently embarked on a search for damped molecular hydrogen (H$_2$) absorption lines at the quasar redshift in low-resolution spectra from the Sloan Digital Sky Survey (SDSS) III (BOSS) Data Release 14. Through this targeted search, we discovered a population of strong proximate H$_2$ absorption systems at $z>2.5$ with $\log N$(H$_2) > 19$ \citep{Noterdaeme2019}. Detailed follow-up observations are currently on-going using the multi-wavelength spectrograph X-shooter on the Very Large Telescope. 

In this paper, we report on the observations of the first object in our sample, the quasar SDSS\,J001514.82$+$184212.34 (hereafter \J) at $z=2.63$. We interpret the observations as evidence of kilo-parsec scale, multi-phase, biconical outflows that are mainly orientated along the line of sight
and also probed in absorption against the nucleus. 
We present the observations and data reduction in Sect.~\ref{s:obs}, the analysis of spatially-resolved emission lines in Sect.~\ref{s:em}, and that of absorption lines and dust extinction towards the nucleus in Sect.~\ref{s:abs}. We discuss our results in Sect.~\ref{s:dis}, along with the derivation of the chemical and physical conditions in the absorbing clouds and the modelling of leaking \lya\ emission. We summarise our findings in Sect.~\ref{s:con}. Throughout this paper, we assume a flat $\Lambda$CDM cosmology with $H_0 = 68$~km\,s$^{-1}$\,Mpc$^{-1}$, $\Omega_{\Lambda} = 0.69$, and $\Omega_{\rm m} = 0.31$ \citep{Planck2016}.

\section{Observations and data reduction \label{s:obs}}

\J\ was observed in stare mode with the X-shooter spectrograph at the Very Large Telescope (VLT) 
at the Paranal Observatory in Chile between July and September 2019. The log of observations is provided in Table~\ref{t:log}.  

\begin{figure}
    \centering
    \includegraphics[width=0.8\hsize]{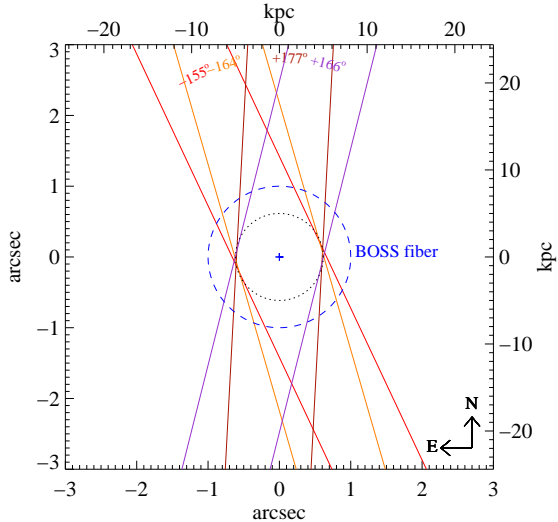}
    \caption{Layout of 1.2$\arcsec$-wide X-shooter slits in the NIR. The small dotted circle represents a projected area of radius 5~kpc (at the redshift of the quasar) with 100\% coverage by all our X-shooter slit positions, and the larger dashed circle represents the BOSS fibre aperture. 
    }
    \label{f:slitpos}
\end{figure}

The X-shooter spectrograph simultaneously covers the wavelength range from 0.3 to 2.5~$\mu$m in three separate spectrographs, which are the so-called arms: UVB ranging from 0.3 to 0.6~$\mu$m; VIS ranging from 0.6 to 1.0~$\mu$m; and NIR ranging from 1.0 to 2.5~$\mu$m.
For all observations, slit widths of 1.0, 0.9, and 1.2~arcsec were used for 
the UVB, VIS, and NIR arms, respectively. The slit was aligned to the parallactic angle at the start {to minimise slit losses} of the exposure, and it was kept fixed on sky during the $\sim1$\,h long exposure (see Fig.~\ref{f:slitpos}).
The observation from August 30 failed due to an error in the atmospheric dispersion corrector (ADC), and exposures from this date were therefore not included in the analysis.
For one other observation (July 29), we noted a strong chromatic suppression of flux in the UVB arm,\footnote{The chromatic slit loss is most likely due to the poor seeing (1.7\arcsec); however, it may also be related to a partial failure of the ADC for the UVB arm.} hence the UVB spectrum from this observation has been discarded from our analysis. As the loss of flux in the VIS and NIR arms is much less severe, we used these spectra in our analysis but weighted them appropriately (using the inverse variance) when combining the individual spectra.

The raw spectra were processed using the official esorex pipeline for X-shooter version 2.6.8. Before passing the raw spectra through the pipeline, we corrected cosmic ray hits using the code {\sc Astroscrappy} \citep{astroscrappy}, which is a Python implementation of the algorithm by \citet{Vandokkum2001}.
The pipeline then performed subtraction of bias and dark levels on the CCD, spectral flat fielding, tracing of the curved echelle orders, and computation of the wavelength solution. The individual curved 2D orders were then rectified onto a straightened grid. The sky background was subsequently subtracted before the individual orders were combined. 
The full spectrum was flux-calibrated using a sensitivity function calculated for each observation using a spectroscopic standard star observed during the same night. The three spectra (UVB, VIS, NIR) were stitched together in the overlapping regions, correcting for any small offsets between the absolute flux calibration of the individual arms.
Lastly, in order to correct for possible slit loss, we scaled the calibrated spectra to match the SDSS $i$ band photometry. Because of the possible variation of the quasar brightness between the time of observation for the SDSS photometry and our data, the absolute flux calibration has a systematic uncertainty of order 20\%, which corresponds to the typical long-term variability of quasars \citep{Hook1994}.

From the 2D spectra produced by the pipeline, we performed an optimal extraction \citep{Horne1986} to obtain 1D spectra. The 1D and 2D spectra were subsequently corrected to vacuum wavelengths and shifted to the heliocentric reference frame before being corrected for Galactic extinction using the maps by \citet{Schlafly2011}. Lastly, we combined the individual observations for each arm into final 1D spectra using a weighted mean. 

{}

\begin{table*}
\centering
\caption{Log of observations and atmospheric parameters \label{t:log} 
}
\begin{tabular}{ccccccc}
    \hline \hline
    Date    &  Slit Position Angle (PA)   & Seeing  & Airmass &  R(H$_2$O)\tablefootmark{a} & R(CO$_2$)\tablefootmark{a} &  FWHM(NIR)\tablefootmark{b}  \\
            & (deg E. of N.)&  (arcsec)         &   &                    &                    &   (pixel)    \\
    \hline
    2019-07-29  & $+166$   &   1.67  & 1.48  & 1.10  &  1.44  &  4.3  \\ 
    2019-08-31  & $-164$   &   0.75  & 1.39  & 0.86  &  1.25  &  3.0  \\ 
    2019-08-31  & $+177$   &   0.65  & 1.41  & 0.91  &  1.30  &  2.9  \\ 
    2019-09-28  & $-155$   &   0.53  & 1.42  & 0.94  &  1.04  &  3.8  \\ 
    \hline
\end{tabular}
\tablefoot{
\tablefoottext{a}{Best-fitting relative abundance compared to the default template in {\tt molecfit}: {$N({\rm H_2 O})=2.725\times 10^{4}$~ppmv and $N({\rm CO}_2)=368.5$~ppmv}. Typical uncertainties of $\sim0.01$.}
\tablefoottext{b}{Best-fit spectral resolution of the NIR spectrum from {\tt molecfit} with typical uncertainties of 0.1~pixel.}
}
\end{table*}

\subsection*{Telluric correction}
The NIR spectra were corrected for telluric absorption using the software tool {\tt molecfit} \citep{Smette2015, Kausch2015}. The code fits a synthetic absorption profile to the telluric absorption features in the data. For the NIR spectra, the dominant molecular species giving strong telluric absorption lines are H$_2$O and CO$_2$. The relative abundances of these two species were fitted together with a fifth-order polynomial model for the intrinsic spectrum in order to reproduce the shape of the wings of the quasar emission lines. 
We fitted the telluric model using seven narrow bands of the spectra dominated by telluric absorption features only. These seven regions were chosen to encompass regions of interest, that is, the regions around the quasar emission lines from oxygen. The wavelength intervals used were as follows: 1.13--1.14,  1.37--1.39, 1.44--1.45, 1.75--1.76, 1.82--1.83, 1.94--1.97, and 2.02--2.03~$\mu$m. For each region, we iteratively masked out pixels affected by skyline residuals or bad pixels. The optimal parameters for each spectrum are given in Table~\ref{t:log}.
The optimised telluric absorption model for each spectrum was then used to correct the corresponding 1D and 2D spectra.

\section{Emission line analysis} \label{s:em}

We analysed the 1D and 2D together spectra to extract information about the spatial and spectral properties of the various emission lines of interest. 
In addition to the 1D spectrum obtained through the optimal extraction method, which mostly samples the unresolved nuclear emission (due to the weighting profile), we also constructed a \emph{total} 1D NIR spectrum directly extracted from the combined 2D data over a wider spatial range of $\pm 9$~pixels (or equivalently 1.9~arcsec) around the quasar trace without weighting.  
This is specifically constructed for the analysis of spatially extended emission lines such as [\OIII]. 
In summary, we therefore consider three kinds of 1D spectra in the following: (i) the \emph{on-trace} spectrum, which is obtained from the trace by optimal extraction; (ii) the \emph{overall} spectrum, which is extracted within wide spatial range; and (iii) the \emph{off-trace} spectrum, which is the difference between the two others.

 \subsection{[\OII]} 
 
\begin{figure}
    \centering
    \includegraphics[width=0.95\hsize]{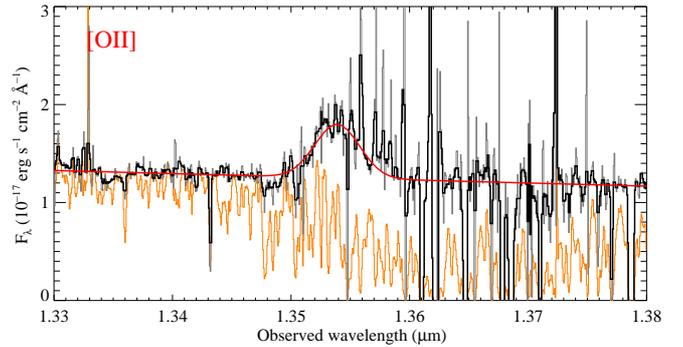}
    \caption{Portion of combined 1D NIR spectrum around the [\OII] doublet emission line. The 1D spectrum with (without) telluric correction is shown in black (orange) together with the fitted profile in red.}
    \label{f:OII1D}
\end{figure}

 The [\OII]$\lambda\lambda3727,3729$ doublet is redshifted into a strong telluric band and not directly visible in the original data. However, applying the telluric correction described in Sect.~\ref{s:obs} allowed us to recover the lines, which, given their large velocity width, are strongly blended and appear as a single line in Fig.~\ref{f:OII1D}. 
 To estimate the intrinsic line width we fitted the [\OII] doublet using the sum of two Gaussians plus a linear continuum in that region. The relative positions of the Gaussians were kept fixed to the doublet wavelength ratio, and their widths were forced to be the same. In principle, the amplitude ratio for the [\OII] doublet can vary between 0.35 and 1.5, depending on the kinetic temperature of the gas and the electron density \citep[e.g.][]{Kisielius2009}. However, because this region of the spectrum is quite noisy, we assumed a fixed ratio of 1.3. This assumption has almost no influence on the total line flux and width.
 The obtained parameters are given in Table~\ref{t:emission}. We measured the [\OII] emission redshift to be $z=2.631$. 

 \subsection{[\OIII] and \Hb} 

\begin{figure}
    \centering
    \includegraphics[width=0.95\hsize]{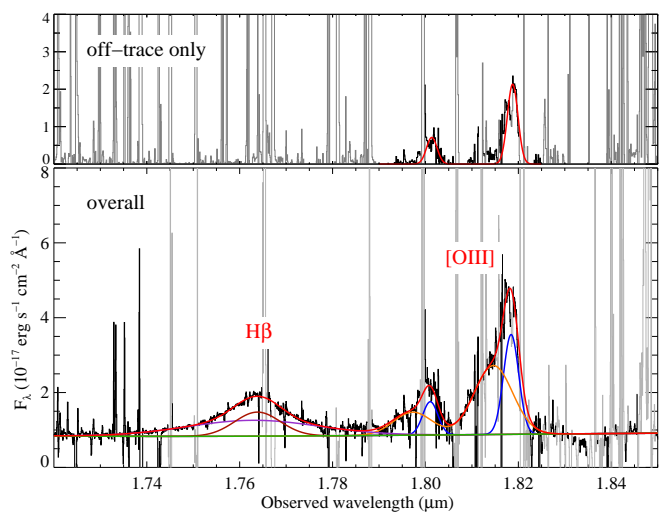} 
    \caption{Portion of combined 1D NIR spectrum around  \Hb\ and [\OIII] emission lines.
    The top panel shows the contribution from the off-trace spatially extended emission, which is apparent in both [\OIII] lines. 
    The bottom panel shows the spectrum collapsed over its full spatial extent at wavelengths $1.790<\lambda (\mu m)<1.825$ and the optimal extraction elsewhere, for presentation purposes. 
    The green line shows our estimate of featureless continuum emission. Individual Gaussians constrained during the fit are shown in different colours, with the total fitted profile shown in red. 
    \label{f:OIII1D}}
\end{figure}

We analysed the \Hb\ and [\OIII]$\lambda\lambda$4960,5008 lines simultaneously, since these lines are located in the same region from 1.74 to 1.84\,$\mu$m. Both the H$\beta$ and the [\OIII] profiles are found to be non-Gaussian and spread over about 2000~\kms. The forbidden [\OIII]$\lambda\lambda4960,5008$ emission lines are clearly asymmetric, and composed of a broad and `narrow' component separated by about 600~\kms. We found that the overall \Hb+[\OIII] complex can be well modelled by the sum of a smooth continuum component (green line in Fig.~\ref{f:OIII1D}) and six Gaussian profiles, two for each spectral line (purple and brown for \Hb; orange and blue for the [\OIII] doublet).
{Because [\OIII] lines can only be deexcited radiatively and since their upper energy levels are only slightly different, the [\OIII] line ratio remains constant.
Therefore,} to fit these lines we tied the redshifts and line widths for the two [\OIII] lines and imposed their amplitude to follow the theoretical 1:3 ratio {\citep[e.g.][]{Storey2000, Dimitrijevic2007}}. Since the narrow [\OIII] component is spatially resolved (see Sect.~\ref{s:spa}), we considered the overall 1D spectrum instead of the on-trace spectrum when fitting this line. We also used the off-trace spectrum to obtain an initial guess for the parameters of the narrow component free from contamination by the broad nuclear component (top panel of Fig.~\ref{f:OIII1D}). We then fitted the overall spectral profile relaxing all parameters and rejecting deviating pixels iteratively. The corresponding line parameters are given in Table~\ref{t:emission}.

The peak of the narrow [\OIII] emission provides us with an estimate of the quasar host's systemic redshift, which also matches that obtained from [\OII].
Given the associated uncertainty of about 100~\kms, we rounded the value to $z=2.631$, which we take as a 
convenient reference redshift to define the zero point of our velocity scale throughout this work. 

 \begin{table}
 \caption{Emission line parameters  
 \label{t:emission}}
     \centering
     \begin{tabular}{l l l l}
     \hline \hline
     {\Large \strut}Line              & $z$      & FWHM   & $L$ \\
                                      &          & (\kms) & (10$^{44}$\,erg\,s$^{-1}$) \\
     \hline
     H$\beta$, comp 1                 & 2.6252   & 5700$\pm$430  &  0.9  \\
     H$\beta$, comp 2                 & 2.6273   & 1960$\pm$150  &  0.5  \\
     $[\OIII]\lambda5008$, broad      & 2.6233   & 1600$\pm$60  &  1.1  \\
     $[\OIII]\lambda5008$, narrow     & 2.6309   & 630$\pm$50   &  0.6  \\
     $[\OII]$                         & 2.6310   & 950$\pm$80  &  0.15 \\     
    \CIV, comp 1                      & 2.6149   & 5620$\pm$170  &  1.3  \\
    \CIV, comp 2                      & 2.6221   & 2050$\pm$50  &  0.6  \\
     Leaking Ly$\alpha$               & 2.634    & 600   &  1.4  \\ 
          \hline
     \end{tabular}
     \tablefoot{The apparent peak and FWHM of the leaking Ly-$\alpha$ emission were estimated visually from the non-Gaussian spectral profile and the luminosity from direct integration. For the other lines, these parameters were obtained from fitting the Gaussian profiles as described in the text.
     Statistical uncertainties on the line luminosities are about 10\% for NIR lines and 5\% for UVB lines. However, the uncertainty on the absolute flux calibration is about 20\%. This has no effect on the line ratios.     }
 \end{table}

\subsection{\CIV\ emission}

\begin{figure}
    \centering
    \begin{tabular}{c}
    \includegraphics[angle=90,width=0.95\hsize]{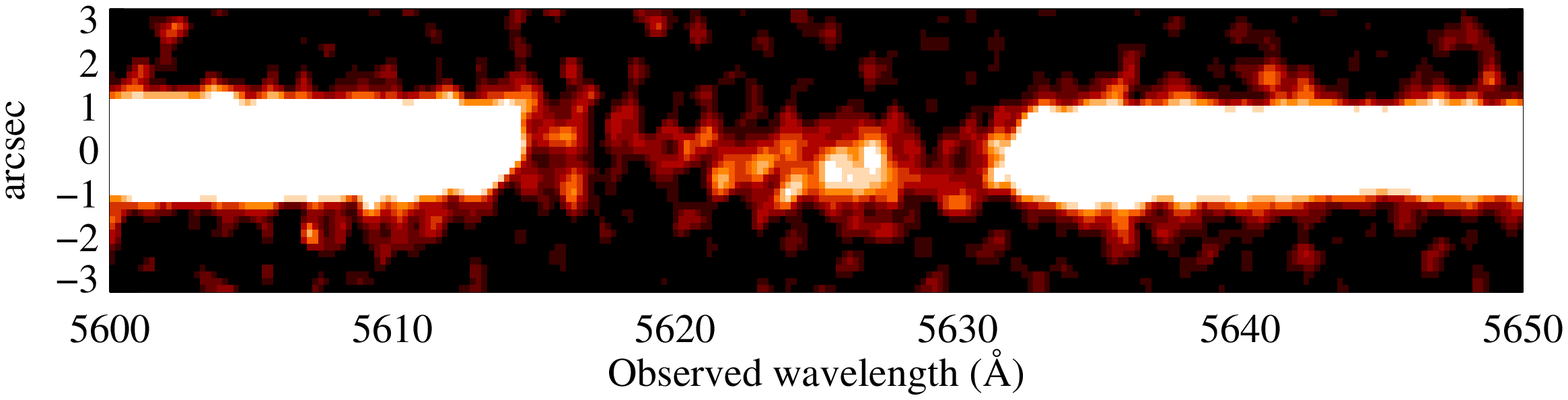}\\
    \includegraphics[width=0.95\hsize]{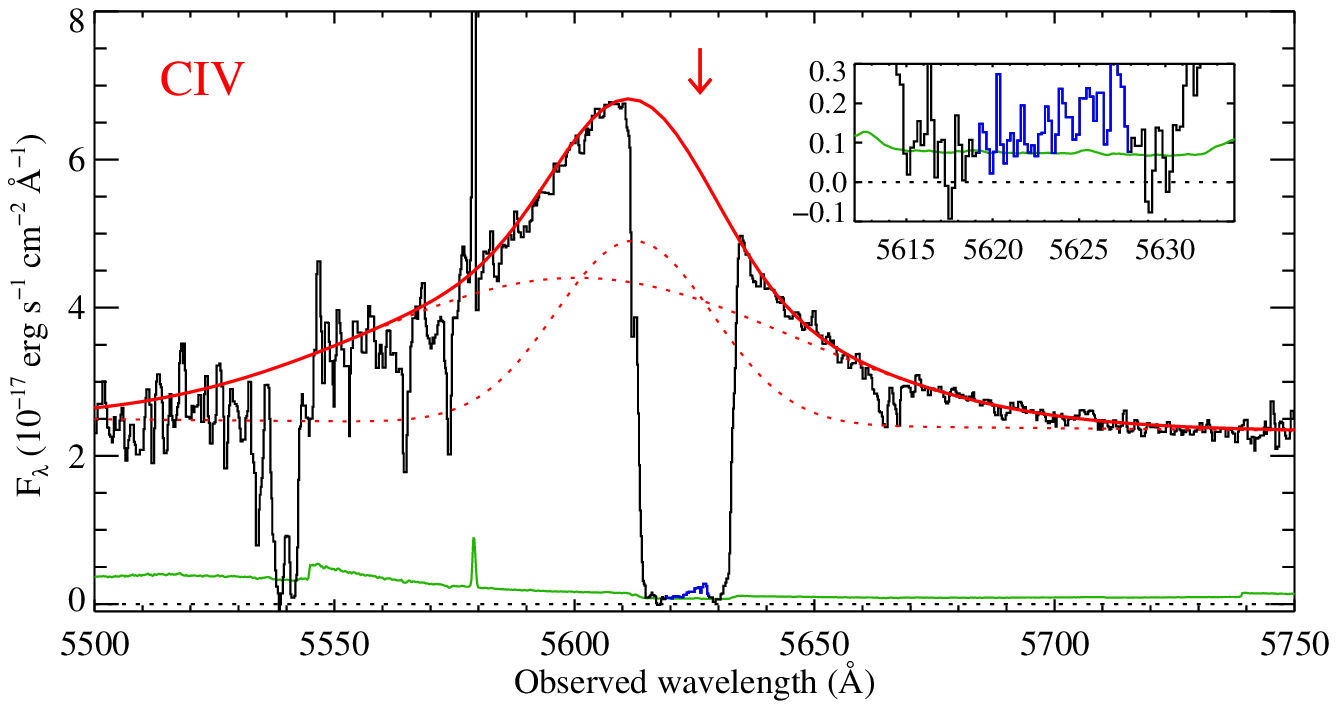}
    \end{tabular}
    \caption{{Bottom:} Portion of {1D} X-shooter spectrum around the \CIV\ emission line, median-smoothed using a 5-pixel sliding window for presentation purposes {(black, with $1\,\sigma$ error level in green)}. The inset shows an unsmoothed zoom-in on the \CIV\ absorption, with the blue region highlighting the residual emission.  The best-fit model is shown by the red solid line and the individual Gaussians as red dotted lines. The downwards arrow indicates the systemic redshift. {Top: 2D spectrum in the region featuring the \CIV\ absorption line as well as residual \CIV\ emission in the trough.}}
    \label{f:CIV1D}
\end{figure}

The region of 1D spectrum around the \CIV\ emission line is shown in Fig.~\ref{f:CIV1D}. Strong absorption from the proximate DLA (PDLA) is observed in the red wing of the emission line. We fitted the emission line using the sum of two Gaussians plus a linear continuum in that region, excluding the region affected by absorption. The \CIV\ emission peaks at $\sim 5611$\,{\AA}, while the \CIV\ absorption is centred at $\sim 5625$\,{\AA}. Furthermore, the \CIV\ emission is shifted by about 700~\kms\ bluewards of the systemic redshift but matches the location of the broad [\OIII] component well. 
Interestingly, while the shape and strength of the \CIV\ absorption profile indicate strong saturation, a residual emission is seen in the trough, with a possible peak at the systemic velocity. Even though the value for each pixel remains relatively close to the noise level, the residual emission is consistently seen above zero for almost all pixels, with integrated flux $F_{\rm resi} \sim (1.3\pm0.1)\times 10^{-17}$~erg\,s$^{-1}$\,cm$^{-2}$, that is, 13\,$\sigma$ significance.
In principle, we cannot rule out that this is due to the absorption profile being composed of a blend of many narrow unsaturated components. We note, however, that reproducing the sharp edges and the width of the profile while keeping the non-zero absorption in the centre is very difficult, even more so with the condition that the \CIV\ column density be bounded by the metallicity. A more likely explanation is that the \CIV\ emission is not fully covered by the \CIV\ absorbing gas. This is corroborated by the spatial extent of the residual emission,
which is discussed in the following sub-section.

{}

\subsection{Spatial extent of line-emitting regions \label{s:spa}}

\begin{figure}
    \centering
    \begin{tabular}{c}
    \includegraphics[bb=70 220 490 400,clip,width=0.8\hsize]{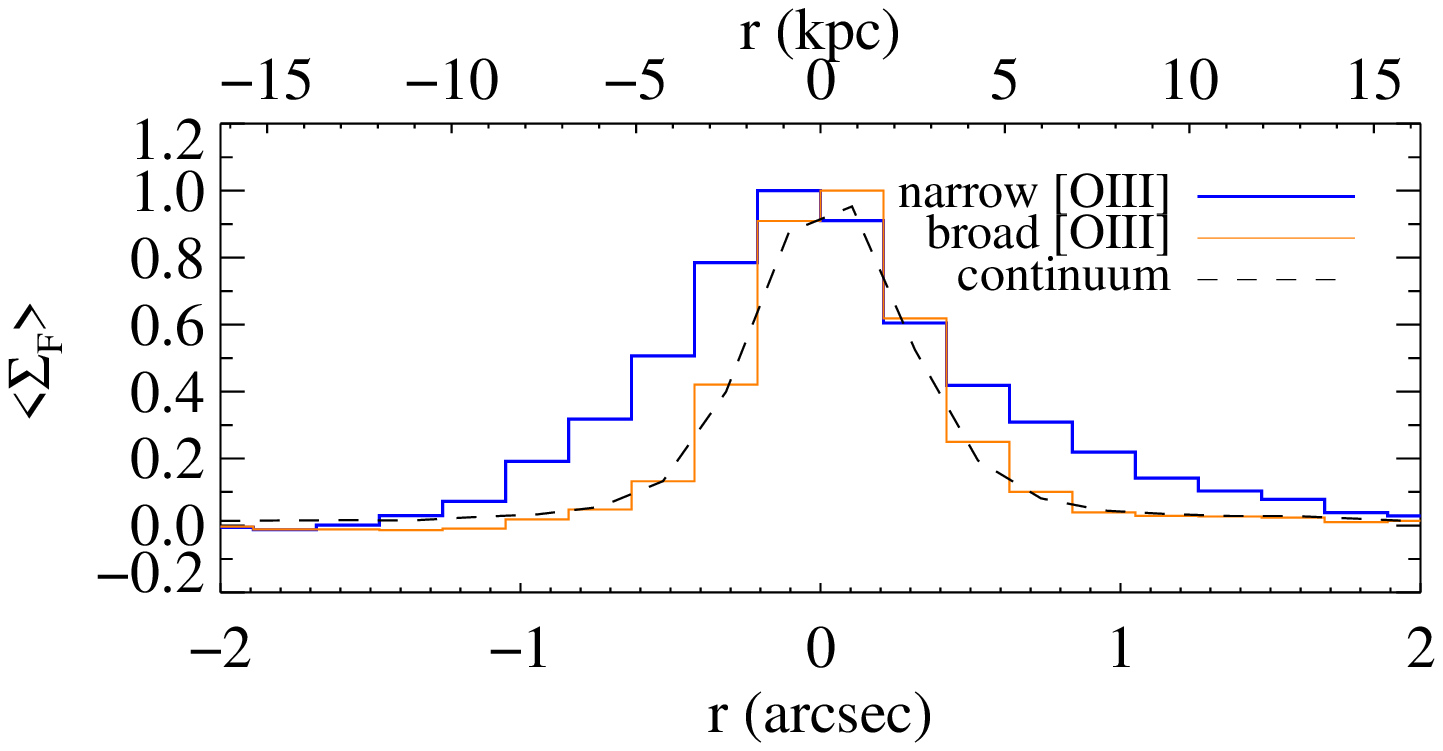}\\
    \includegraphics[bb=70 220 490 360,clip,width=0.8\hsize]{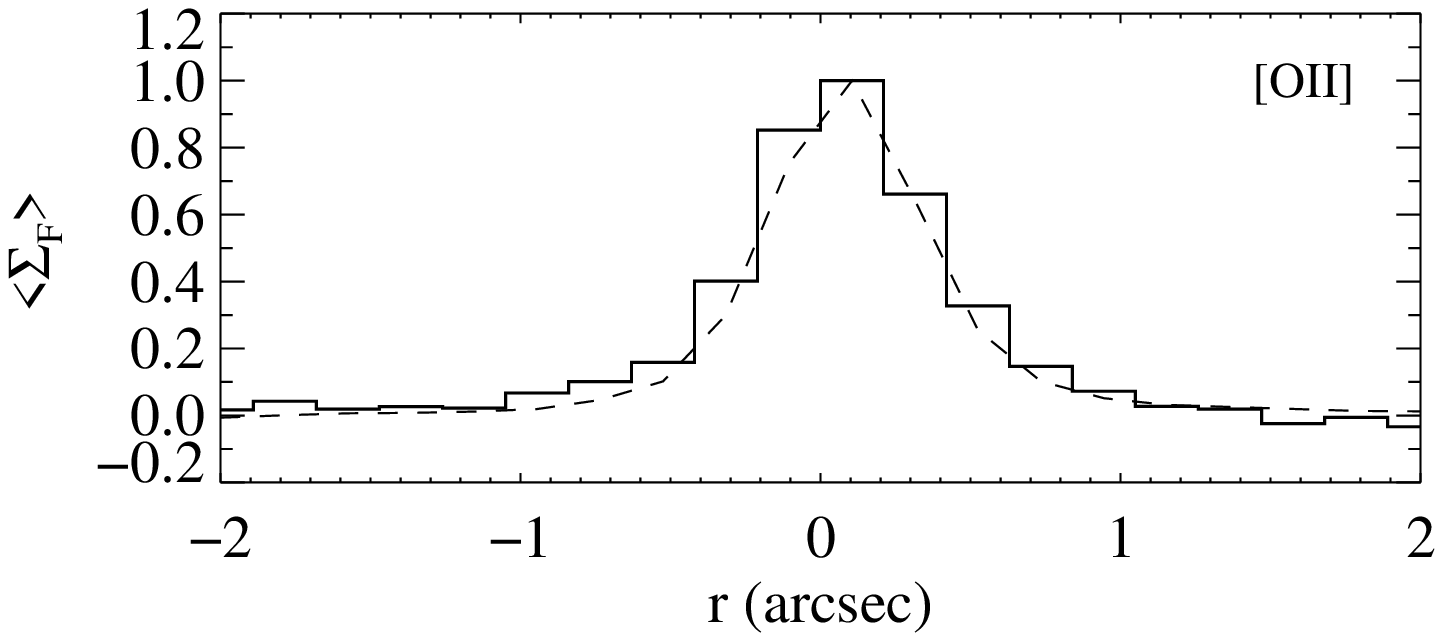}\\
    \includegraphics[bb=70 220 490 360,clip,width=0.8\hsize]{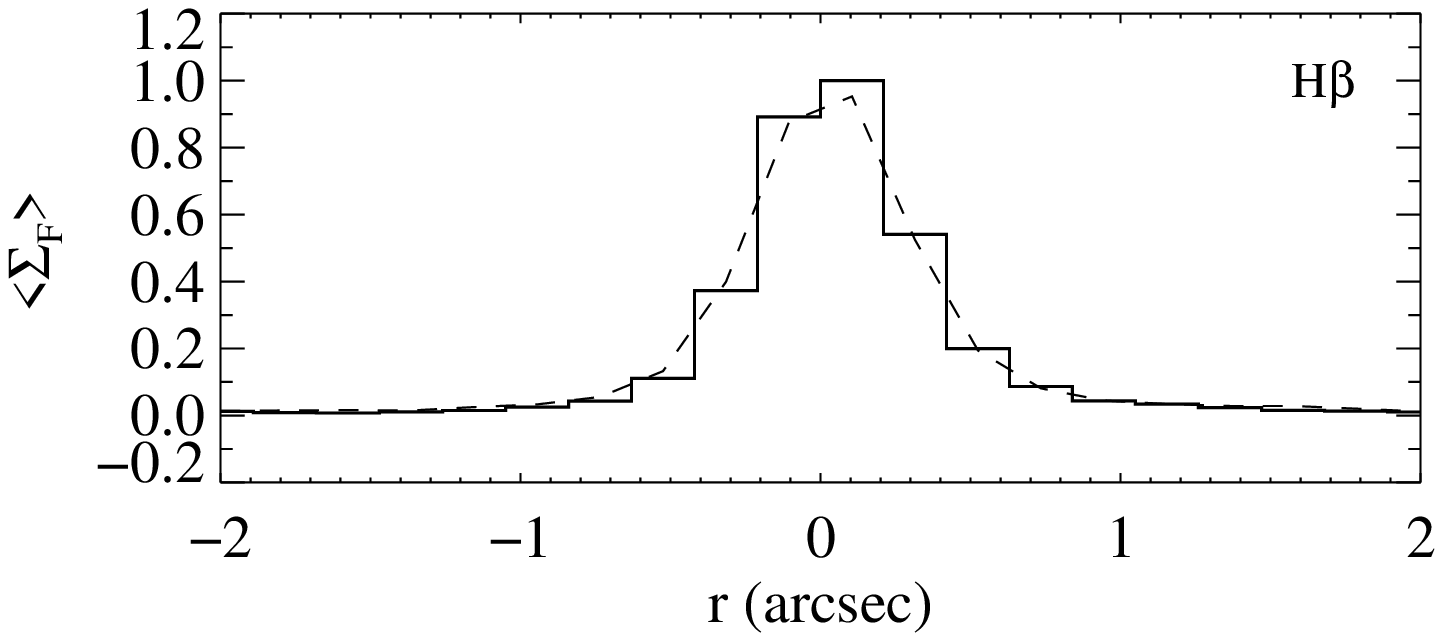}\\
    \includegraphics[bb=70 175 490 360,clip,width=0.8\hsize]{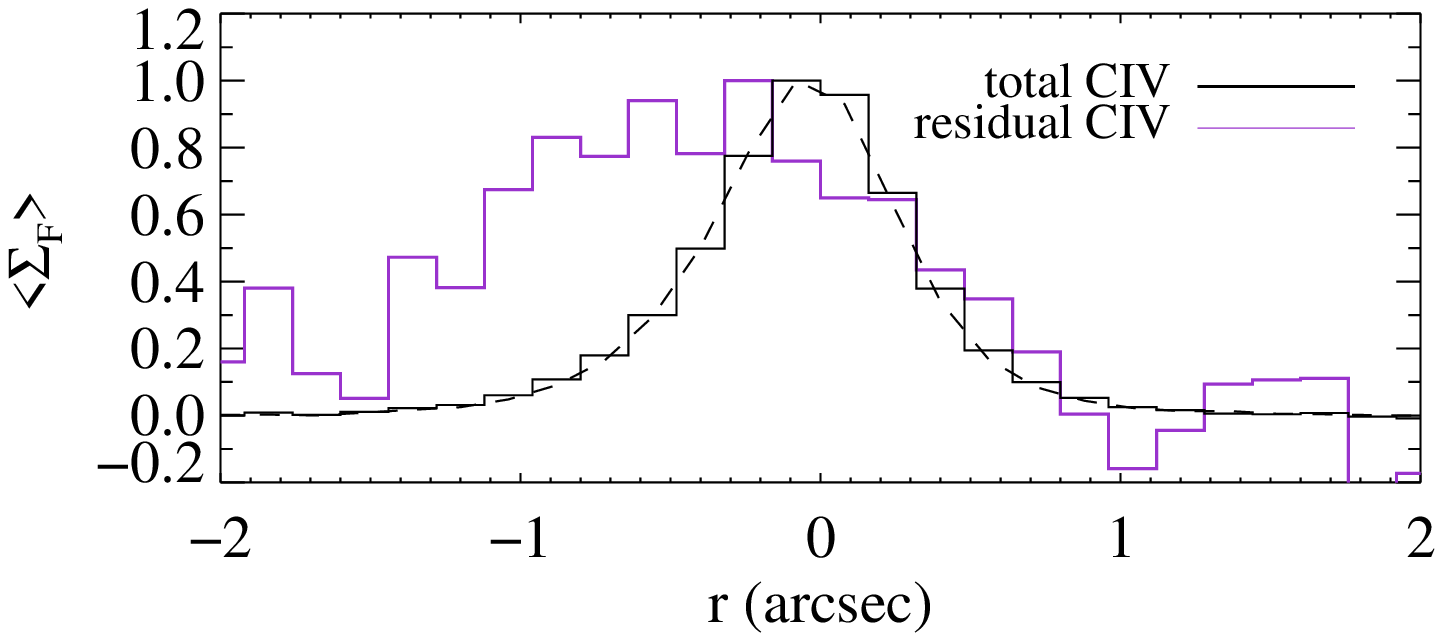}\\
    \end{tabular}
    \caption{Average observed spatial profiles at the location of different emission lines. In each panel, the dashed curve shows the spatial point spread function obtained from the quasar continuum in a nearby region. We note that the profile around the narrow [\OIII] emission still contains a contribution from the continuum and the broad [\OIII] emission. For the sake of easy comparison, each profile is normalised by its maximum value. 
    \label{f:spa}}
\end{figure}

 The spatial profiles of the [\OII], [\OIII], \Hb\ and \CIV\ emission lines are compared to that of the continuum in Fig.~\ref{f:spa}. These were obtained by collapsing the combined 2D spectrum along the wavelength axis around each line, while rejecting outlying pixels (due to telluric residuals or bad pixels). The regions for collapsing the data were chosen as follows. 
 For [\OII] and H$\beta$, we collapsed the data in a region centred at the peak of each line using a width matching the measured FWHM.
 Since the [\OIII] profile is very asymmetric, we considered {two regions representative of} the broad and narrow components separately.
 {To investigate the spatial profile around} the narrow [\OIII] component, we again used a region centred at the peak position within one FWHM. However, to assess whether the photons from the broad component come from compact (unresolved) or spatially extended regions, we restricted the region to 1.805--1.815 $\mu$m. This avoids contamination by the narrow component.  In the case of \CIV, we collapsed the data over the 5580--5650~{\AA} region, which roughly corresponds to the FWHM centred at the line peak. We also collapsed the data in the 5619--5628~{\AA} region, corresponding to the region of the \CIV\ absorption trough where some residual emission is seen {(Fig.~\ref{f:CIV1D})}. 
 
 We found that the spatial profiles of [\OII], \Hb\  the broad [\OIII], and the overall \CIV\ emission are unresolved, perfectly matching the spatial profile of the continuum and hence mostly originating from the nuclear region. 
 In turn, the narrow [\OIII] line is significantly extended, over several kpc, on both sides of the trace of the continuum. 
 The \CIV\ absorption trough acts like a natural coronagraph, removing the bright quasar glow at these wavelengths. This provides easier access to extended but fainter \CIV\ emission. {While it is hard to appreciate the spatial profile of the residual (weak) \CIV\ emission from the 2D data (top panel of Fig.~\ref{f:CIV1D}), when collapsing along the wavelength axis, and albeit still noisy,} this profile clearly presents a spatial kpc-scale extension below the trace.
 
 {}

\begin{figure*}
    \centering
    \addtolength{\tabcolsep}{-10pt}
    \begin{tabular}{cc}
      \includegraphics[bb=335 50 558 750,angle=90,width=0.52\hsize]{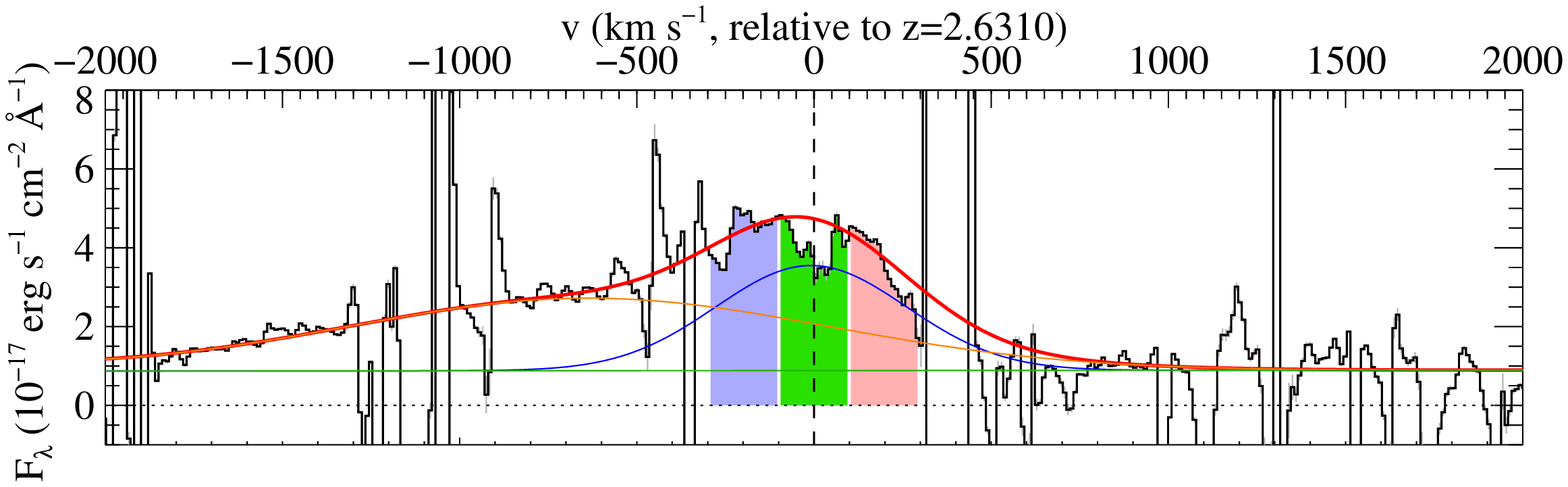} &\\
      \includegraphics[bb=425 50 558 750,clip=,angle=90,width=0.52\hsize]{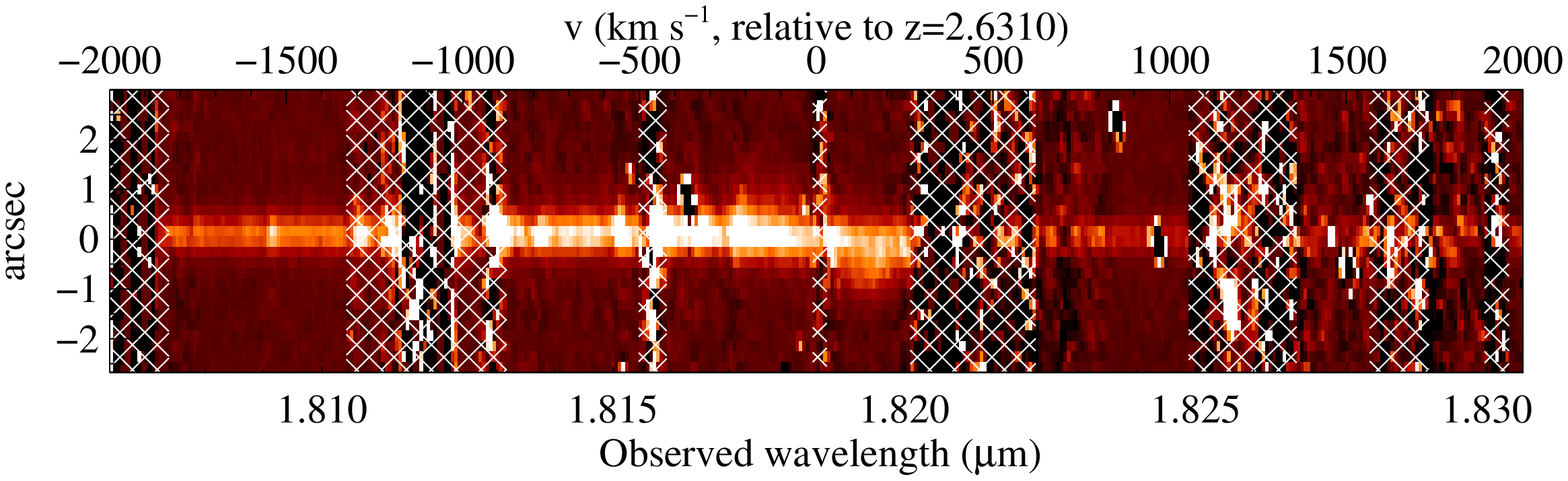}   &   
      \includegraphics[bb=425 30 558 730,clip=,angle=90,width=0.52\hsize]{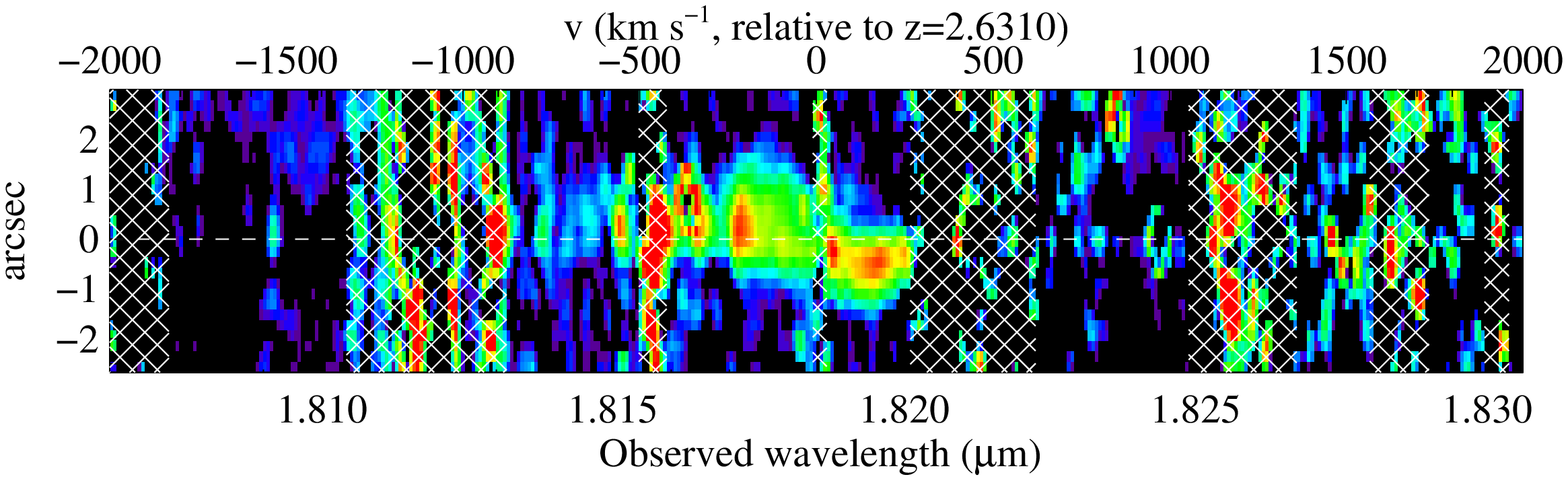} \\
      \includegraphics[bb=425 50 558 750,clip=,angle=90,width=0.52\hsize]{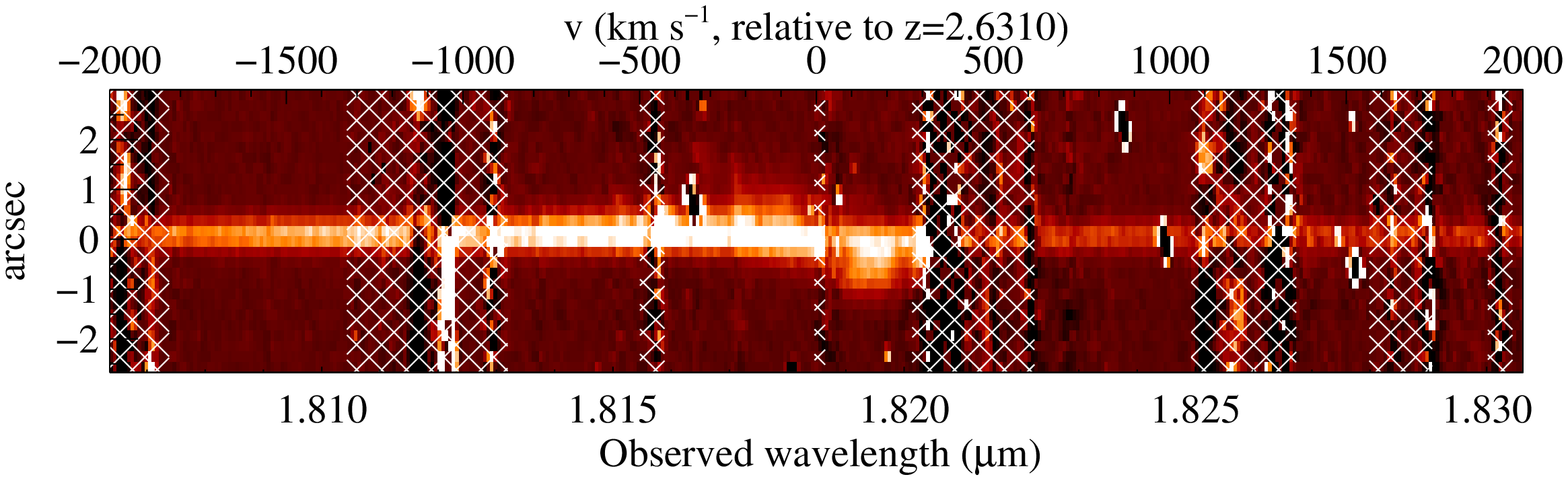}   &   
      \includegraphics[bb=425 30 558 730,clip=,angle=90,width=0.52\hsize]{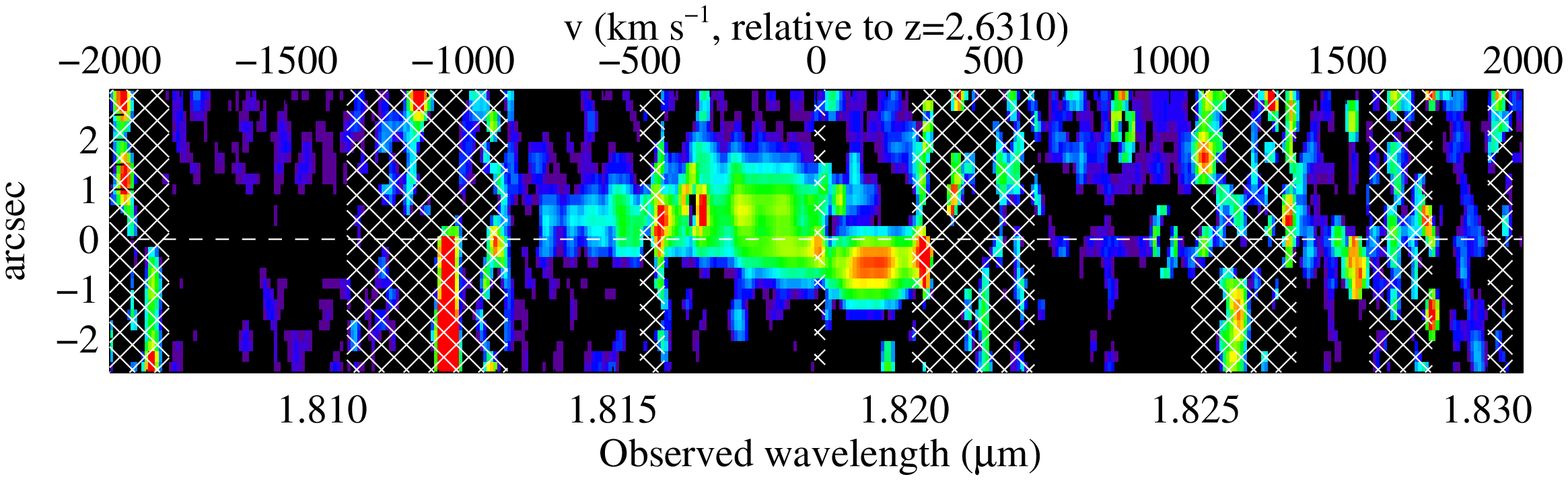} \\
      \includegraphics[bb=425 50 558 750,clip=,angle=90,width=0.52\hsize]{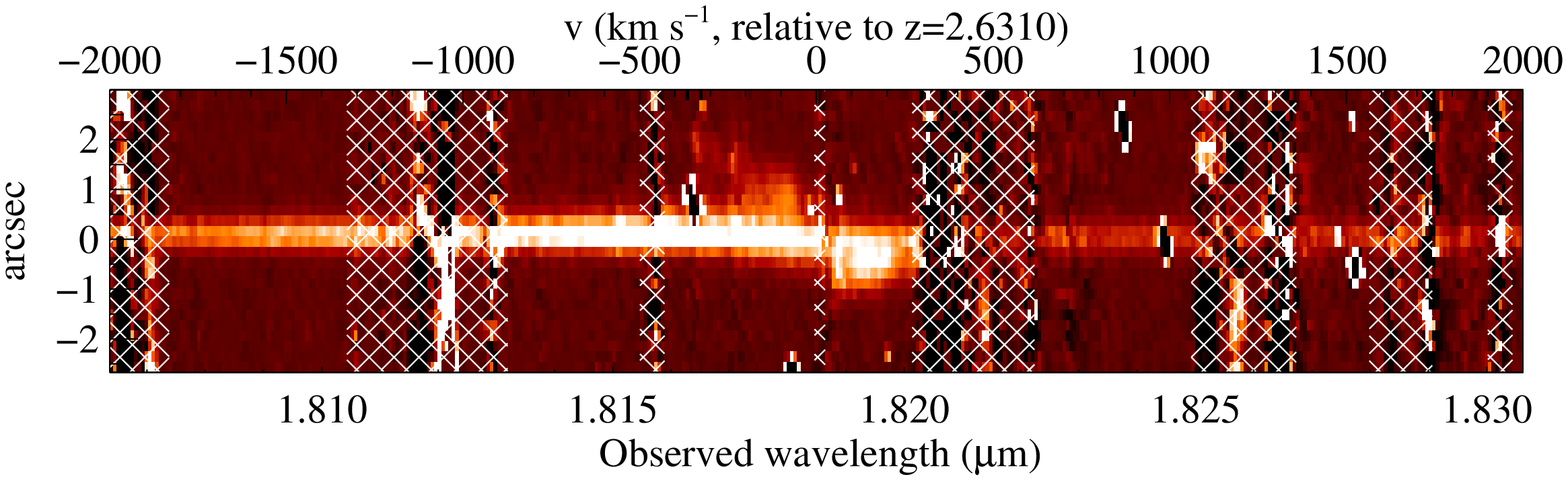}   &   
      \includegraphics[bb=425 30 558 730,clip=,angle=90,width=0.52\hsize]{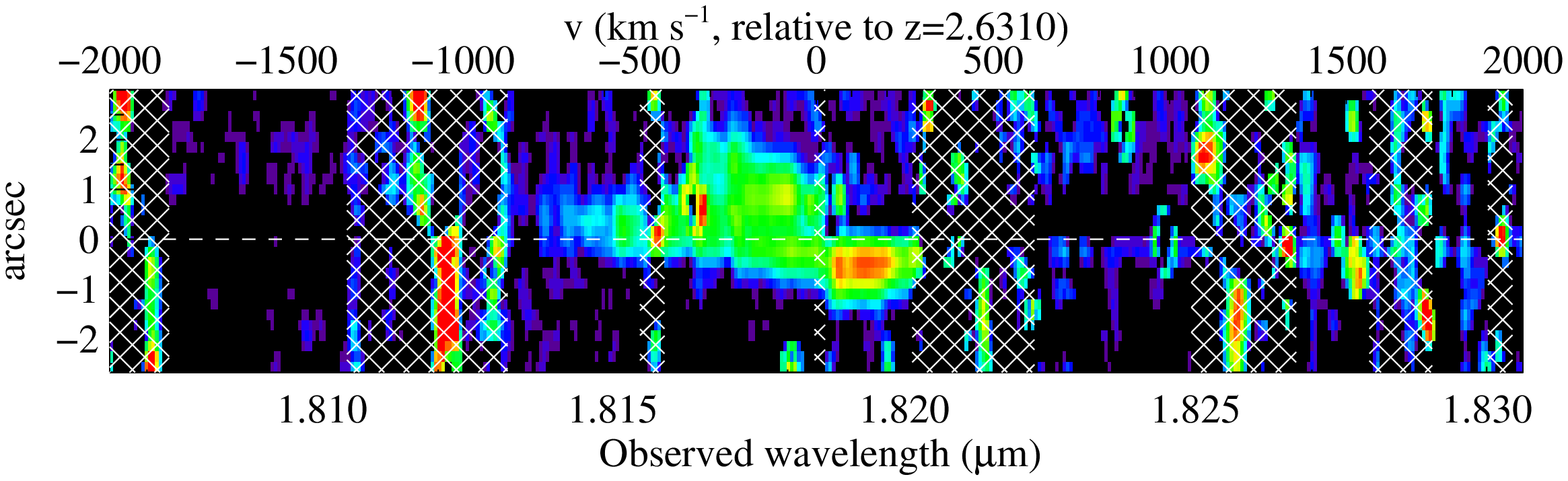} \\
      \includegraphics[bb=388 50 558 750,clip=,angle=90,width=0.52\hsize]{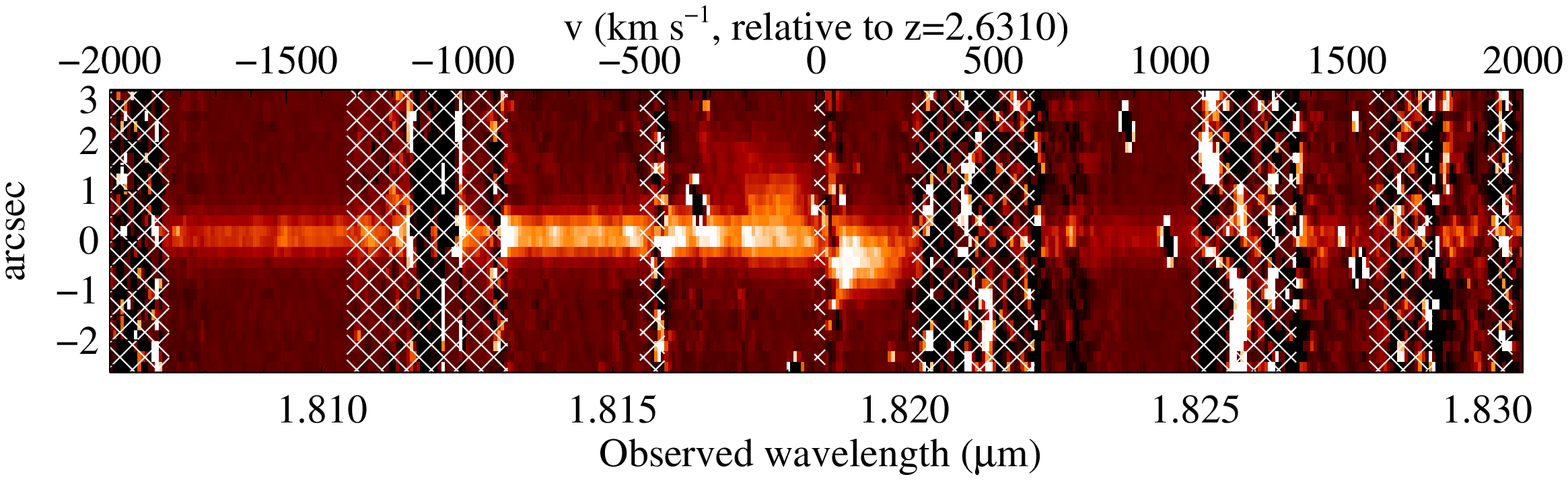}   &   
      \includegraphics[bb=388 30 558 730,clip=,angle=90,width=0.52\hsize]{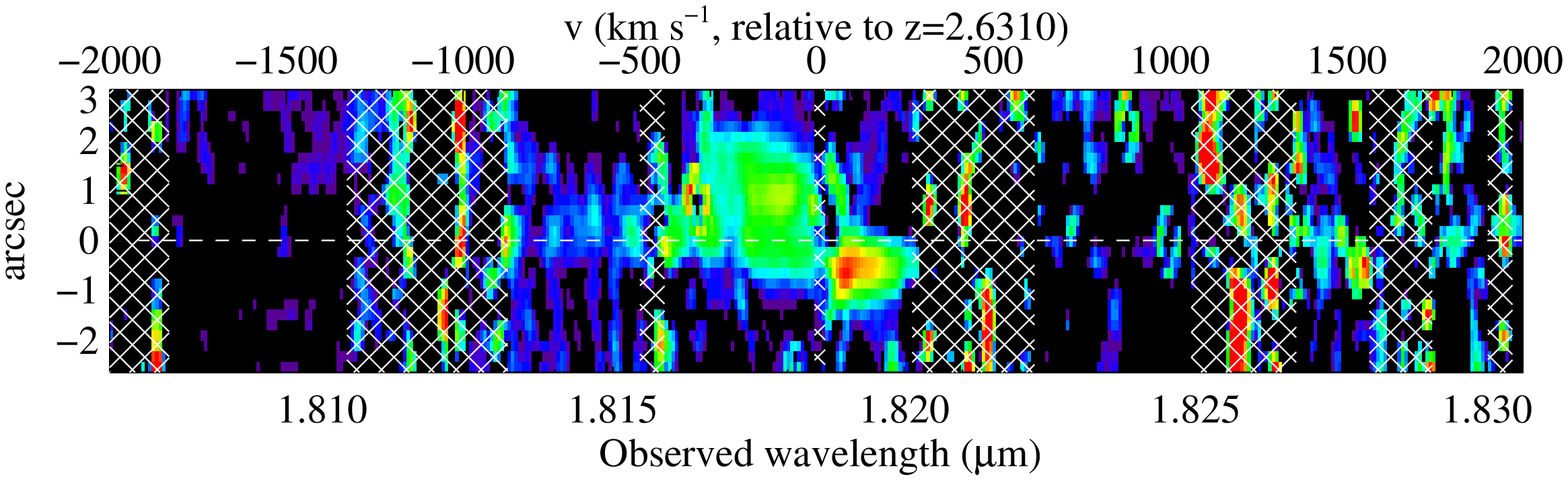} \\
    \end{tabular}{}
        \addtolength{\tabcolsep}{10pt}
    \caption{
    Individual 2D spectra around the [\OIII]$\lambda$5008 region (from top to bottom: PA~=~$-155\deg$, $-164\deg$ , $+177\deg$, and $+166\deg$ east of north, i.e. in these figures, north is down and south is up). 
    The total observed emission is shown in the left panels, while the unresolved emission (continuum+broad component) has been subtracted in the right panels to better highlight the spatially-extended narrow [\OIII] emission alone. Each panel is 4000~\kms\ wide. The main skyline residuals are masked out by a white hashed grid. The top-left panel shows the total extracted 1D spectrum, with the corresponding Gaussian fits. The coloured regions show the velocity bins used in Fig.~\ref{f:OIIIspa}.
    }
    \label{f:OIII2D}
\end{figure*}

In order to further characterise the extension of the narrow [\OIII] emission line, we show the individual 2D spectra in Fig.~\ref{f:OIII2D}, with and without subtraction of the smooth continuum and broad emission. 
The extended narrow [\OIII] emission appears as blobs both above and below the quasar trace, with very similar locations and extents for all position angles (PA). This is not very surprising given the small differences in PAs (due to the observations always being taken with the slit aligned to parallactic angle; see Fig.~\ref{f:slitpos}).  
However, we noticed that the observed fluxes are higher for PA~$=+177\deg$ and PA~$=-164\deg$ than for the other position angles. This suggests that slit losses could be at play for $\mathrm{PA} = 155 \deg$ and $+166 \deg,$ and hence the extended emission is oriented close to the north-south direction. 

{}

\begin{figure}
    \centering
    \includegraphics[bb=90 178 487 408,width=0.95\hsize]{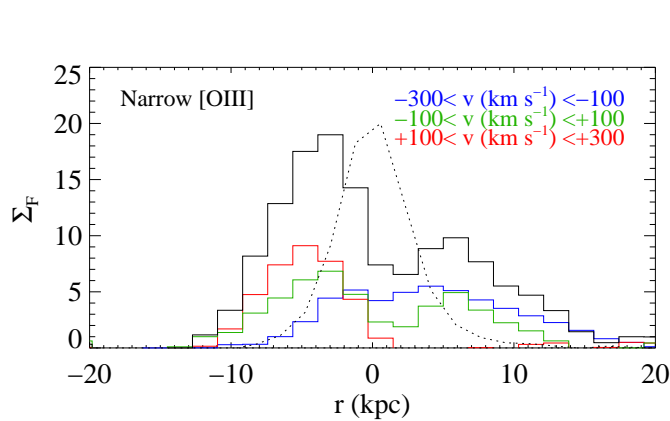}
    \caption{Total spatial profile of narrow [\OIII] emission (black) and in three velocity bins (with colours indicated in the legend) after subtracting the unresolved broad [\OIII] and continuum emission. The dotted line shows the spectral PSF obtained from the trace of the quasar continuum. 
    \label{f:OIIIspa}}
\end{figure}

In Fig.~\ref{f:OIIIspa}, we show the average spatial profile in three velocity bins around our reference redshift ($v\,(\kms) =[-300,-100]; [-100,+100]; [+100,+300]$) after removing the unresolved continuum+broad [\OIII] emission. 
The {centroids} of the both blue and red blobs are at projected distances $\sim$4--6~pkpc from the nuclear emission. In addition, the 
centre and the blue wing of the narrow [\OIII] emission extend from the nucleus up to large distances. 
While part of these spatial extents are due to the smearing effect of the seeing, it clearly indicates that [\OIII] emission extends over at least $\sim$10~pkpc from the central engine.
We remark that the [\OIII] emission extends above and below the trace for roughly negative and positive velocities (according to our reference redshift), respectively.

{}

\subsection{The Ly-$\alpha$ emission}
\label{s:lyaem}

\begin{figure}
    \centering
    \includegraphics[bb=90 178 487 408,width=0.95\hsize]{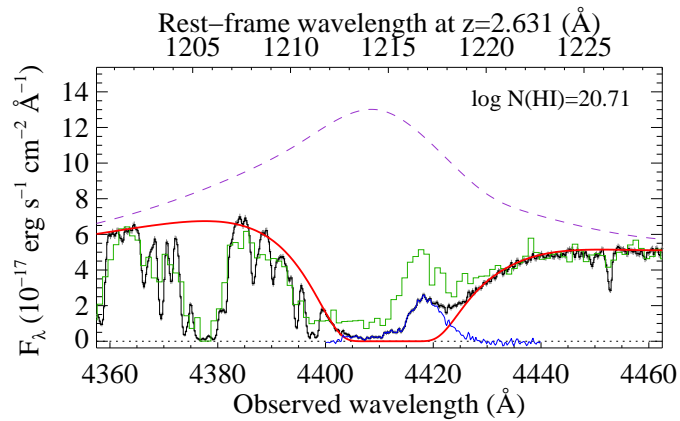}
    \caption{Decomposition of rest-frame \lya\ region. The observed on-trace X-shooter spectrum is shown in black with uncertainties as grey error bars. 
    The reconstructed unabsorbed quasar emission line is shown as the dashed purple curve. {The red curve is the combination of this unabsorbed emission together with the modelled DLA absorption profile with $\log N(\HI)=20.71$}. Subtracting the absorption profile from the intrinsic emission profile provides us with the residual emission shown in blue, over the 4400--4440~{\AA\ wavelength interval}. The BOSS fibre spectrum is shown in green.
    }
    \label{f:Lya}
\end{figure}

The spectrum of \J\ in the \lya\ region is the sum of the
quasar emission absorbed by the proximate DLA and the leaking \lya\ emission superimposed on 
the DLA trough, as already mentioned in our discovery work \citep{Noterdaeme2019}. 
One difficulty with the low-resolution, low-S/N BOSS fibre spectrum was to 
ascertain the combined model of absorption and emission since the quasar redshift was not easily determined and the wings of the DLA were not clearly detected. The higher quality X-shooter spectrum confirms our previous decomposition (see Fig.~\ref{f:Lya}). The absorption profile is well constrained by the damping wings of the DLA as well as other lines from the Lyman series, which were fitted simultaneously with H$_2$ lines (Fig.~\ref{f:H2}). The intrinsic quasar emission profile was reconstructed simultaneously using a smooth spline profile, {aided by the quasar template matching (see Sect.~\ref{s:dust})}; however, the exact shape of the peak is highly uncertain since it is positioned at the same place as the DLA trough. Nonetheless, we note that the exact shape and strength of the intrinsic, broad \lya\ emission line has little influence on the analysis presented in this paper. 

{}

\begin{figure*}
    \centering
    \addtolength{\tabcolsep}{-10pt}
    \begin{tabular}{cc}
            &   \includegraphics[bb=300 30 595 745,clip=,angle=90,width=0.5\hsize]{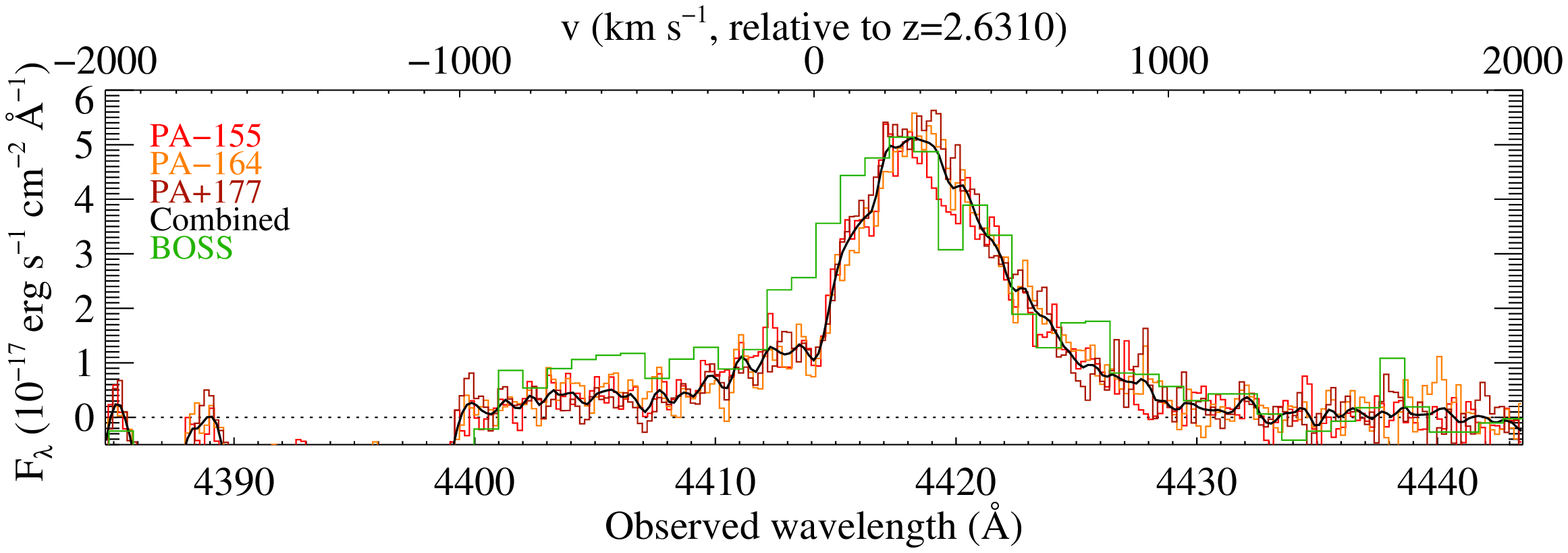} \\
      \includegraphics[bb=425 35 558 750,clip=,angle=90,width=0.5\hsize]{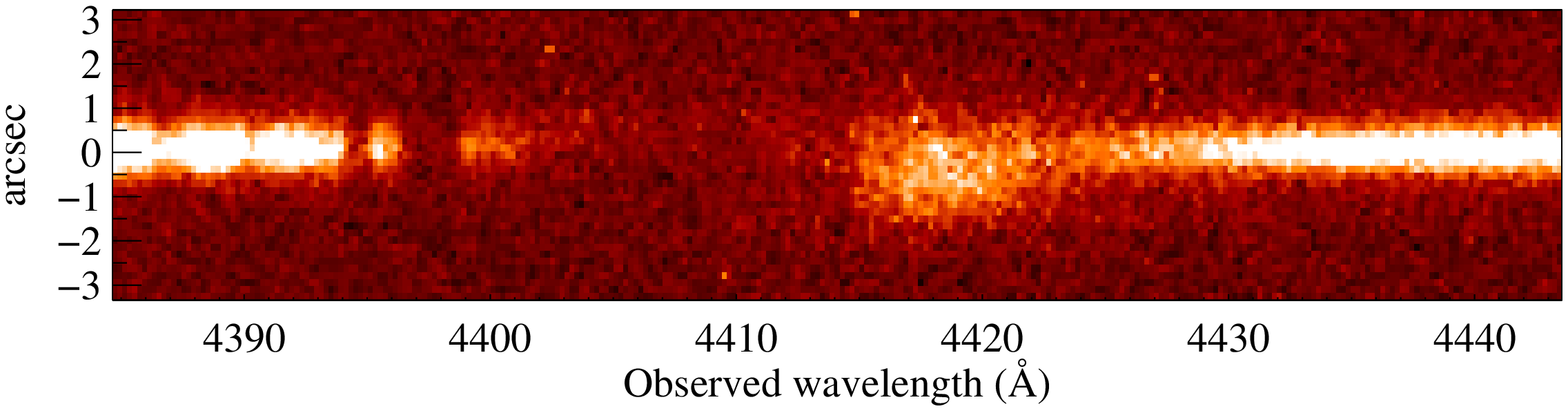}   &   
      \includegraphics[bb=425 30 558 745,clip=,angle=90,width=0.5\hsize]{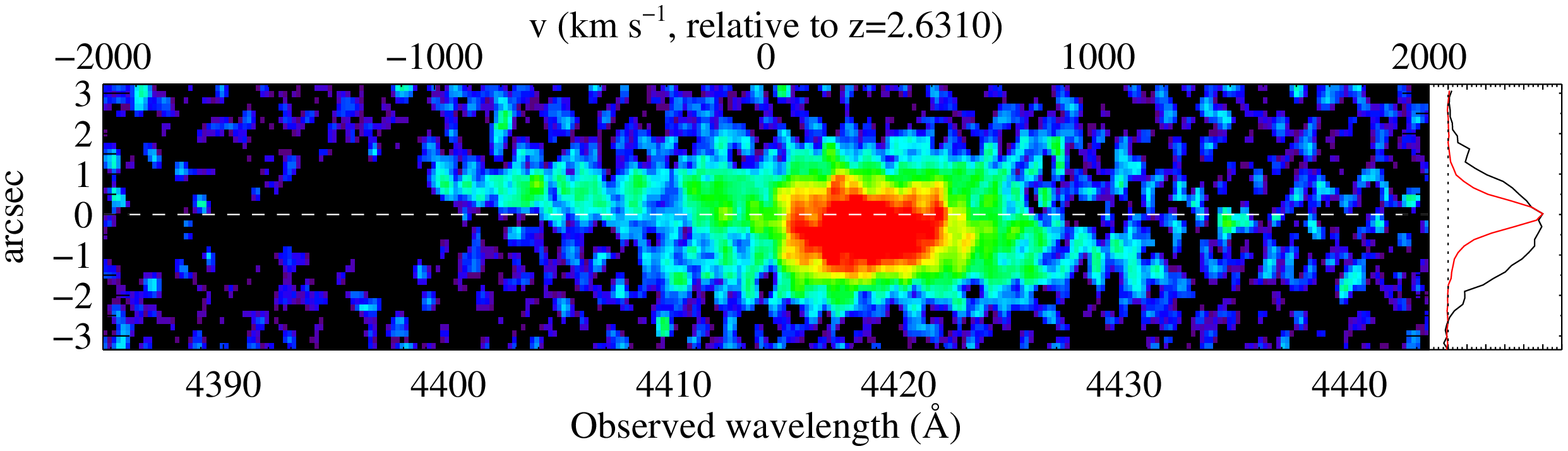} \\
      \includegraphics[bb=425 35 558 750,clip=,angle=90,width=0.5\hsize]{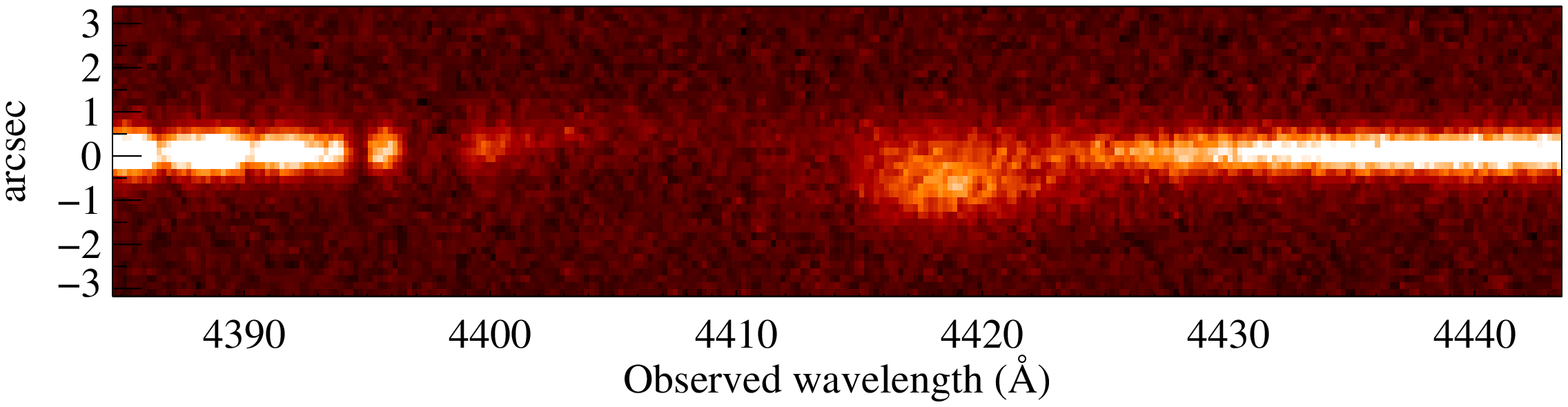}   &   
      \includegraphics[bb=425 30 558 745,clip=,angle=90,width=0.5\hsize]{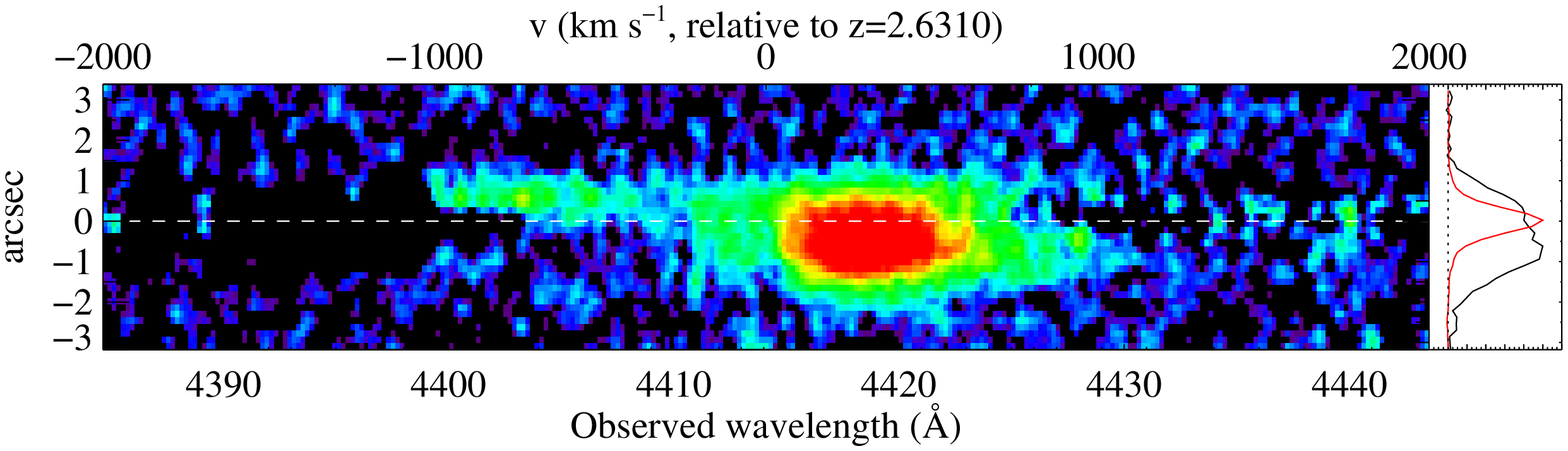} \\
      \includegraphics[bb=388 35 558 750,clip=,angle=90,width=0.5\hsize]{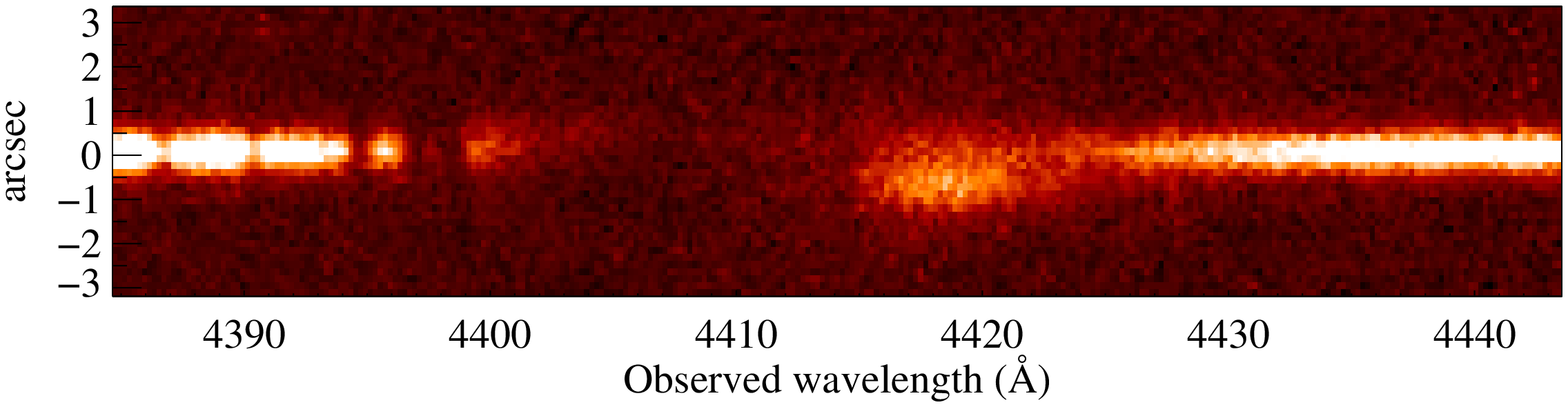}   &   
      \includegraphics[bb=388 30 558 745,clip=,angle=90,width=0.5\hsize]{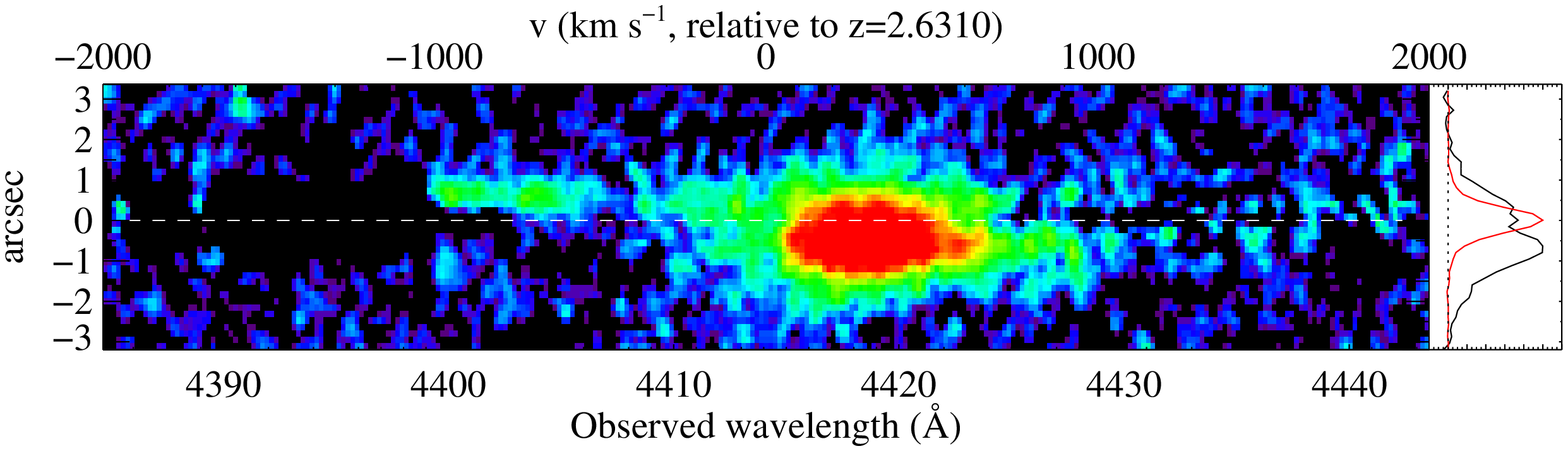} \\
    \end{tabular}{}
        \addtolength{\tabcolsep}{10pt}

    \caption{\lya\ emission in PDLA core before (left) and after (right) subtraction of the unresolved quasar continuum emission. From top to bottom: PAs~=~$-155$, $-164$ and $+177\deg$.
    The top-right panel shows the spectral profile of the extended \lya\ emission for each position angle as well as their average, collapsing over the full spatial range shown in the 2D panels (hence an aperture of $6\times1$ arcsec). The SDSS-III/BOSS spectrum is shown in green. The small side panels show the spatial profiles of the unresolved quasar continuum emission (red) and the extended \lya\ emission (black). 
    }
    \label{f:Lya2D}
\end{figure*}

In Fig.~\ref{f:Lya}, we also show the original BOSS spectrum. The residual \lya\ emission appears nearly two times stronger in the BOSS spectrum than in the X-shooter spectrum. 
The reason is that the optimal extraction method used for the X-shooter spectrum gives lower weight to the extended and spatially offset \lya\ emission. 
Instead, the 2D spectra clearly show that the residual \lya\ emission is spatially offset and extended (Fig.~\ref{f:Lya2D}), {and the overall (non-optimal) X-shooter extraction does match the SDSS data (top-right panel of that figure) much better. This will be discussed in the following.}

Since the system was identified through its H$_2$ absorption lines 
imprinted on the quasar continuum, the absorbing gas intercepts the line of sight to the central engine, removing the unresolved emission and 
leaving us with a perfect PSF subtraction of the {\sl nuclear} emission over the 
saturated region ($\sim 4405-4420$~{\AA} observed) of the DLA trough. In other words, the DLA acts as a natural coronagraph \citep[see e.g.][]{Finley2013,Jiang2016}.
However, the leaking Ly-$\alpha$ emission remains partly blended with light from the quasar nucleus in the wings of the DLA (mostly in the red wing, $\lambda_{\rm obs} \sim 4420$--$4430~{\AA}$). 
We therefore subtracted the nuclear emission over a wider region in order to isolate the extended emission. For this purpose, we constructed a 2D model of the nuclear emission using a Moffat profile to describe the spatial point spread function (SPSF). The shape and position of the SPSF is constrained by the continuum emission far from the \lya\ region. The amplitude of the spatial profile is then multiplied by the combined 1D spectral model for the nuclear emission and Ly-$\alpha$ absorption (i.e. the red line in Fig.~\ref{f:Lya}). 
The resulting core-subtracted 2D spectra are shown in the right-hand panels of Fig.~\ref{f:Lya2D}. The right-most panels indicate the corresponding spatial profiles for the nuclear (red) and residual (black) emission.

In the top-right panel of Fig.~\ref{f:Lya2D}, we show the collapsed 1D spectra of the extended Ly-$\alpha$ emission over the full spatial range presented in the 2D panels (i.e. $\sim 6$ arcsec). The individual spectra for the different position angles are all consistent with one another, indicating that differential slit losses are negligible.
Given the observed north-south extension of the emission, the narrow slit width (1~arcsec in UVB), and the position angles of the slits (see Fig.~\ref{f:slitpos}), this suggests that most of the extended \lya\ emission falls within the region covered by all three slits, with an offset of $\sim$0.7~arcsec from the nuclear emission, as seen from the side panels of Fig.~\ref{f:Lya2D}.
Most of the extended \lya\ emission then corresponds to the same location as the red component of the narrow [\OIII] emission. However, the \lya\ flux remains slightly higher in the BOSS spectrum (2~arcsec diameter fibre), in particular bluewards of the \lya\ peak. We note that shifts in the zero level have been observed in BOSS spectra, but on much lower levels. Furthermore, other saturated lines in the BOSS spectrum do not show any residual flux. This means that the excess emission at $\sim 4400$--$4420$~{\AA} in the BOSS spectrum compared to the X-shooter one is likely real, suggesting that some extended emission is not covered by the X-shooter slits. The emission could then extend beyond 5~kpc from the nucleus in the direction perpendicular to the slits.

{}

\section{Absorption line analysis} \label{s:abs}

We analysed the absorption lines using standard multi-component Voigt profile fitting and 
vpfit v10.3 \citep{Carswell2014} to obtain redshifts, Doppler parameters, and column densities of different species. We used both the UVB and VIS spectra with resolving powers of about 5400 and 8900, respectively. While we fitted all absorption lines simultaneously, the description of the analysis is split into sub-sections for convenience. 

\subsection{Neutral atomic hydrogen}

{The total \HI\ column density was obtained not only from fitting the damped Ly-$\alpha$ 
absorption together with the unabsorbed quasar continuum (see Sect.~\ref{s:lyaem}), but also by including other lines from the Lyman series, the central redshift of the overall \HI\ absorption also being  
constrained by the metal lines.}
We obtained a total column density $\log N(\HI)=20.71\pm 0.02$, consistent with the previously estimated value from the low-resolution, low-S/N BOSS data. The region where extended emission is seen was ignored during the fitting process.
The result of the fit is shown in Fig.~\ref{f:Lya}.
\subsection{Molecular hydrogen content}
\label{s:abs:h2}

\begin{figure*}
    \centering
    \includegraphics[bb=20 19 595 770,clip=,width=0.97\hsize]{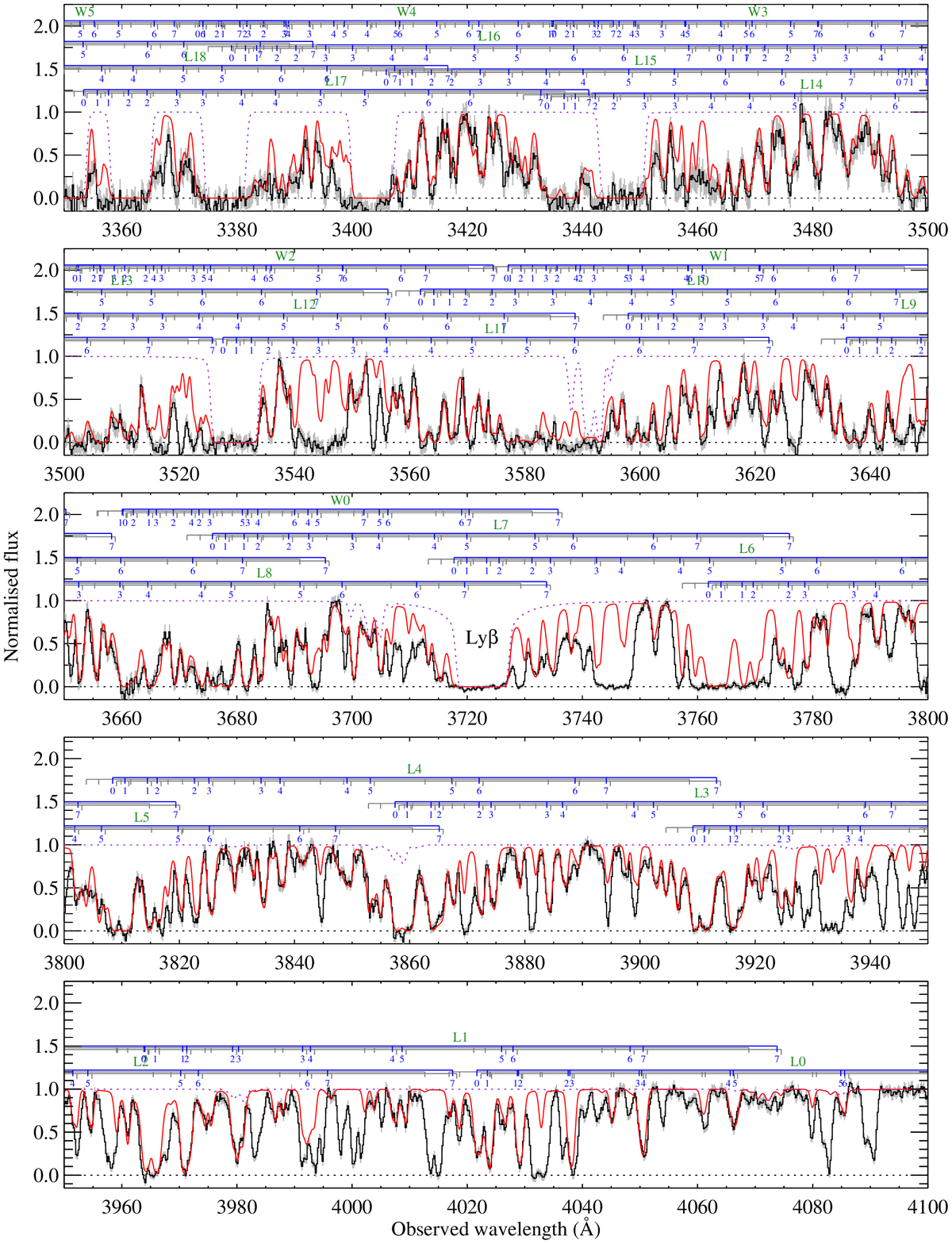}
    \caption{Portion of X-shooter UVB spectrum of J0015+1842 (black, with uncertainties in grey). Horizontal blue segments connect rotational levels (short tick marks) from a given Lyman (L) or Werner (W) band of H$_2$ for the central H$_2$ component. The same is shown in grey for the two other components, but they are not labelled for visibility. The overall model (MCMC-based) spectrum is over-plotted in red, with contribution from non-H$_2$ lines (\HI\ and metal lines) shown as a dotted purple line. 
    }
    \label{f:H2}
\end{figure*}

\begin{figure}
    \centering
         \includegraphics[bb=165 230 445 630,clip=,angle=90,width=0.55\hsize]{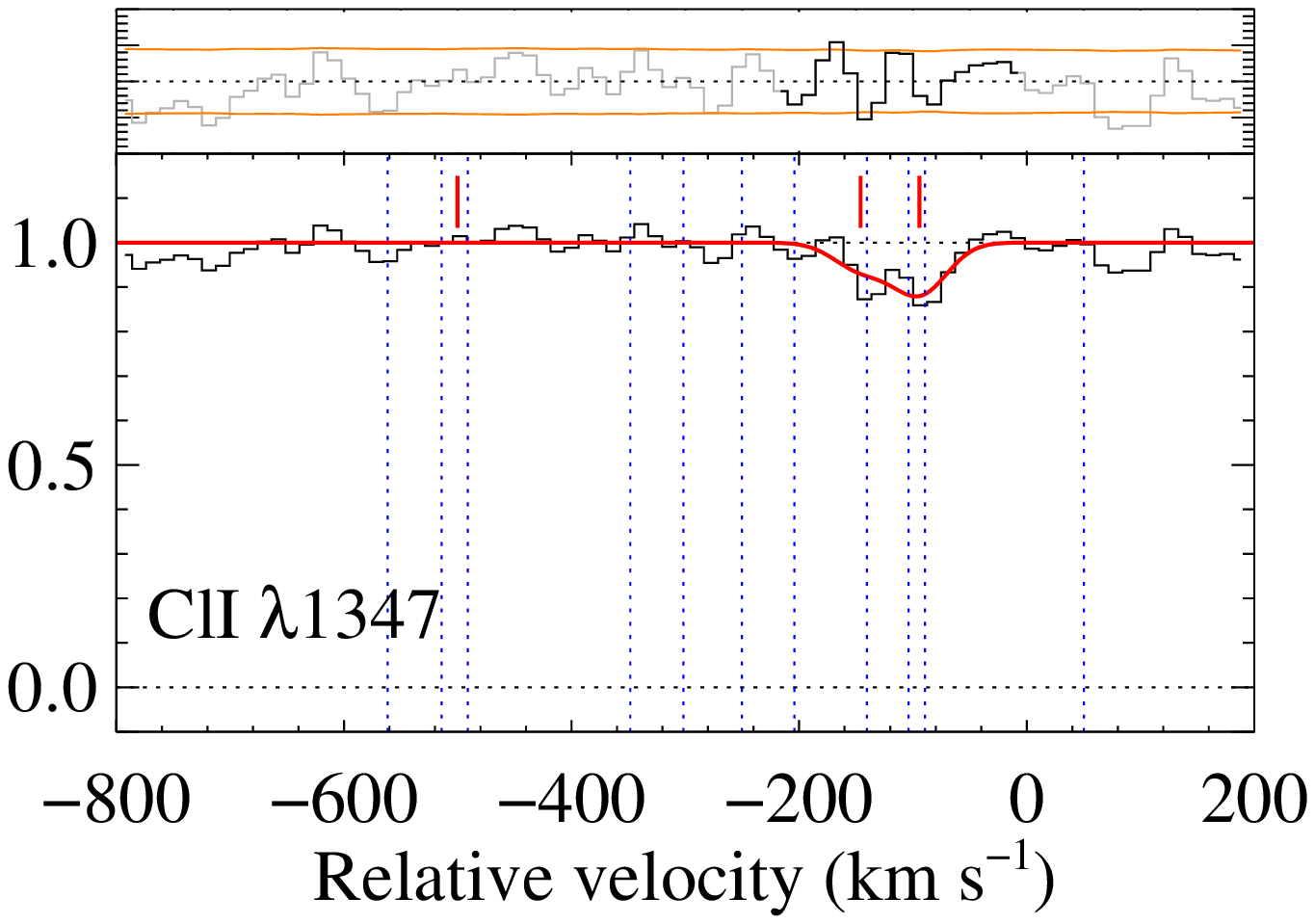} 
    \label{f:ClI}
    \caption{Fit to neutral chlorine absorption line. The observed data are shown in black, with the best-fit profile shown in red. \textit{Top panel:} residuals (black) and 1\,$\sigma$ error level (orange). The short tick marks show the location of the H$_2$ components, while long dashed blue lines show those of other metal lines. As for all similar figures in this work, the origin of the velocity scale is set to $z_{ref}=2.631$. 
    }
\end{figure}

Bluewards of $\lambda_{\rm obs}=4100~{\AA}$, the spectrum is crowded with numerous H$_2$ lines from different rotational levels (Fig.~\ref{f:H2}), confirming the detection from \citet{Noterdaeme2019} based on a low-resolution, low-S/N BOSS spectrum. 
We modelled the H$_2$ absorption profile using three components (which we label A, B, and C from blue to red). 
The strongest two components (B and C) are located close to each other in velocity space, so their respective absorption lines are strongly blended. Notwithstanding, these two components are also seen in the neutral chlorine profile (with total $\log N(\ClI)\approx 13.4\pm0.1$, see Fig.~\ref{f:ClI}), which is known to be chemically linked to H$_2$
(see e.g. \citealt{Balashev2015} and references therein).

We first used standard multi-component fitting using vpfit, including rotational levels up to $J=5$.  Because of the decreasing S/N and increasing line blending towards the blue, only the data at $\lambda > 3600$~{\AA} was used to constrain the fit of the molecular hydrogen lines. We tied together the redshifts and Doppler parameters between rotational levels for a given component. This implicitly assumes that H$_2$ in different rotational levels is well mixed, which is not necessarily the case. Indeed, an increase of $b$ with $J$ has been observed with VLT/UVES in a few cases of relatively low H$_2$ column densities \citep[e.g.][]{Noterdaeme2007b, Balashev2009, AlbornozVasquez2014}. However, in this work, the fit to the lower rotational levels is independent of the exact $b$ value since the lines are damped. 

Given the complexity of the H$_2$ absorption spectrum and to account for the possibility of different Doppler parameters, which can have an effect on the high-$J$ levels, we also fitted the H$_2$ lines with a Markov chain Monte Carlo (MCMC) procedure. 
During this fit, we again tied the Doppler parameters of $J=0$ and $J=1$, as these levels are mostly co-spatial, but those from other rotational levels were allowed to vary independently. However, we added a penalty function to the likelihood function to favour the solutions where Doppler parameters increase with $J$. We also included the H$_2$ absorption from $J=6$ and $J=7$ levels: these are albeit marginally detected.
To get posteriors on fitting parameters, we used an affine sampler \citep{Goodman2010} with a flat prior on $\log N$, $b,$ and $z$.
 The result from the MCMC fit is shown in Fig.~\ref{f:H2}, but we note that the standard vpfit procedure provides a visually almost indistinguishable profile. Indeed, higher spectral resolution remains necessary to ascertain the velocity structure and widths of the lines. Hence, we provide the best-fit parameters from both fitting methods in Table~\ref{t:H2}.

{}

\begin{table*}
\begin{center}
\caption{Result of Voigt profile fitting to molecular hydrogen lines.\label{t:H2}}
\begin{tabular}{cccccccc}
\hline \hline
{\Large {\strut}}    &  & \multicolumn{2}{c}{A, $z=2.6249$} & \multicolumn{2}{c}{B, $z=2.6292$} & \multicolumn{2}{c}{C, $z=2.6299$} \\
& & vpfit & MCMC & vpfit & MCMC & vpfit & MCMC \\
\hline
{\Large \strut}$J$=0 & $\log N $ & 17.91$\pm$0.04 &  $18.15^{+0.04}_{-0.06}$& 18.90$\pm$0.05 & $18.97^{+0.06}_{-0.04}$ & 18.76$\pm$0.10 & $18.94^{+0.06}_{-0.10}$ \\
{\Large \strut}& $b~(\kms)$ & 2.0$\pm$0.4 &  $1.3^{+0.1}_{-0.2}$ & 2.7$\pm$0.3 & $1.3^{+0.1}_{-0.2}$ & 5.5$\pm$0.3 &  $1.4^{+0.1}_{-0.3}$ \\
{\Large \strut}$J$=1 & $\log N $ & 18.49$\pm$0.04 & $18.67^{+0.03}_{-0.03}$ & 19.27$\pm$0.05 & $19.25^{+0.05}_{-0.05}$ & 19.00$\pm$0.07 & $19.12^{+0.07}_{-0.05}$ \\
{\Large \strut}& $b~(\kms)$ & $\tablefootmark{a}$ & $\tablefootmark{a}$ & $\tablefootmark{a}$ & $\tablefootmark{a}$ & $\tablefootmark{a}$ & $\tablefootmark{a}$\\
{\Large \strut}$J$=2 & $\log N $ & 17.91$\pm$0.08 & $18.18^{+0.03}_{-0.10}$ & 18.32$\pm$0.07 &  $18.20^{+0.11}_{-0.13}$ & 18.11$\pm$0.12 & $18.42^{+0.09}_{-0.06}$\\
{\Large \strut}& $b~(\kms)$ & $\tablefootmark{a}$ & $1.6^{+0.8}_{-0.3}$ & $\tablefootmark{a}$ &  $4.9^{+0.7}_{-0.4}$ & $\tablefootmark{a}$ & $4.9^{+0.6}_{-0.4}$ \\
{\Large \strut}$J$=3 & $\log N $ & 17.72$\pm$0.10 &  $18.11^{+0.06}_{-0.09}$ & 17.90$\pm$0.08 & $17.73^{+0.10}_{-0.28}$ & 18.17$\pm$0.08 & $18.30^{+0.08}_{-0.07}$ \\
{\Large \strut}& $b~(\kms)$ & $\tablefootmark{a}$ &  $2.4^{+0.7}_{-0.4}$ & $\tablefootmark{a}$ & $4.9^{+0.6}_{-0.4}$ & $\tablefootmark{a}$ & $5.3^{+0.5}_{-0.6}$\\
{\Large \strut}$J$=4 & $\log N $ & 17.16$\pm$0.25 & $17.30^{+0.24}_{-1.20}$ & 17.36$\pm$0.13 & $16.12^{+0.32}_{-0.13}$ & 16.11$\pm$0.17 & $16.18^{+0.48}_{-0.17}$\\
{\Large \strut}& $b~(\kms)$ & $\tablefootmark{a}$ & $3.0^{+1.3}_{-0.4}$ & $\tablefootmark{a}$ & $5.3^{+0.3}_{-0.6}$ & $\tablefootmark{a}$ & $6.7^{+0.6}_{-1.0}$ \\
{\Large \strut}$J$=5 & $\log N $ & 16.60$\pm$0.53 & $16.1^{+0.6}_{-0.4}$ & 16.08$\pm$0.34 & $15.51^{+0.25}_{-0.16}$ & 16.22$\pm$0.13 & $16.13^{+0.09}_{-0.14}$ \\
{\Large \strut}& $b~(\kms)$ & $\tablefootmark{a}$ &  $4.4^{+0.8}_{-0.7}$ & $\tablefootmark{a}$ & $5.5^{+0.6}_{-0.7}$ & $\tablefootmark{a}$ & $7.3^{+0.4}_{-0.6}$ \\
{\Large \strut}$J$=6 & $\log N $ & -- & $14.55^{+0.05}_{-0.08}$ & -- & $14.49^{+0.10}_{-0.12}$ & -- & $14.69^{+0.05}_{-0.08}$ \\
{\Large \strut}& $b~(\kms)\tablefootmark{b}$ & -- &  $23^{+6}_{-5}$ & -- & $20^{+5}_{-8}$ & -- & $15^{+6}_{-5}$\\
{\Large \strut}$J$=7 & $\log N $ & -- & $14.24^{+0.22}_{-0.24}$ & -- & $14.29^{+0.21}_{-0.25}$ & -- & $14.69^{+0.08}_{-0.06}$\\
{\Large \strut}& $b~(\kms)\tablefootmark{b}$ & -- & $38^{+4}_{-13}$ & -- & $24^{+5}_{-6}$ & -- & $25^{+2}_{-9}$\\
{\Large \strut}Total & $\log N$ & 18.73$\pm$0.03 & $18.95^{+0.03}_{-0.02}$ & 19.47$\pm$0.03 & $19.47^{+0.05}_{-0.04}$ & 19.27$\pm$0.05 &  $19.44^{+0.04}_{-0.05}$ \\
\hline
\hline
\end{tabular}
\tablefoot{
\tablefoottext{a}{Doppler parameter tied for all rotational levels (vpfit) and for J=0 and J=1 (MCMC).}
\tablefoottext{b}{The absorption lines corresponding to these levels are little sensitive on the Doppler parameter.}
}
\end{center}
\end{table*}

In all three components, the column densities of high rotational levels ($J\gtrsim2$) are significantly enhanced compared to what is seen in typical H$_2$-bearing DLAs at high redshifts. 
The observed $T_{01}$ temperatures ($\sim 100$~K for components B and C and closer to $\sim$200~K for component A) remain consistent with what is seen for intervening systems, albeit with a small tendency to be in the upper range for the observed column densities \citep[e.g.][]{Balashev2019}.  This could result from an enhanced heating of the gas through photoelectric effect of UV photons on dust grains. We discuss the observed H$_2$ excitation diagrams quantitatively in Sect.~\ref{s:phys}.

\subsection{Chemical enrichment}

Metal absorption lines are seen spread over 600~\kms\, mostly bluewards of the systemic redshift. Since the \HI\ column density is large enough for self-shielding from ionising photons, we focused our analysis on the low-ionisation species in order to estimate the gas phase metallicity, since these are in their dominant ionisation stage in the neutral gas.

The initial guess for the velocity decomposition was obtained from the absorption lines of singly ionised silicon and iron that have several transitions with a wide range of oscillator strengths. Singly ionised sulphur and zinc were subsequently added to the model, and finally, all parameters were fitted simultaneously under the usual assumption that \ZnII, \SII, \SiII, and \FeII\ are located in the same gas (i.e. we tied their Doppler parameters and redshifts). Since \ZnII$\lambda$2026 is blended with $\MgI\lambda2026$, we also included this line in the fit with an independent Doppler parameter and redshift. We note that \SiII$\lambda1526$ is detected in the overlap between the UVB and VIS spectra. We therefore included both these spectral regions to constrain the fit.

\begin{figure}
    \centering
    \addtolength{\tabcolsep}{-5pt}    
    \begin{tabular}{c c}

         \includegraphics[bb=220 230 445 630,clip=,angle=90,width=0.48\hsize]{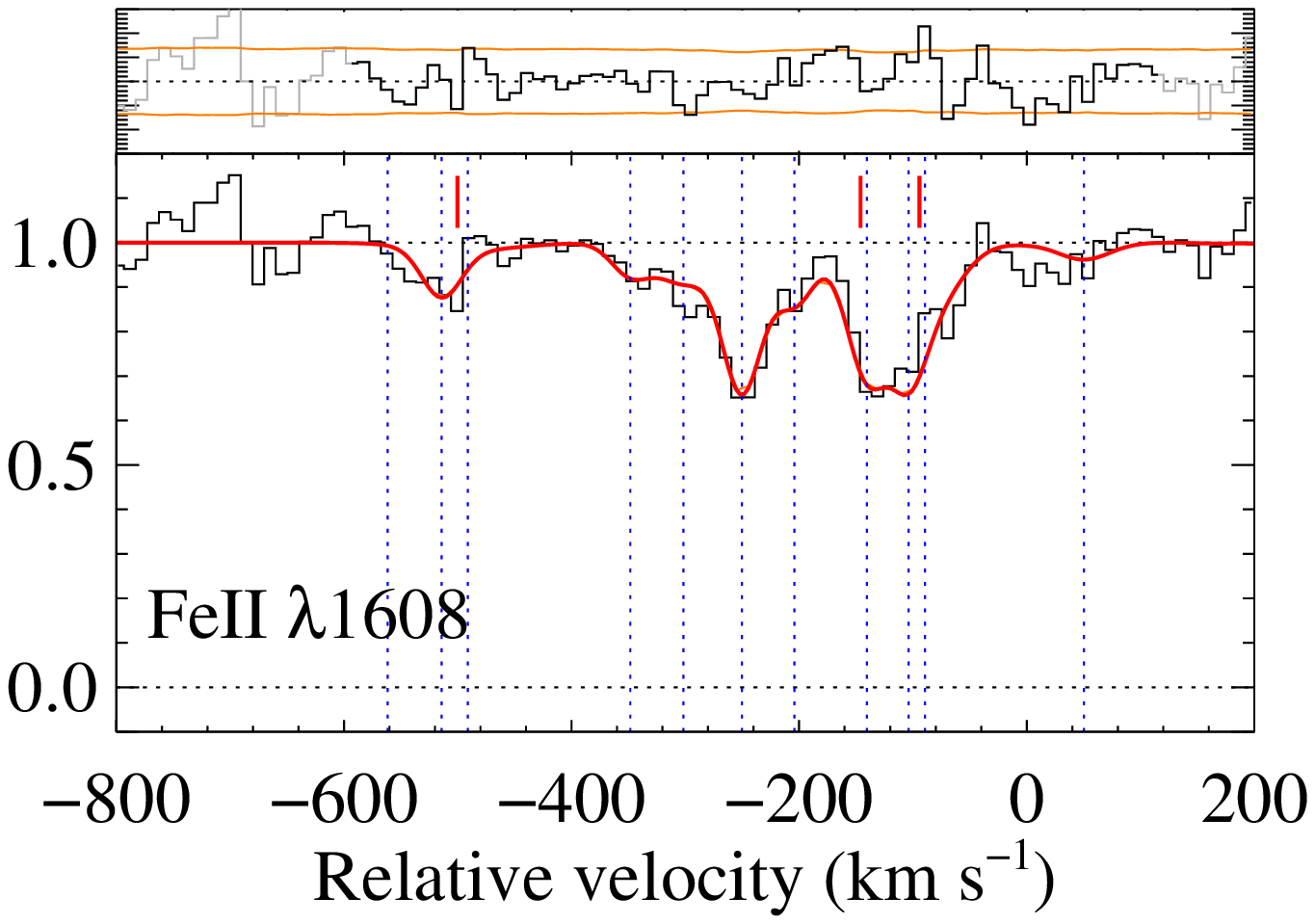} &  \includegraphics[bb=220 230 445 630,clip=,angle=90,width=0.48\hsize]{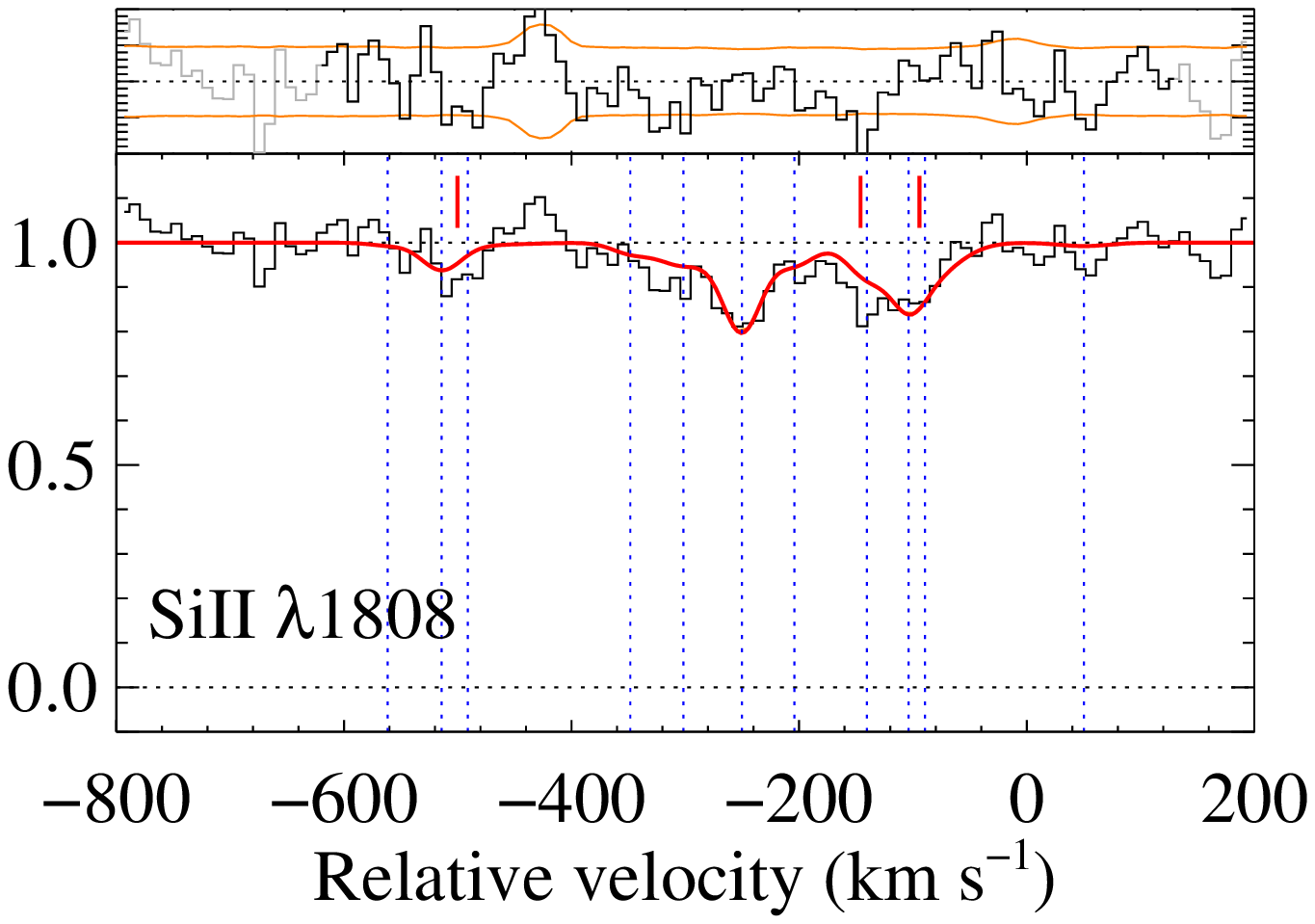} \\
         \includegraphics[bb=220 230 445 630,clip=,angle=90,width=0.48\hsize]{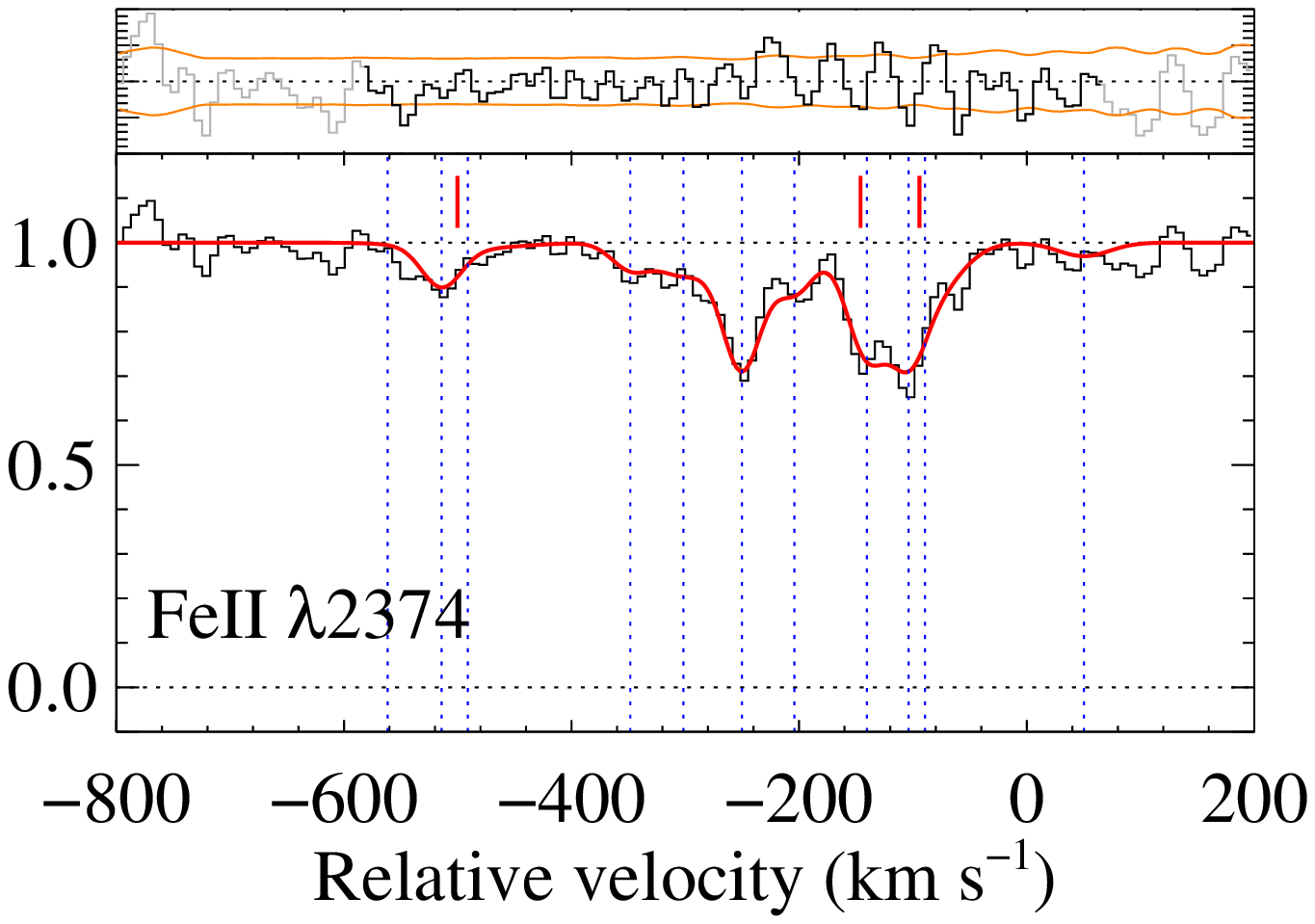} & \includegraphics[bb=220 230 445 630,clip=,angle=90,width=0.48\hsize]{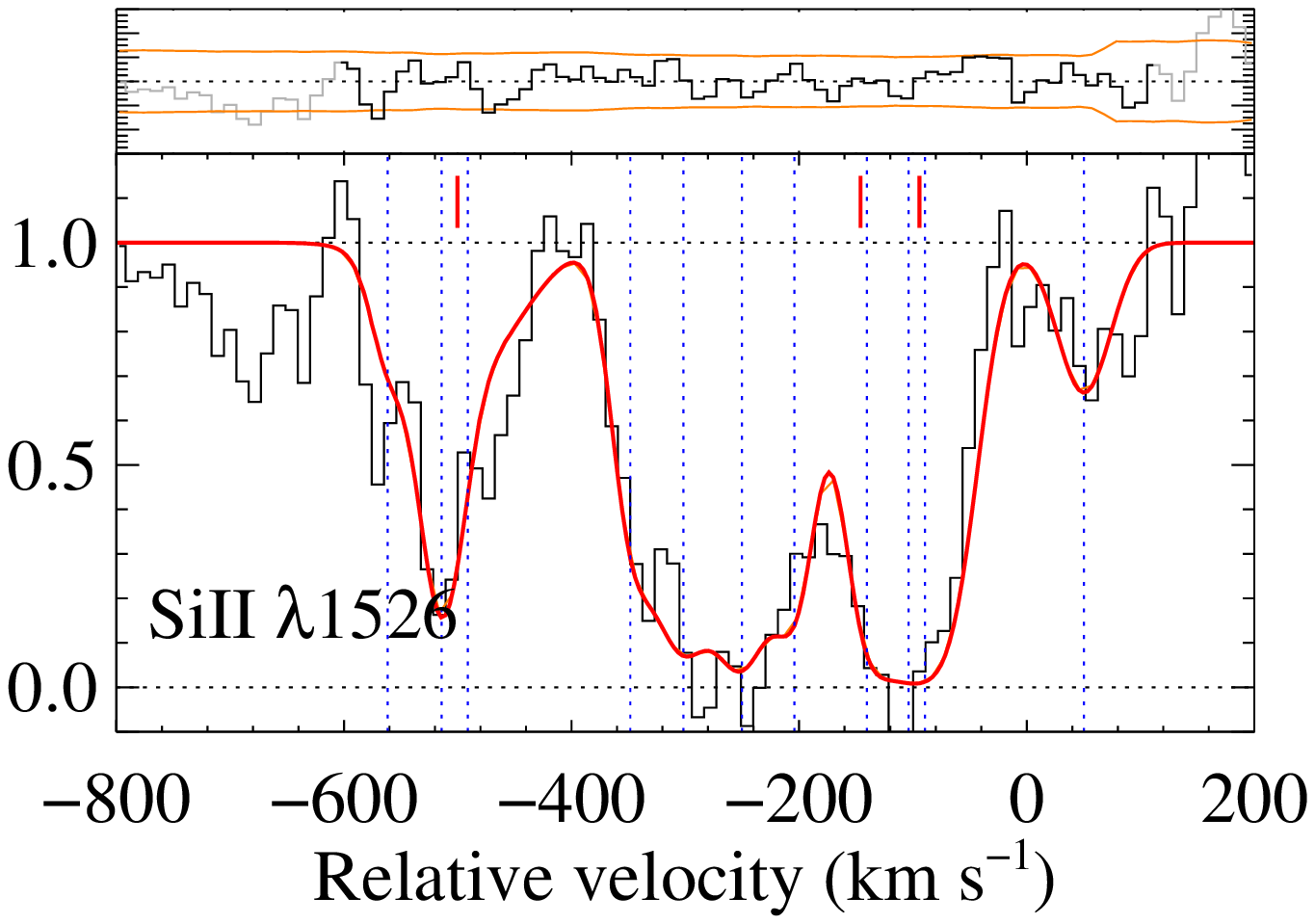} \\
         \includegraphics[bb=220 230 445 630,clip=,angle=90,width=0.48\hsize]{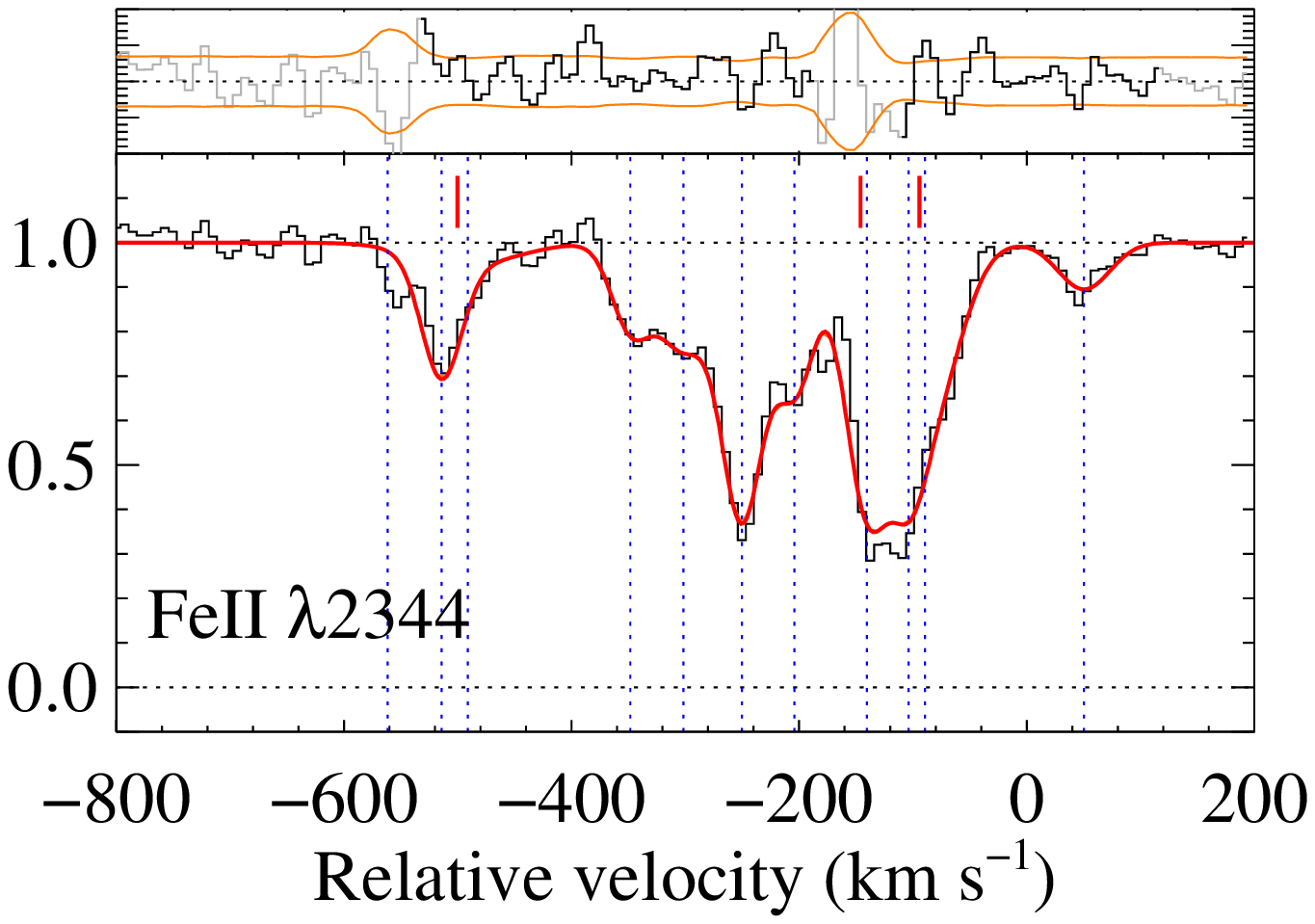} &  \includegraphics[bb=220 230 445 630,clip=,angle=90,width=0.48\hsize]{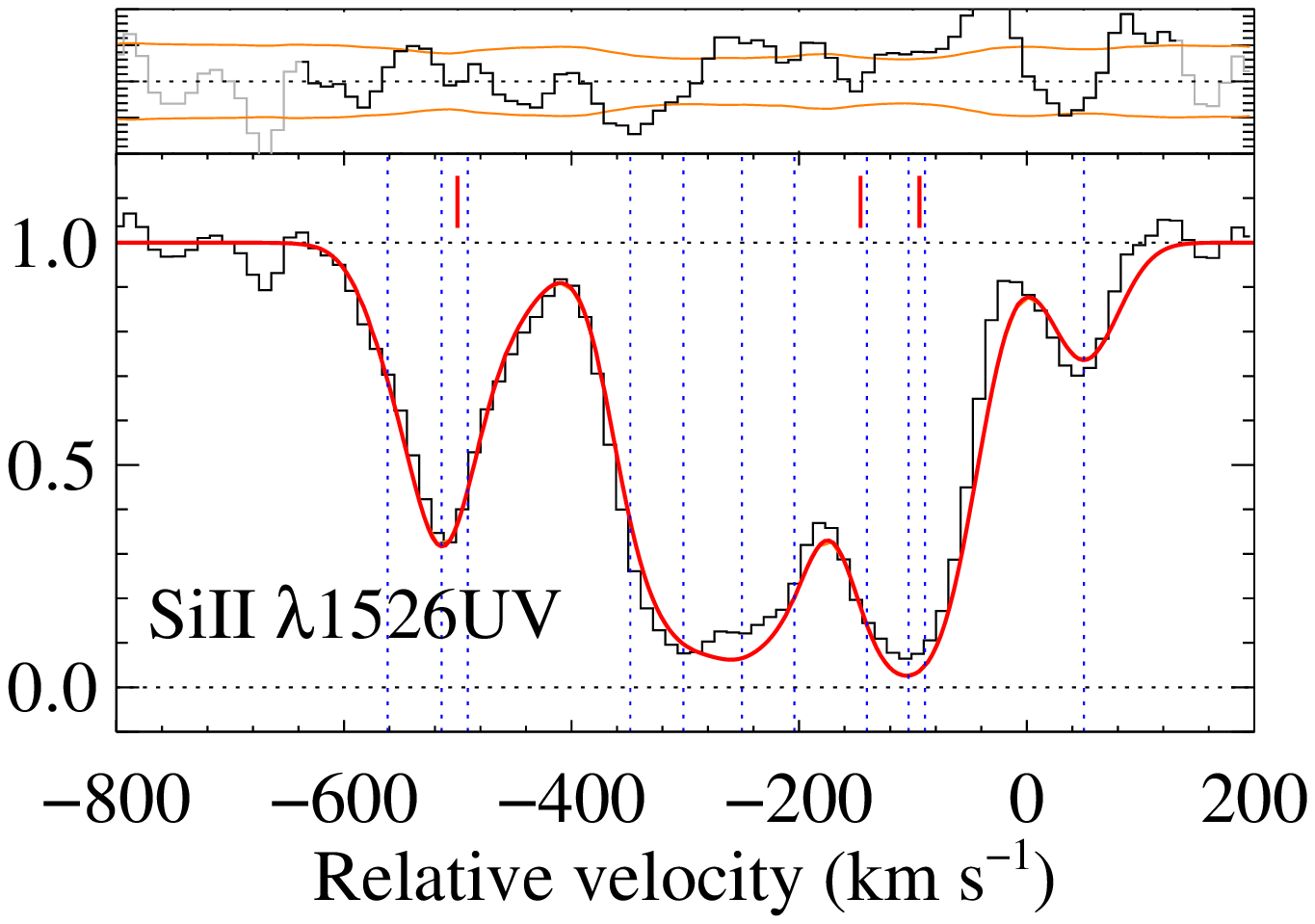} \\
         \includegraphics[bb=220 230 445 630,clip=,angle=90,width=0.48\hsize]{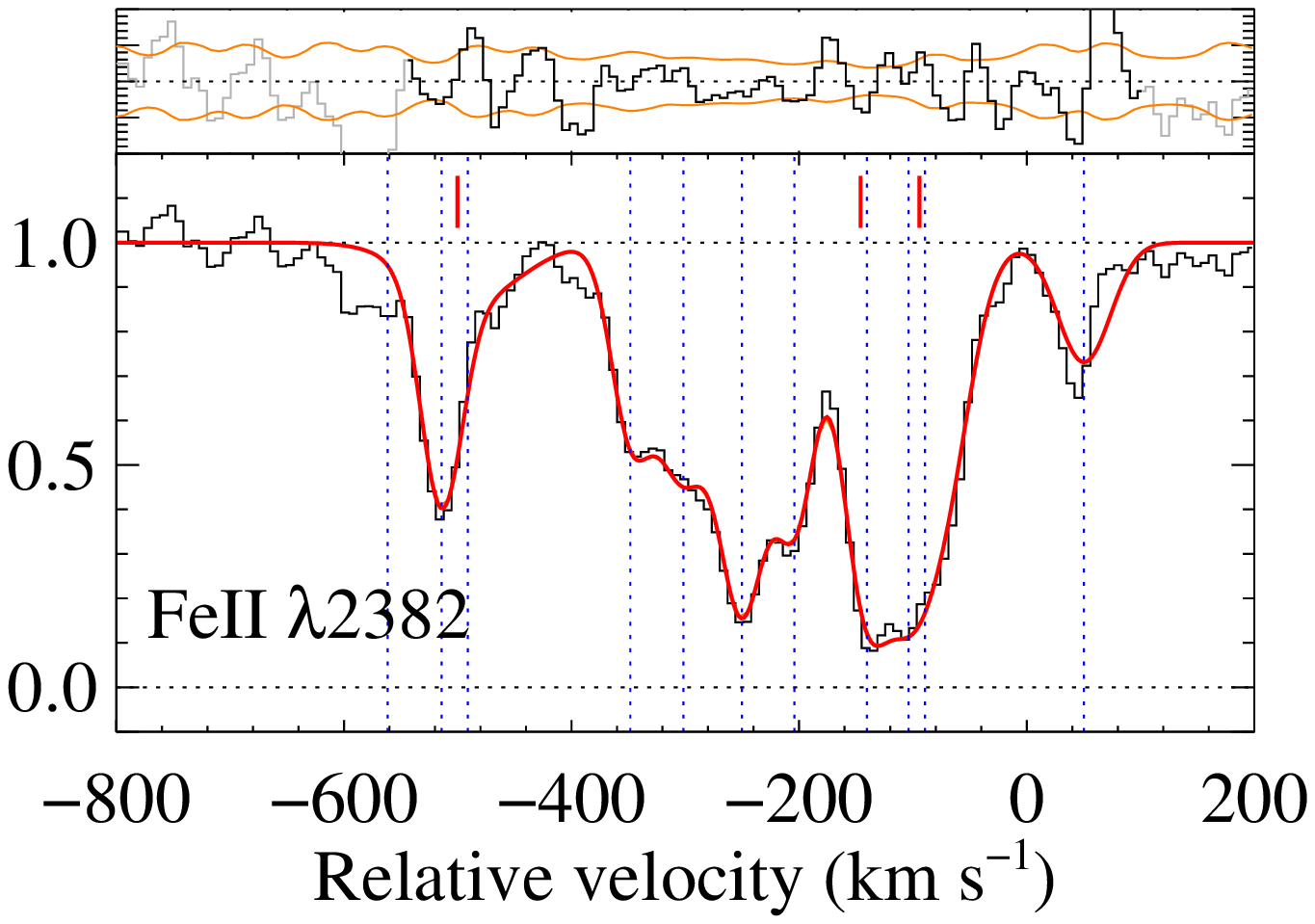} & \includegraphics[bb=220 230 445 630,clip=,angle=90,width=0.48\hsize]{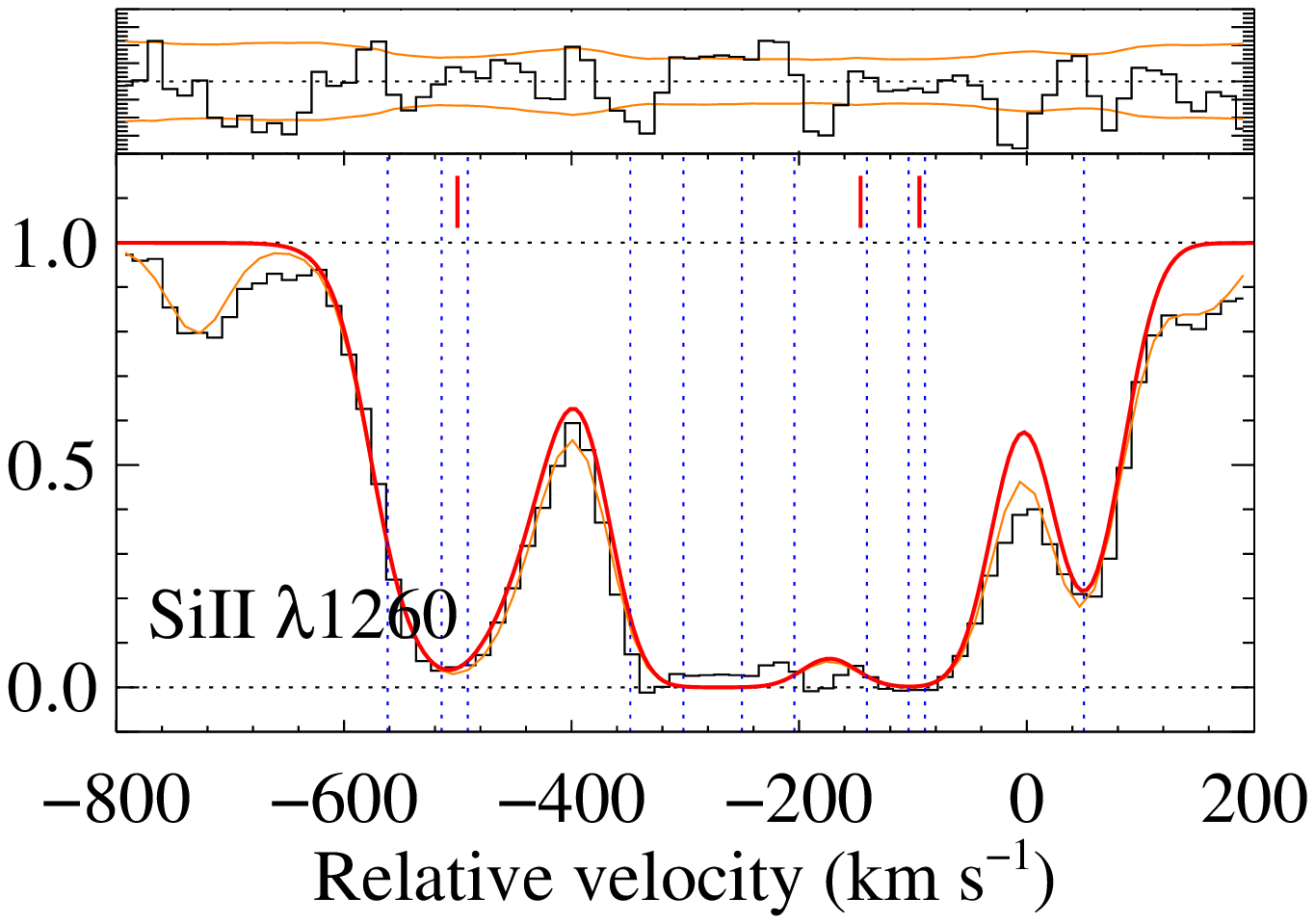} \\
    
         \includegraphics[bb=220 230 445 630,clip=,angle=90,width=0.48\hsize]{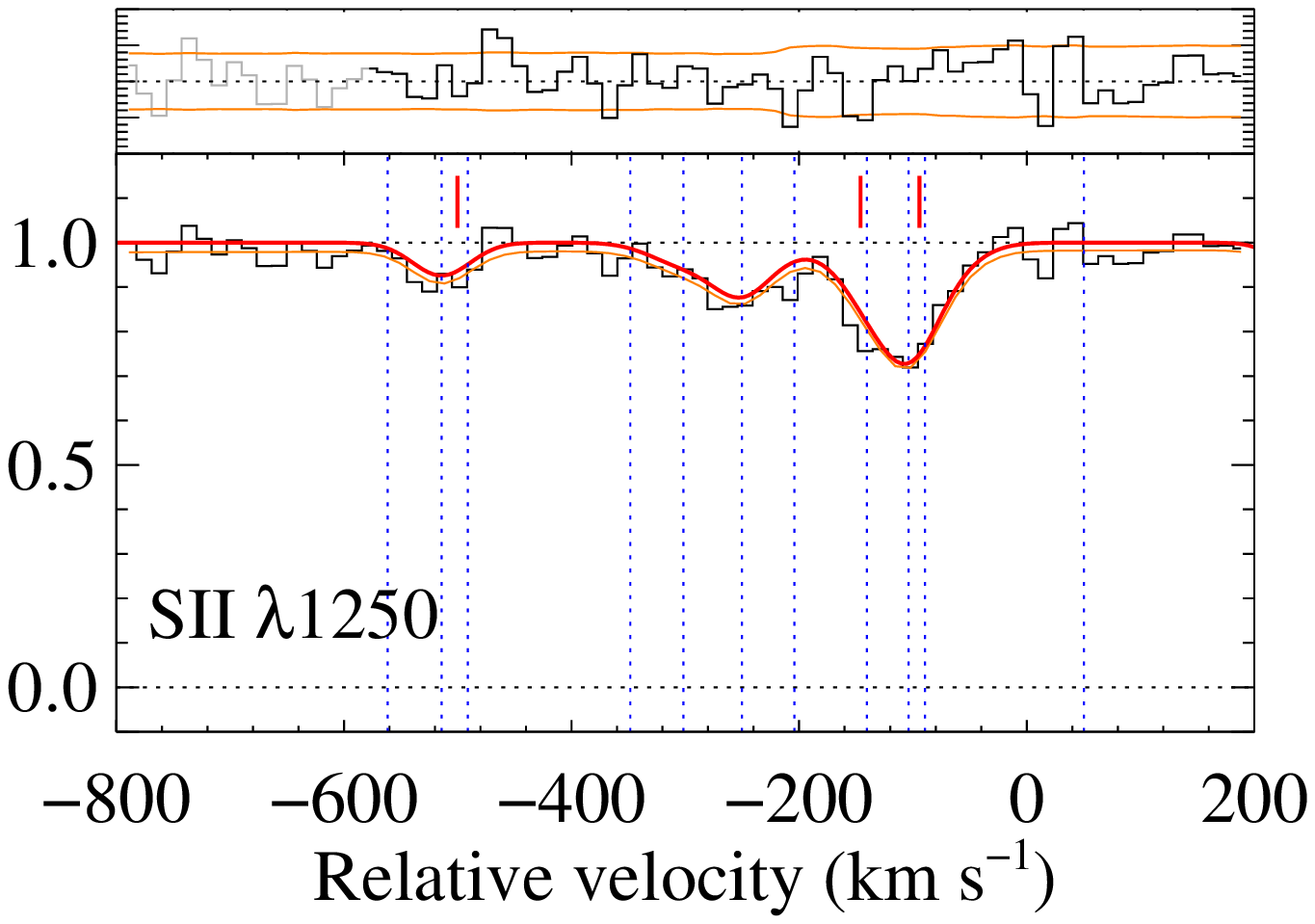} & \includegraphics[bb=220 230 445 630,clip=,angle=90,width=0.48\hsize]{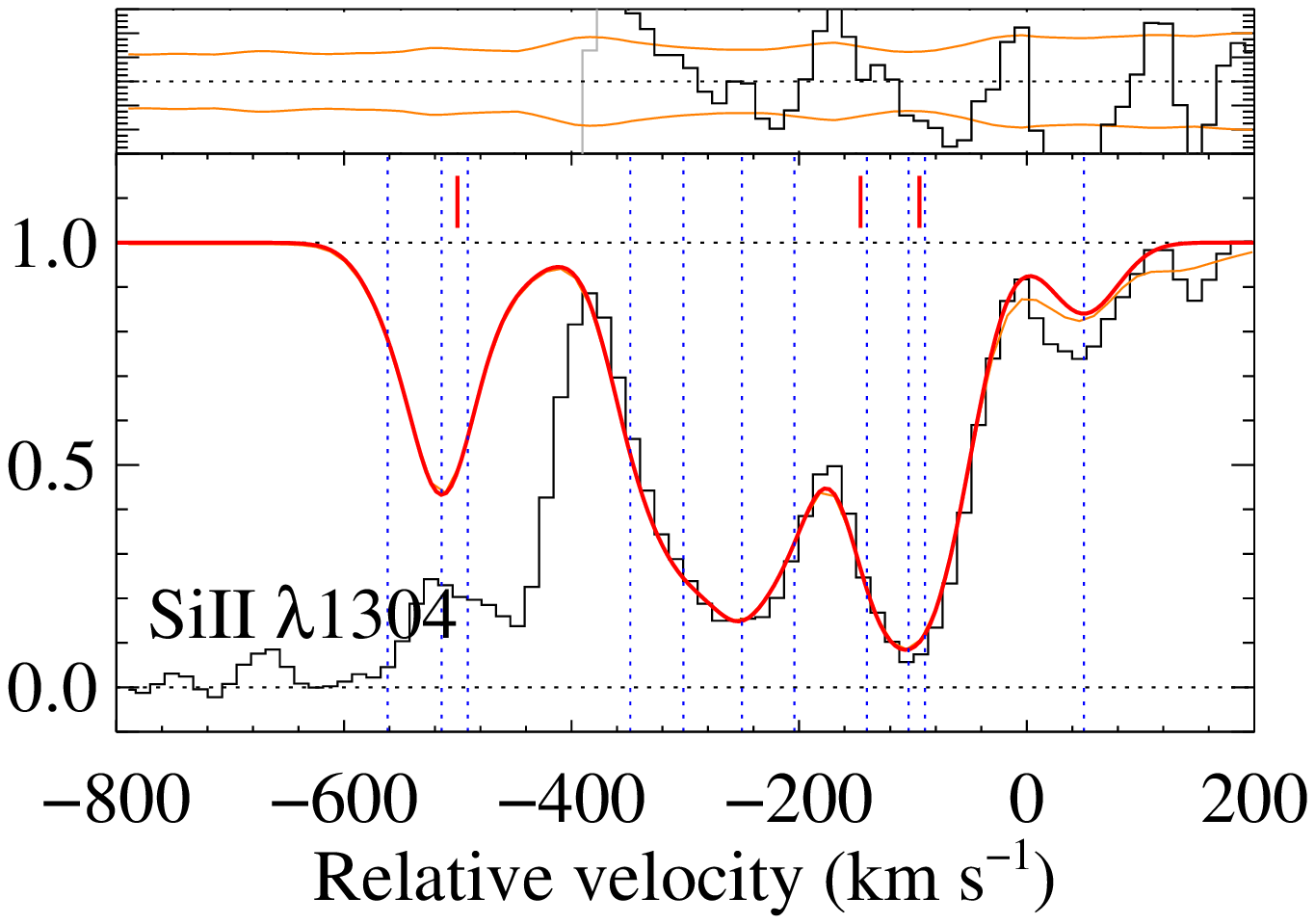} \\
         \includegraphics[bb=220 230 445 630,clip=,angle=90,width=0.48\hsize]{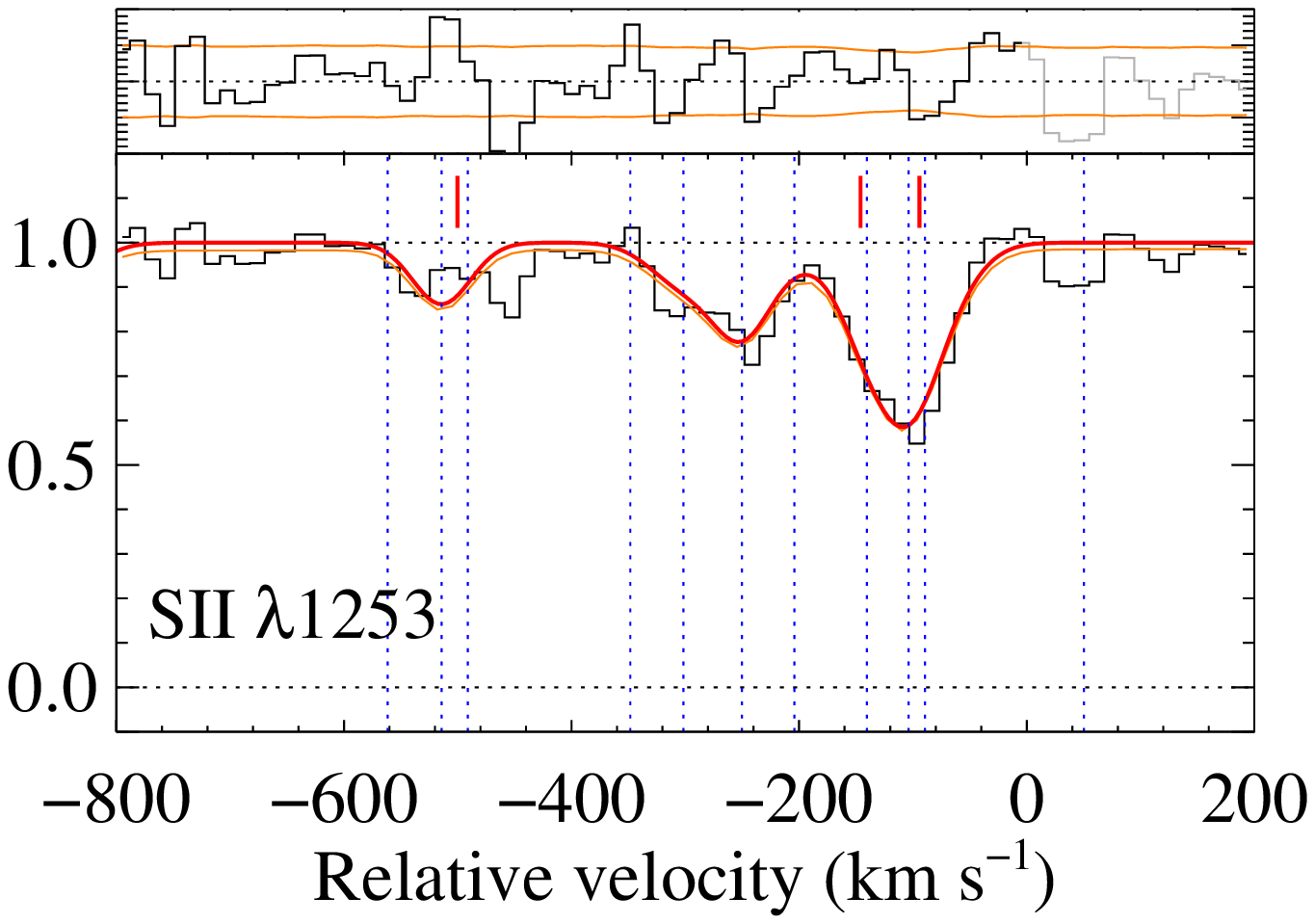} & \includegraphics[bb=220 230 445 630,clip=,angle=90,width=0.48\hsize]{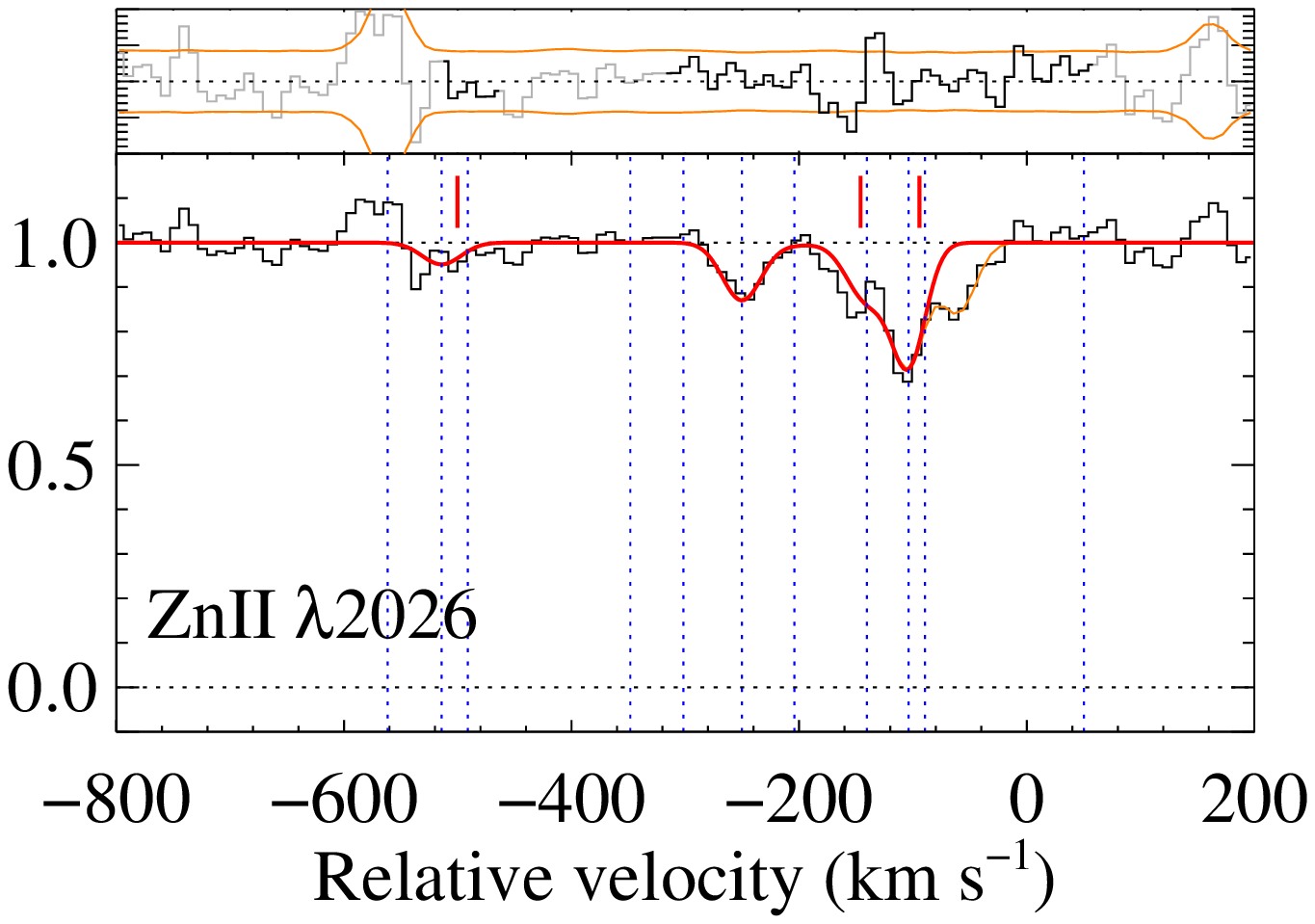}\\
         \includegraphics[bb=165 230 445 630,clip=,angle=90,width=0.48\hsize]{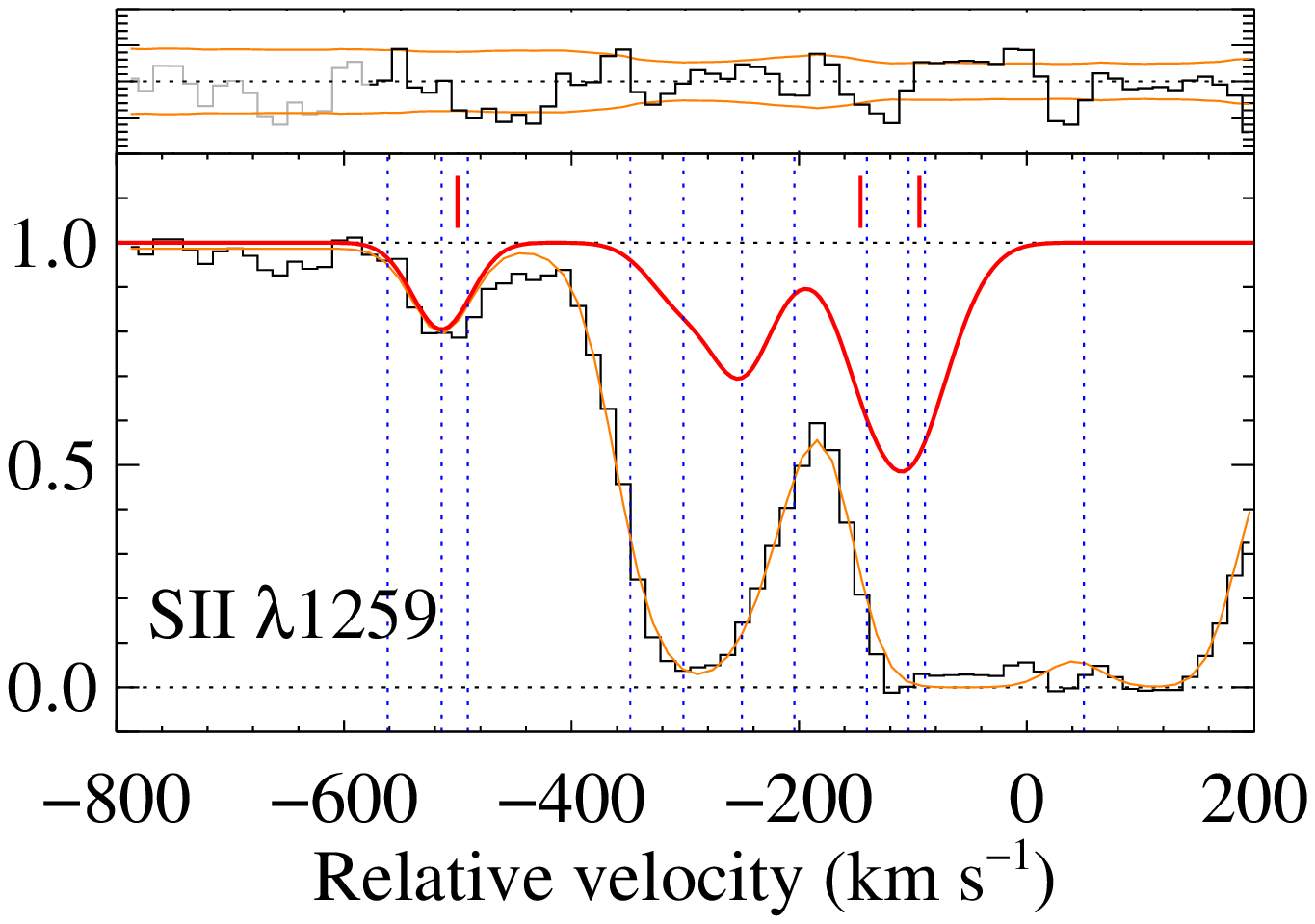} & \includegraphics[bb=165 230 445 630,clip=,angle=90,width=0.48\hsize]{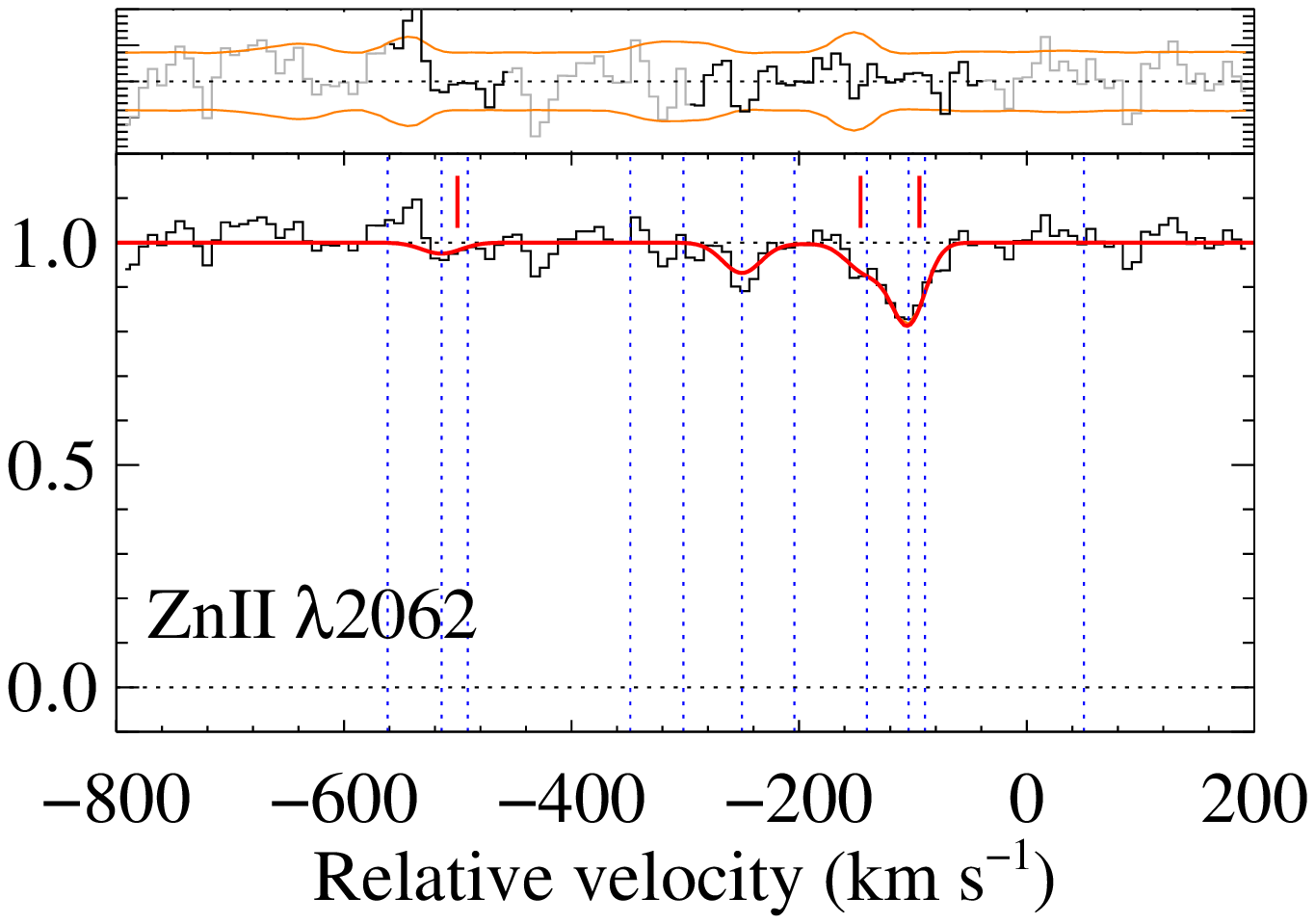}
    \end{tabular}{}
        \addtolength{\tabcolsep}{+5pt}    
    \caption{Fit to metal absorption lines. The observed data are shown in black with the synthetic profile for each given species overplotted in red. In case of several lines contributing to the observed profile (see e.g. \SII$\lambda$1259 and \SiII$\lambda$1260), the contribution for the labelled line alone is shown in red, while the total synthetic profile is shown in orange. 
    Residuals are shown in the small panel above each line, with the orange line showing the 1\,$\sigma$ error level. 
    The location of the different components 
    of the model are indicated by blue vertical dashed lines, with short red marks showing the position of H$_2$ components. The zero of the velocity scale is set to $z_{ref}=2.631$.}
    \label{f:metals}
\end{figure}

The result of the fits are shown in Fig.~\ref{f:metals} and the corresponding parameters are provided in Table.~\ref{t:metals}. The total column densities are $\log N(\cmsq)= 15.52\pm0.03$ (\SiII), $14.46\pm0.02$ (\FeII), $15.39\pm0.08$ (\SII), and $13.34\pm0.07$ (\ZnII). Using the total H (\HI + H$_2$) column density, $\log N($H)$=20.75\pm0.02$, these values translate to overall gas-phase abundances of [Si/H]~=~$-0.74 \pm 0.04$, [Fe/H]~=~$-1.77 \pm 0.03$, [S/H]~=~$-0.49 \pm 0.08,$ and [Zn/H]~=~$-0.39 \pm 0.05$, where ionisation corrections were assumed to be negligible, and the solar reference values were taken from \citet{Asplund2009} following the recommendation of \citet{Lodders2009} on whether to use photospheric, meteoritic, or average values. Both zinc and sulphur are known to be volatile species and hence provide a good estimate of the metallicity, which is about 40\% solar. In turn, silicon and iron tend to deplete more onto dust grains \citep[e.g.][]{DeCia2016}. Indeed, we measured depletion by a factor of two for silicon, while 96\% of iron is locked into dust grains.

\begin{table*}
\centering
\caption{Result of Voigt profile fitting to singly ionised metal lines. 
\label{t:metals}}
\begin{tabular}{c c c c c c c}
\hline \hline
{\Large \strut} z & b (\kms) & \multicolumn{5}{c}{$\log N (\cmsq)$} \\
  &          & \SiII\ & \FeII & \SII & \ZnII  & \SiII*  \\ 
\hline
2.624197 &  6.6$\pm$2.2 &     13.30 $\pm$ 0.13 &                    - &                    - &                    - &                    - \\ 
2.624772 & 13.1$\pm$1.4 &     14.41 $\pm$ 0.10 &     13.32 $\pm$ 0.03 &     14.48 $\pm$ 0.05 &     11.81 $\pm$ 0.20 &     11.81 $\pm$ 0.27 \\ 
2.625051 & 58.9$\pm$3.2 &     13.83 $\pm$ 0.05 &     12.88 $\pm$ 0.08 &                    - &                    - &                    - \\ 
\smallskip\\
2.626781 & 12.3$\pm$1.6 &     13.98 $\pm$ 0.10 &     13.10 $\pm$ 0.06 &                    - &                    - &                    - \\ 
2.627347 & 24.6$\pm$4.9 &     14.48 $\pm$ 0.04 &     13.34 $\pm$ 0.05 &     14.32 $\pm$ 0.11 &                    - &     12.66 $\pm$ 0.05 \\ 
2.627968 & 12.4$\pm$0.9 &     14.99 $\pm$ 0.05 &     13.89 $\pm$ 0.02 &     14.66 $\pm$ 0.07 &     12.25 $\pm$ 0.06 &     12.97 $\pm$ 0.06 \\ 
2.628528 & 14.7$\pm$1.1 &     14.37 $\pm$ 0.04 &     13.44 $\pm$ 0.02 &     13.81 $\pm$ 0.28 &     10.84 $\pm$ 1.44 &     11.49 $\pm$ 0.67 \\ 
\smallskip\\
2.629299 & 14.9$\pm$1.2 &     14.52 $\pm$ 0.08 &     13.82 $\pm$ 0.04 &     14.65 $\pm$ 0.08 &     12.27 $\pm$ 0.08 &     12.44 $\pm$ 0.09 \\ 
2.629742 &  5.1$\pm$1.2 &     14.59 $\pm$ 0.15 &     13.63 $\pm$ 0.07 &     14.99 $\pm$ 0.20 &     12.73 $\pm$ 0.08 &     12.62 $\pm$ 0.32 \\ 
2.629918 & 32.3$\pm$2.2 &     14.76 $\pm$ 0.07 &                    - &                    - &                    - &                    - \\ 
\smallskip\\
2.631610 & 24.5$\pm$1.1 &     13.64 $\pm$ 0.02 &     12.95 $\pm$ 0.03 &                    - &                    - &                      - \\
\multicolumn{2}{c}{Total}   &     15.52 $\pm$ 0.03 &     14.46 $\pm$ 0.02 &     15.39 $\pm$ 0.08 &     12.99 $\pm$ 0.05 &     13.34 $\pm$ 0.07 \\ \hline
\end{tabular}
\end{table*}

\subsection{Excited carbon and silicon}

\begin{figure}
    \centering
    \begin{tabular}{c}
    \includegraphics[bb=188 20 451 728,clip=,angle=90,width=0.9\hsize]{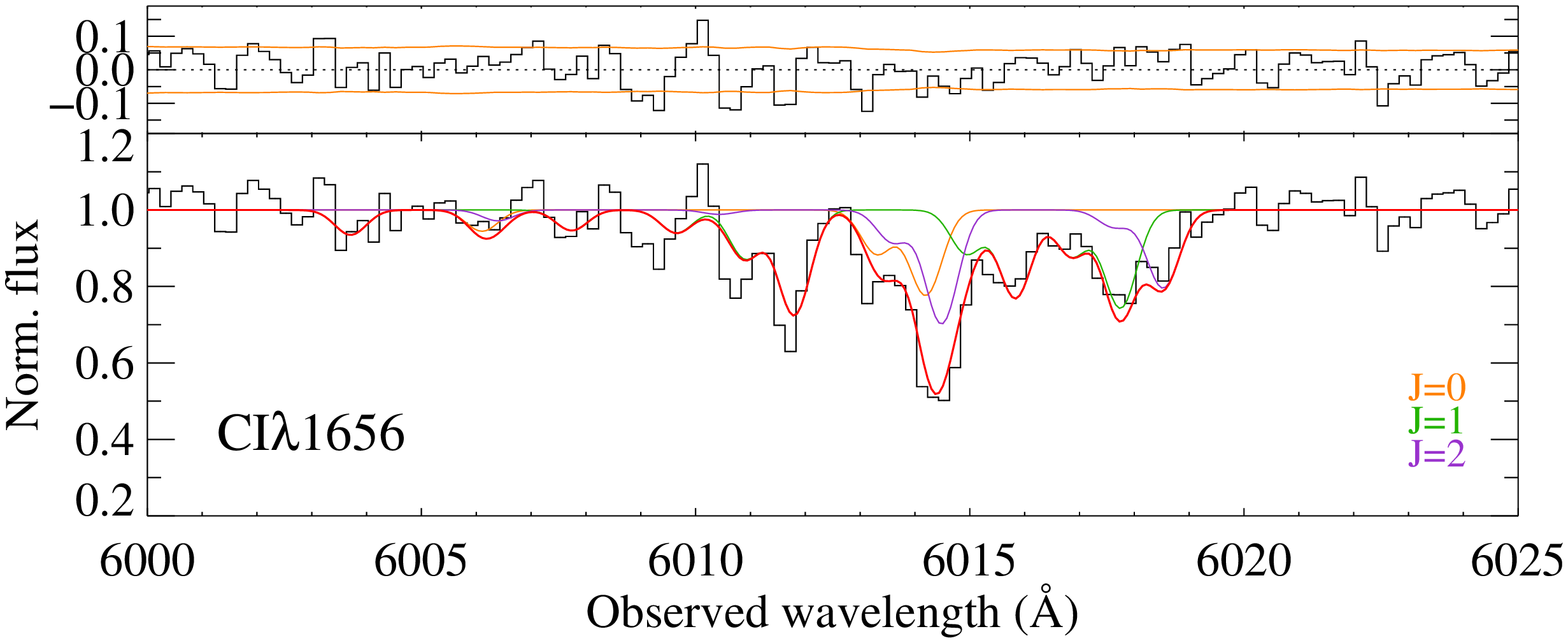}\\
    \includegraphics[bb=188 20 451 728,clip=,angle=90,width=0.9\hsize]{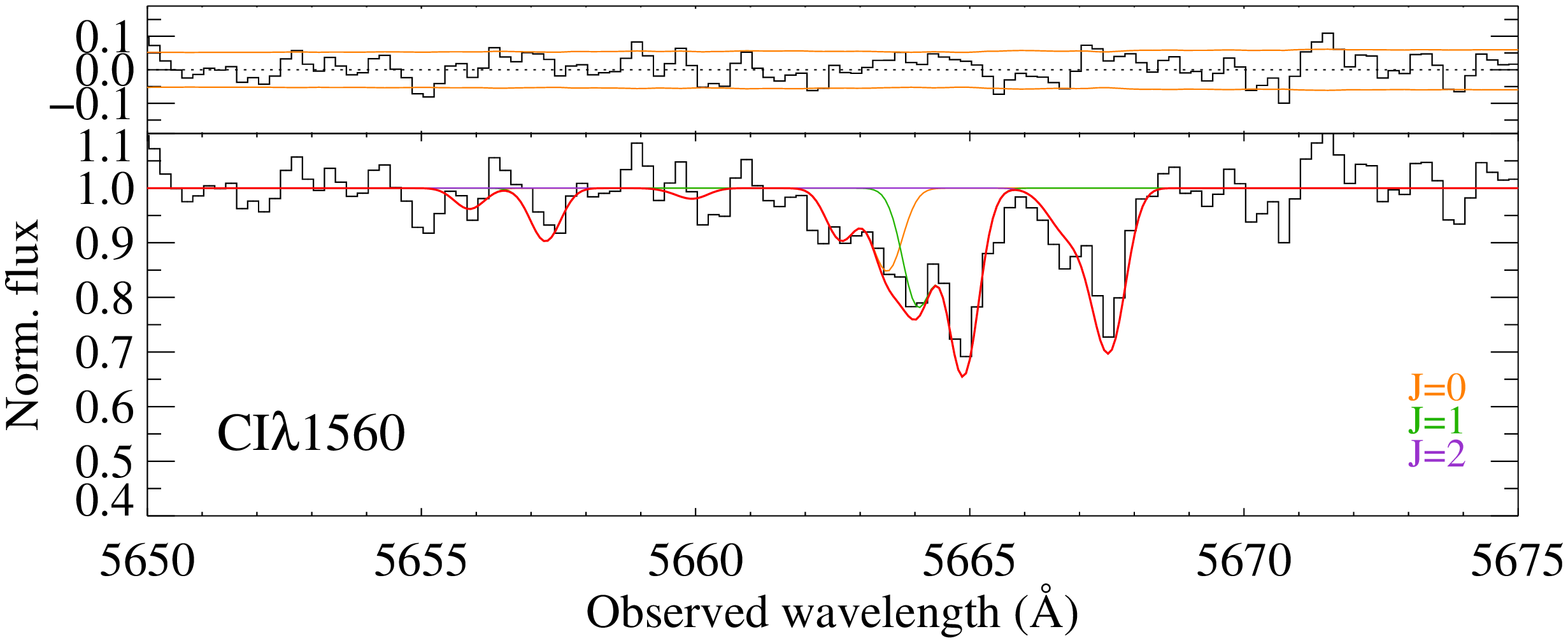}\\
    \includegraphics[bb=188 20 451 728,clip=,angle=90,width=0.9\hsize]{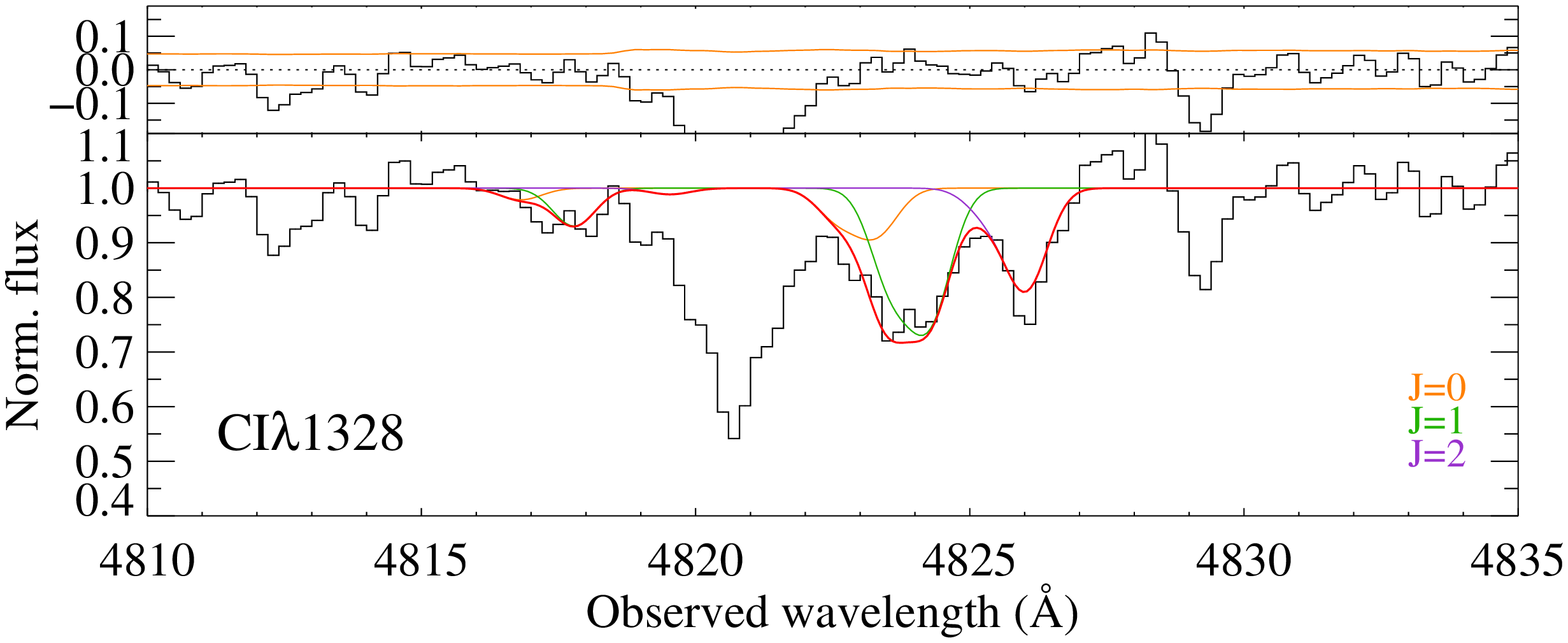}\\
    \includegraphics[bb=165 20 451 728,clip=,angle=90,width=0.9\hsize]{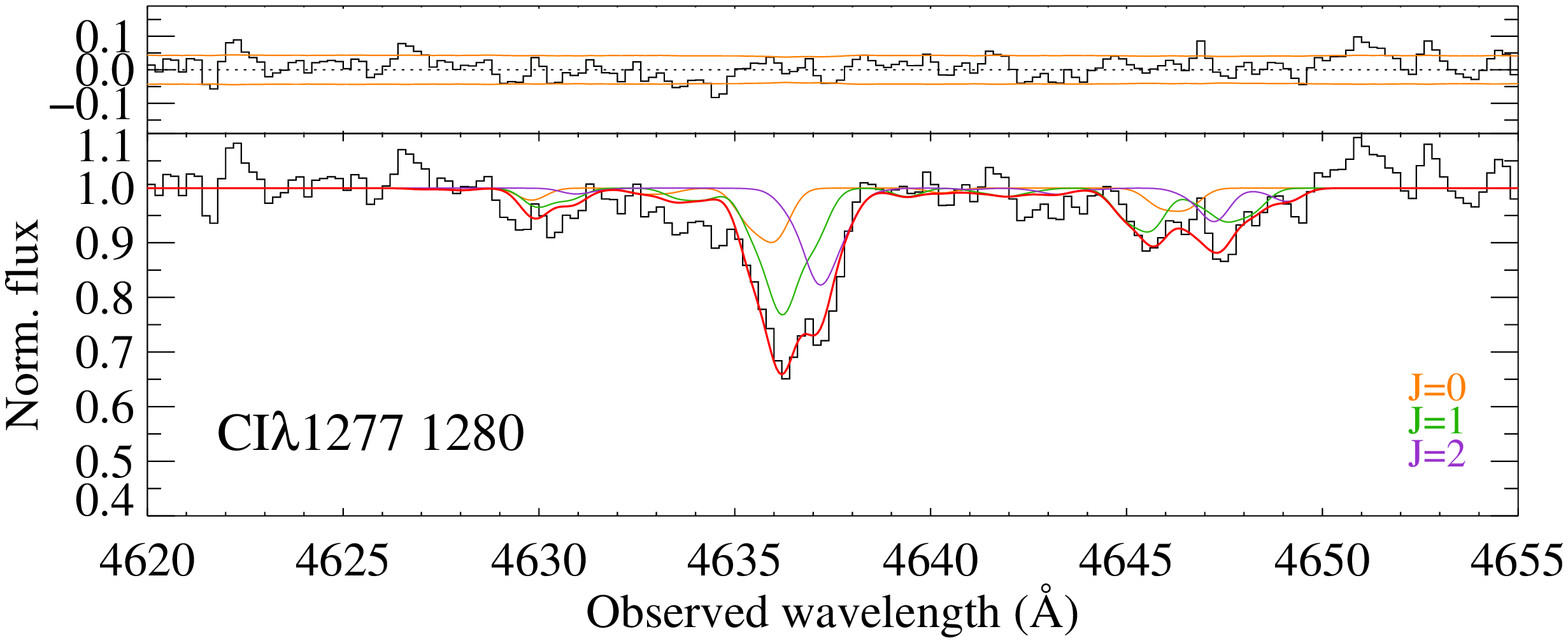}\\
    \end{tabular}{}
    \caption{Fit to \CI\ absorption lines. The data are shown in black with the best-fit over-plotted in red. The contribution from individual fine-structure levels are shown in orange (\CI), green (\CI*), and purple (\CI**). The top panels show the residuals.}
    \label{f:CI}
\end{figure}

Neutral carbon absorption lines are detected in the spectrum of \J. This is not surprising given the presence of H$_2$ and the high metallicity of the system \citep{Noterdaeme2018}.
Three components are detected, matching those seen in H$_2,$ and we detect all three fine structure levels of the ground state triplet ($2s^22p^2\,^3P_{J=0,1,2}$). All \CI\ (J=0), \CI* (J=1), and \CI** (J=2) lines were first fitted together using the bands at 1277, 1280, 1328, 1560, and 1656~{\AA}, avoiding regions affected by blends (in particular for the 1328~{\AA} band) and tying the redshifts and Doppler parameters for each component. The result of the fit is shown in Fig.~\ref{f:CI}, and the corresponding parameters are summarised in Table~\ref{t:CI}. The Doppler parameters are found to be as small as 1~\kms. While such small values are similar to what is seen in H$_2$, and not uncommon for \CI-absorbers \citep[e.g.][]{Noterdaeme2017}, here they are observed far below the spectral resolution, implying large uncertainties on the column densities. We therefore repeated the fit using the MCMC method, which suggested that the uncertainties on the $b$-values provided by vpfit were likely underestimated. Finally, we performed a final test assuming a fixed, somehow higher $b=3.5~\kms$ value. The column densities obtained from the different methods remain similar, although differing in some cases by up to $\sim$0.5~dex. Nevertheless, all results indicate higher excited-to-ground, fine-structure level ratios than usually seen in intervening systems. We caution, however, that high spectral resolution observations remain vital to accurately measuring the corresponding column densities.

\begin{table}
\centering
\caption{Result of Voigt profile fitting to neutral carbon lines. \label{t:CI}}
\addtolength{\tabcolsep}{-3pt}
\begin{tabular}{ccccc}
\hline \hline
{\Large \strut}z & b (\kms) & \multicolumn{3}{c}{$\log N (\cmsq)$} \\
                 &          & \CI,J=0  & \CI,J=1   & \CI,J=2 \\
\hline
\multicolumn{5}{l}{\sl vpfit, free $b$} \\ 
2.62484 &  0.9$\pm$0.5 & 12.82$\pm$0.43 & 13.39$\pm$0.29 & 12.45$\pm$0.48 \\
2.62919 &  1.3$\pm$0.2 & 13.47$\pm$0.28 & 14.06$\pm$0.19 & 13.13$\pm$0.20 \\
2.62973 &  3.6$\pm$0.4 & 13.41$\pm$0.11 & 14.05$\pm$0.07 & 13.93$\pm$0.07 \\
\multicolumn{5}{l}{\sl MCMC, free $b$}  \\ 
2.62482 & $1.7^{+2.0}_{-0.7}$ & $12.82^{+0.30}_{-0.35}$ & $13.30^{+0.10}_{-0.14}$ & $12.72^{+0.27}_{-0.65}$ \\
2.62914 & $2.3^{+7.0}_{-1.0}$ & $13.11^{+0.15}_{-0.15}$ &  $13.50^{+0.08}_{-0.09}$ & $13.07^{+0.16}_{-0.13}$ \\ 
2.62973 & $2.3^{+0.3}_{-0.3}$ &  $13.75^{+0.34}_{-0.26}$ & $14.66^{+0.30}_{-0.15}$ & $14.27^{+0.18}_{-0.18}$ \\
\multicolumn{5}{l}{\sl vpfit, fixed $b$}  \\ 
2.62483 &  3.50        & 12.59$\pm$0.30 & 13.15$\pm$0.10 & 12.22$\pm$0.72 \\
2.62914 &  3.50        & 13.15$\pm$0.13 & 13.54$\pm$0.08 & 13.03$\pm$0.15 \\
2.62972 &  3.50        & 13.44$\pm$0.11 & 14.15$\pm$0.07 & 13.96$\pm$0.07 \\
\hline
\end{tabular}
\addtolength{\tabcolsep}{-3pt}
\end{table}

Singly ionised carbon and silicon are also detected in excited fine-structure states (\CII* and \SiII*, respectively). While \CII*\ is in principle useful in estimating the cooling rate of the gas through [\CII]158$\mu$m emission \citep{Pottasch1979}, the corresponding absorption lines here are saturated, impeding the measurement of the \CII*\ column density. 

\begin{figure}
    \centering
    \addtolength{\tabcolsep}{-5pt}   
    \begin{tabular}{c c}
             \includegraphics[bb=165 228 445 628,clip=,angle=90,width=0.49\hsize]{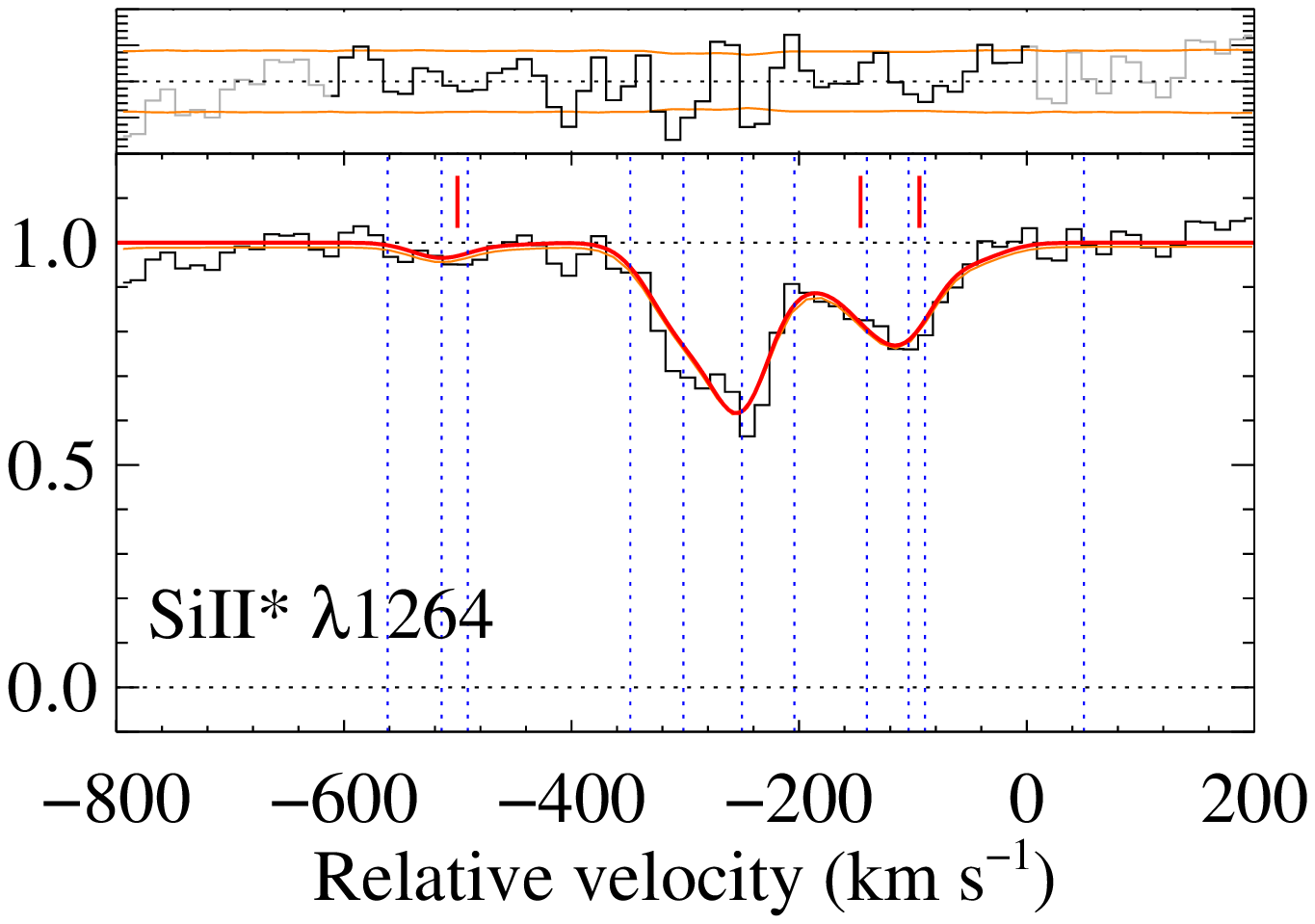} &
         \includegraphics[bb=165 228 445 628,clip=,angle=90,width=0.49\hsize]{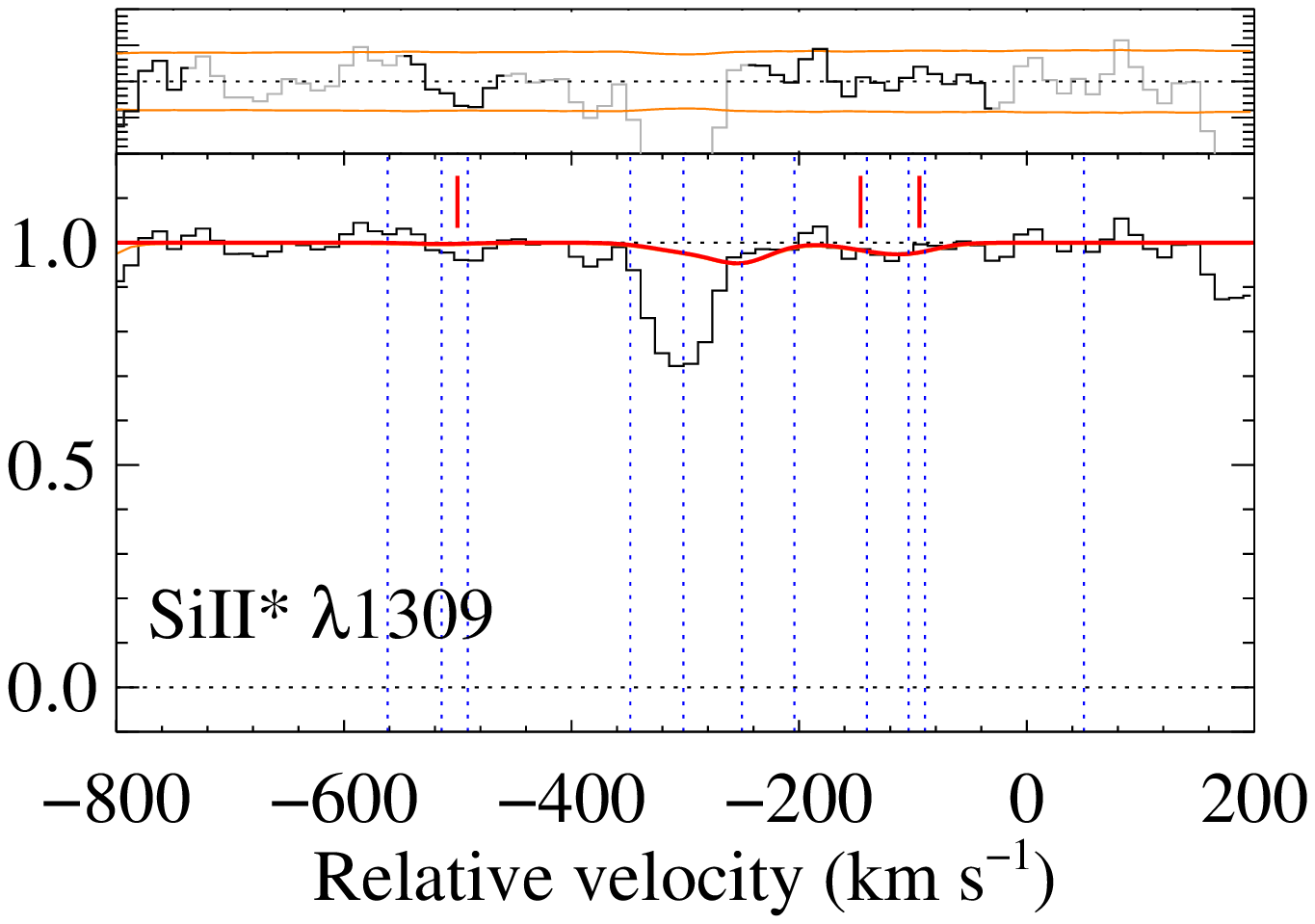} \\
    \end{tabular}
        \addtolength{\tabcolsep}{+5pt}    
    \caption{Same as Fig.~\ref{f:metals} for the absorption lines from excited fine-structure of singly-ionised silicon. }
    \label{f:SiIIs}
\end{figure}

The \SiII$^*$ absorption is in turn much weaker but clearly detected in its strongest line at 1264~{\AA}. We identify three main components that correspond to those seen in the metal lines. We also used the weaker 1309~{\AA} line to ascertain the detection and constrain the fit, shown in Fig.~\ref{f:SiIIs}. The corresponding parameters are given in Table~\ref{t:metals}.

\subsection{Dust extinction}
\label{s:dust}

\begin{figure*}
    \centering
    \includegraphics[width=0.95\textwidth]{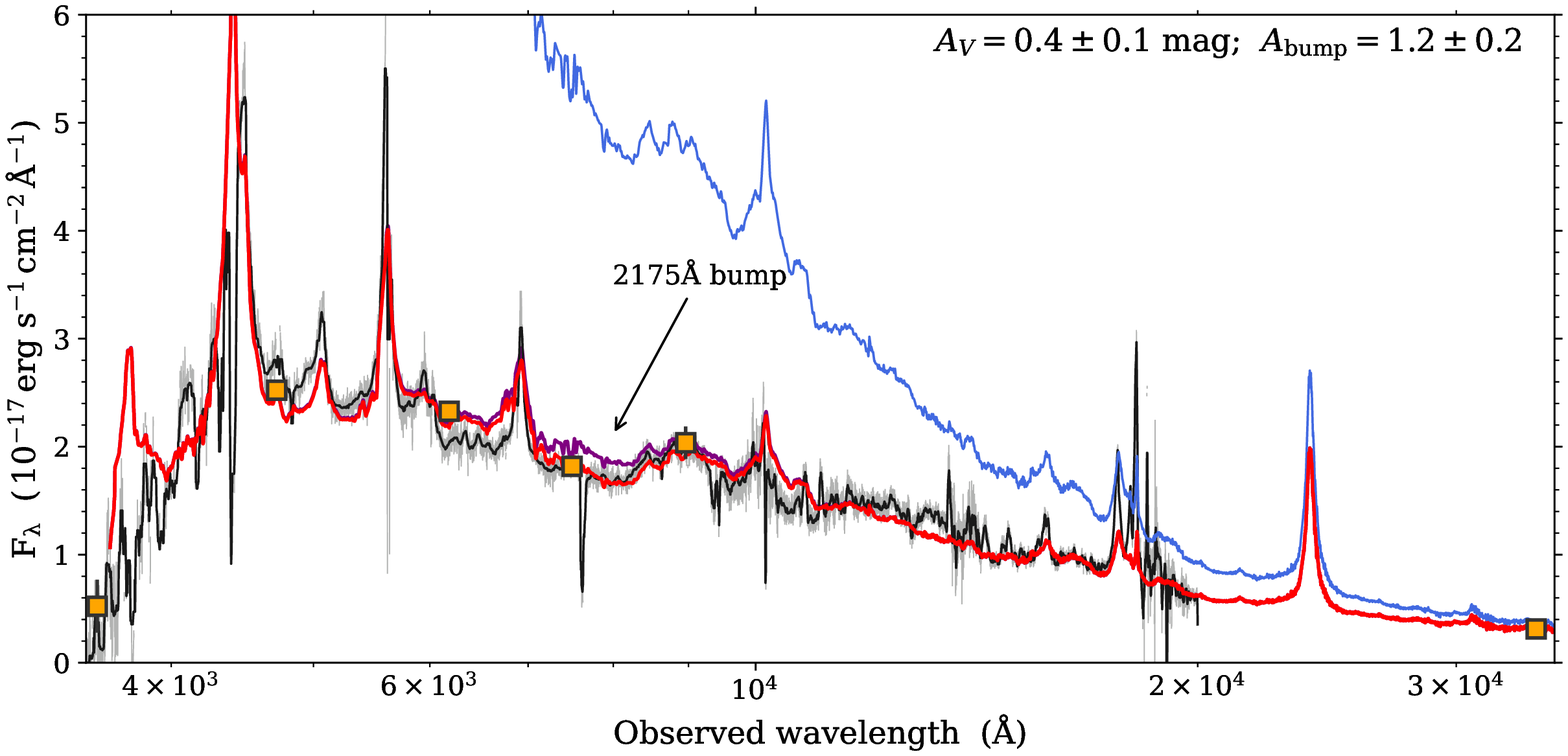}
    \caption{Combined X-shooter spectrum of J0015+1842 in black with best-fit SED corrected for dust extinction shown in red. For visual purposes, the data have been filtered (rejecting 4-$\sigma$ outliers; light grey) and smoothed (black) in order to remove the strong skyline residuals in the NIR data. The position of the 2175\AA\ extinction feature is indicated by an arrow, and the best-fit model without the 2175\AA\ feature is indicated by the purple line. The unreddened quasar template is shown in blue. Photometric data from SDSS ($u$, $g$, $r$, $i$, and $z$ bands) and WISE (band 1) are shown as orange squares.}
    \label{f:dustfit}
\end{figure*}

We fitted the dust reddening using the formalism by \citet{FM2007} assuming various fixed extinction laws \citep{Gordon2003} to identify the overall shape of the dust reddening law applied to the composite X-shooter spectrum from \cite{Selsing2016}.
The details of the method are presented by \citet{Noterdaeme2017}. We found a best match using a rather shallow extinction curve like the one observed towards the Large Magellanic super-shell (LMC2). However, we needed to vary the strength of the extinction bump at 2175~\AA\ in order to properly match the observed spectrum. {The spectrum exhibits significant variations of the iron emission lines in the region between the \CIV\ and \MgII\ lines. We included a model of these iron lines using the template given by \citet{Vestergaard2001}. While this template improves the overall fit, the exact variations of the myriad of iron lines cannot be captured by a fixed line template.}
Taking systematic uncertainties due to variations in the intrinsic quasar shape into account, we found a best-fit solution of $A(V) = 0.4\pm0.1$~mag, Fe$_2 = 0.5\pm0.2$, Fe$_3=-1.5\pm0.5$, $c_3 = 0.7 \pm 0.1$, where $A(V)$ is the amount of extinction and Fe$_2$ and Fe$_3$ multiplicative factors parametrising the strengths of the \FeII\ and \ion{Fe}{iii} lines with respect to the template. The inferred strength of the 2175~\AA\ bump is $A_{\rm bump} = \pi c_3 /(2 \gamma) = 1.2\pm0.2$, where $\gamma=0.95$ is the width of the bump \citep{Gordon2003}. The best-fit solution is shown in Fig.~\ref{f:dustfit}.
Based on the inferred extinction law from the spectral fitting, we constructed a dust model to reproduce this extinction law by varying the dust-grain size distribution and the ratio of silicate to graphite grains (Si/C). The grain size distribution was assumed to be a power-law distribution with a slope of -3.5 and a maximum grain size of 0.25~$\mu$m similar to the standard ISM size distribution \citep[e.g.][]{Nozawa2013}. The best match to the extinction law was found for a minimum dust grain size of 0.01~$\mu$m (compared to 0.005~$\mu$m for standard ISM dust grains) and a silicate-to-graphite ratio of Si/C = 2 relative to the standard ISM abundance ratio.

\section{Results and discussions} \label{s:dis}

In this section, we first discuss the implications of the emission line properties and why this suggests that we are observing galactic-scale outflows, and we show that the absorbing gas is likely a different manifestation of the same galactic-scale outflows, intercepted by the line of sight to the nucleus. We then derive the physical conditions in the absorbing gas and its distance to the central engine.  
Finally, we substantiate the overall picture by modelling the \lya\ absorption-emission profile both spatially and spectrally.

\subsection{Evidence of galactic-scale outflows}

\subsubsection{The extended emission from ionised gas} 

According to the standard paradigm of AGN, the broad emission lines of quasars arise from dense regions 
with high velocities close to the central engine. Since these regions are far too small  \citep[$\sim 1$\,pc, e.g.][]{Kaspi2007} to be resolved at cosmological distances, information about their sizes is generally obtained through reverberation mapping \citep{Blandford1982}. However, valuable constraints can also be obtained from partial coverage by a foreground absorber \citep[e.g.][]{Balashev2011}. 

The narrow emission lines are in turn expected to arise further away from the obscuring torus, from gas at sufficiently low density where the emission is dominated by forbidden-line transitions. While initially thought to be less than kpc-sized, evidence of the presence of extended narrow line regions (NLR) around diverse types of AGN has increased. This is 
particularly evident for low redshift Type 2 AGN, where the observations are facilitated by larger angular sizes and obscuration of the nucleus emission \citep[e.g.][]{Sun2017}. However, evidence for kpc-scale emission is also seen in high-redshift quasars \citep[e.g.][]{Rupke2017}. 
This also includes the ubiquity of giant ($\gtrsim$ 100 kpc) Ly$\alpha$ nebulae around luminous quasars at $3<z<4$ \citep{Borisova2016}.
The narrow emission lines therefore contain 
important information for the study of ionised gas on large scales.

If the energy from accretion disc winds propagates to the host galaxy ISM, then we expect to observe some connection between the properties of the different types of emission lines.
For example, \citet{Coatman2019} observed a correlation between the blueshifts of 
the overall \CIV\ and [\OIII] emissions, as well as an anti-correlation between [\OIII] equivalent width and \CIV\ blueshift. This suggests that high-velocity accretion disc winds, arising in a compact region very close ($<1$~pc) to the supermassive black hole 
\citep{Netzer2015} can influence the host-galaxy on kpc-scales, assuming [\OIII] is extended over such scales.\footnote{This was actually not constrained by the data obtained by \citeauthor{Coatman2019}} While our observation here concerns a single object, the observed kinematics and equivalent width are in agreement with those observed by these authors. 
More importantly, we also confirm that most of the \CIV\ emission arises from the broad nuclear region, but we observe a weak \CIV\ emission at least as extended as the narrow [\OIII] emission, supporting the above picture.
Finally, \citet{Tombesi2015} found evidence for accretion-disc winds driving galactic-scale molecular outflows, as later confirmed by \citet{Veilleux2017}. Taken together, these works suggest that the H$_2$ absorption system observed here could be linked to the same outflow process.

The commonly-used \citet*[][BPT]{Baldwin1981} diagram provides a useful diagnostic to discriminate between photoionisation by AGN or star formation. 
While we do not have access to the [N\,{\sc ii}] or H$\alpha$ lines, the fact that [\OIII] is detected outside the unresolved nuclear region while $H\beta$ is not (see Figs.~\ref{f:OIII1D} and \ref{f:spa}) suggests a very high [\OIII]/\Hb\ ratio, as expected in the case of ionisation by AGN \citep[e.g.][]{Kewley2001, Kauffmann2003}. This ionisation mechanism is further supported by the large velocity dispersion, FWHM~$\sim$~600~\kms, of the extended [\OIII] emission  \citep[see e.g.][]{Zhang2018}.

As shown in Sect.~\ref{s:spa}, the narrow [\OIII] emission presents two blobs on both sides of the nucleus (at a distance of $\sim$5 kpc projected on the plane of the sky) and with opposite line-of-sight velocities with respect to our reference redshift. 
The bulk velocities of the emission are of several hundred \kms. These are at the limit between what is expected from rotational and non-gravitational velocities \citep[e.g.][]{Comerford2018}. However, the emission is well detached from the nucleus (in particular the red lobe, see Fig.~\ref{f:OIIIspa}) which is better explained by an outflow. This is also consistent with the presence of blueshifted broad [\OIII] emission closer to the nucleus, following the picture of energy propagation from the inner to the outer regions \citep[e.g.][]{Fabian2012}.
Since the [\OIII] emissivity scales with the square of the density, the observed [\OIII] emission traces the densest ionised gas clouds, which, being detached, could indicate compressed regions at the front of the outflow. 
Furthermore, a careful look at Fig.~\ref{f:OIII2D} reveals that as the distance from the nuclear emission increases, the line width of the narrow emission decreases, and the centroid of the emission approaches the systemic velocity (see right panels of that figure). 
These kinematic signatures are consistent with a decelerating outflow and a possible compression at the front of the outflow.  
However, this interpretation remains tentative and requires confirmation through deeper observations with higher spatial resolution, ideally using integral field spectroscopy.

\subsubsection{Kinematics: Absorption and emission as different manifestations of the same wind}

\begin{figure}
    \centering
    \begin{tabular}{c}
    \includegraphics[bb=215 15 393 755,clip=,angle=90,width=\hsize]{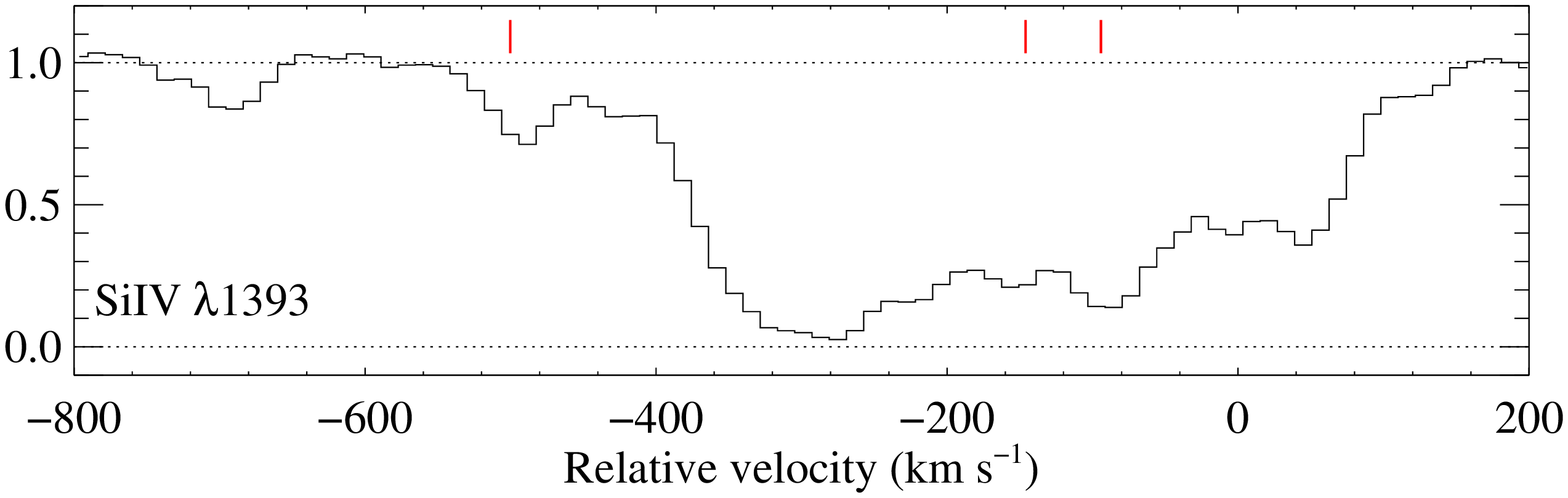} \\
    \includegraphics[bb=215 15 393 755,clip=,angle=90,width=\hsize]{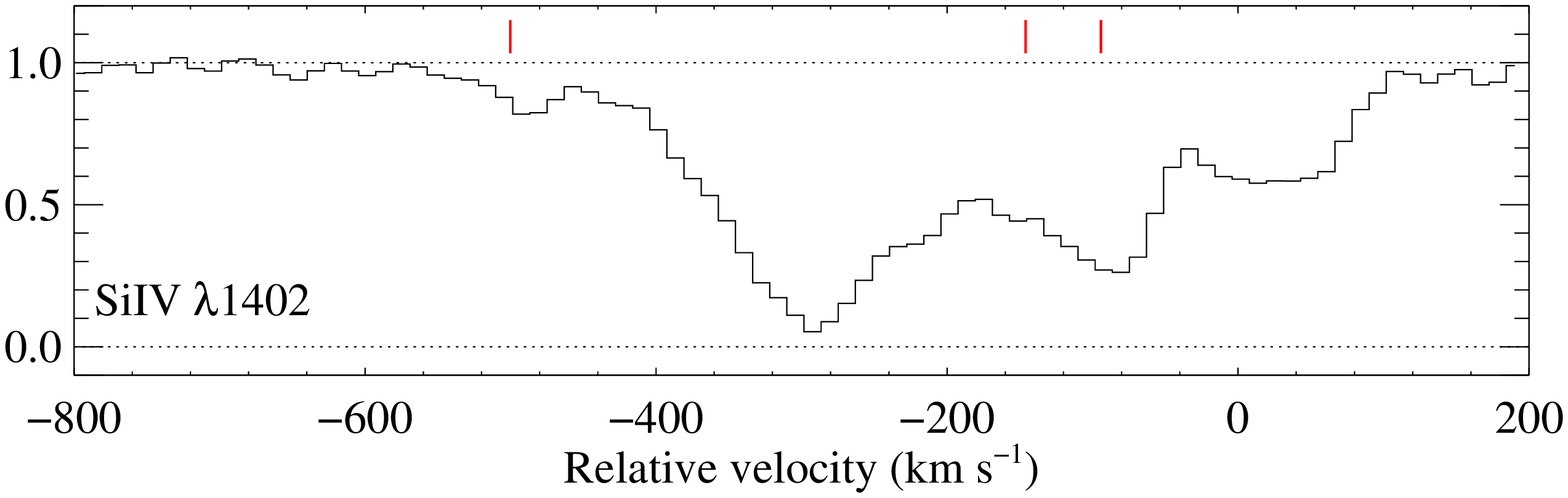}\\
    \includegraphics[bb=165 15 393 755,clip=,angle=90,width=\hsize]{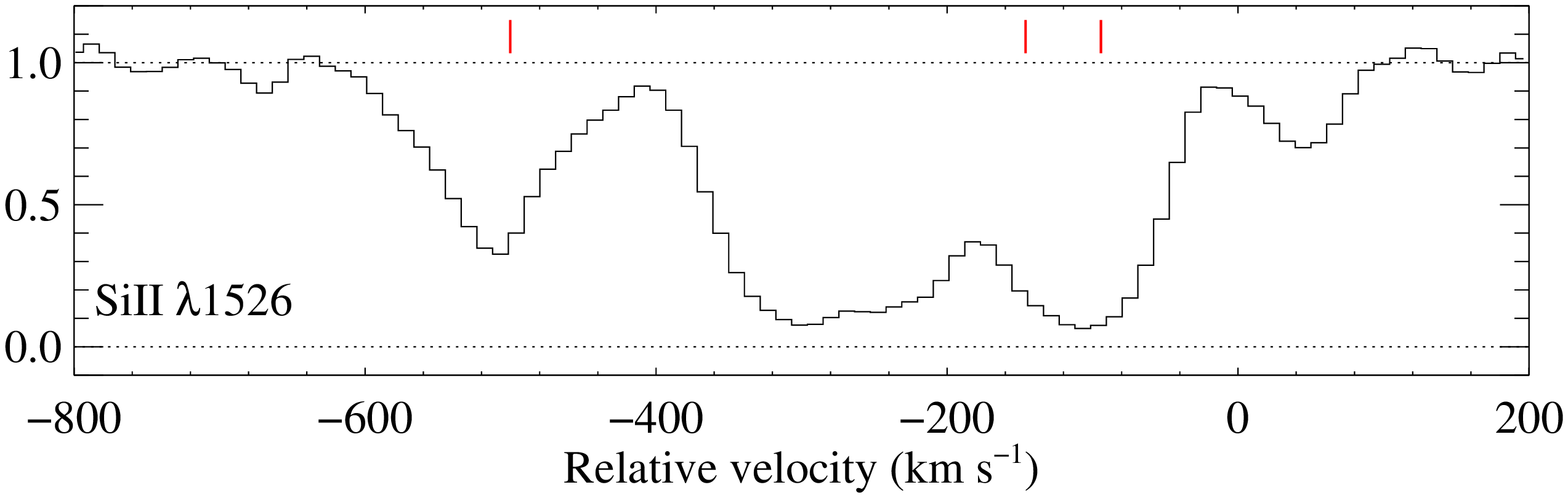} \\ 
    \end{tabular}{}
    \caption{Velocity profile of thrice-ionised silicon (top and middle panels), compared to that of singly-ionised silicon (bottom panel).  Short tick marks indicate the position of H$_2$ components. The zero of the velocity scale is set to $z_{ref}=2.631$.}
    \label{f:SiIV} 
\end{figure}

While the velocity profile of neutral hydrogen seen in absorption is not directly accessible because of saturation of \HI\ lines, 
from Fig.~\ref{f:SiIV} it is clear that the profiles of low-ionisation and high-ionisation metal species share the same overall structure, with several 
components spread over about 600~\kms, almost completely blueshifted with respect to the assumed systemic redshift. 
Such a broad, blueshifted structure is consistent with the expected signature of outflowing gas. The kinematics contrast what is usually seen in intervening DLAs, where the velocity spread of low-ionisation metals are generally much smaller \citep{Ledoux2006} and systematically less extended than more ionised species (see Fig.~4 of \citealt{Fox2007}). 
The profiles observed here suggest that ionised and neutral phases could be well mixed up within the gross structure of the outflowing gas. 

The absorption line kinematics observed in the spectrum of \J\ also corresponds well to the kinematics of the blueshifted, extended, ionised gas seen in emission (the blue component of the extended [\OIII] emission), indicating that we are looking at the same phenomenon both in emission and absorption and that the {kpc-scale} outflowing gas likely contains a mixture of different phases. 
A similar interpretation has recently been put forward by \citet{Xu2019}. Based on the kinematics of [\OIII] emission and \CIV\ absorption in seven quasars, the authors suggest that outflows seen in emission and absorption are likely different manifestations of the same wind.

{}

\subsection{Physical conditions in the absorbing gas}
\label{s:phys}

If the absorbing gas is also part of the outflow, which can typically span around a few tens of kpc from the AGN, then we expect to see effects of the quasar proximity on the physical conditions. 
The presence and excitation of atomic and molecular species provides a unique opportunity to investigate these physical conditions.

Singly-ionised silicon is an interesting species since its excited fine-structure level has been detected in only a few high-$z$ intervening DLAs to date \citep[e.g.][]{Kulkarni2012, Noterdaeme2015} and most likely in regions of high pressure \citep{Neeleman2015}. However, it appears to be more conspicuous in proximate DLAs, where the excitation of silicon fine structure could be enhanced by both high densities and strong UV flux \citep[e.g.][]{Ellison2010, Fathivavsari2018}. Hence, its presence may indicate close proximity to the quasar central engine. However, in the absence of other information, it is hard to discriminate between the case of high density or high UV flux. 

Molecular hydrogen can provide important information, since both the presence of H$_2$ and the excitation of its rotational levels depend on the strength of the UV field, the gas density and its temperature \citep[e.g.][]{Balashev2019}. Owing to formation on dust grains and destruction through line absorption in the Lyman and Werner bands, the presence of molecular hydrogen mainly depends on the ratio between the UV intensity and the gas density \citep[e.g.][]{Sternberg2014}.
This implies that the density required for an atomic-to-molecular transition to occur follows $n\propto d^{-2}$, where $d$ is the distance to the central UV source. For the \HI\ column density and metallicity (or equivalently dust-to-gas ratio) observed for \J, the normalisation of this relation\footnote{We note that this simple estimate assumes the transition to occur and does not exclude the presence of H$_2$ at lower densities (or distances) if the transition is not complete.} is about $\log n\,$(cm$^{-3}) \sim 4$ at $d \sim10$~kpc distances from the nucleus \citep{Noterdaeme2019}. 

\begin{figure}
    \centering
    \includegraphics[width=0.95\hsize]{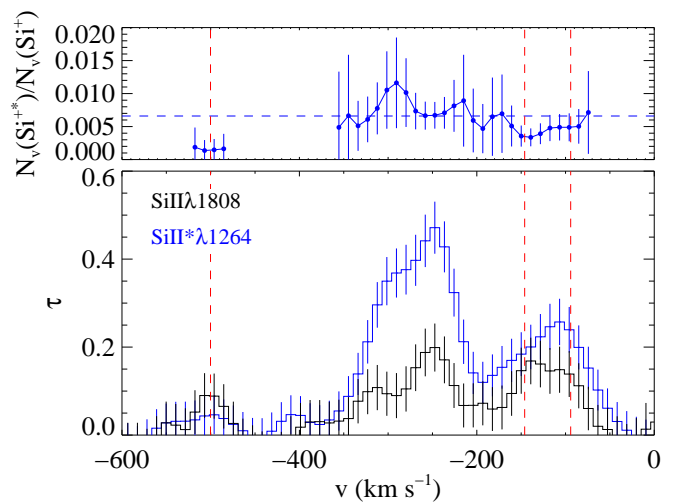}
    \caption{Bottom: Pixel optical depths of \SiII$\lambda$1808 and \SiII*$\lambda$1264 lines. Top: Apparent column density ratio per unit velocity bin (i.e. the ratios of $\tau/(f \lambda_0$)). 
    The location of H$_2$ components is marked by the red dashed lines. The horizontal dashed line shows the overall column density ratio as derived from the Voigt-profile fitting. 
    Values for which the optical depth of either line is below 0.03 lead to large errors and are not plotted. 
    The zero point of the velocity scale is set to $z_{ref}=2.631$.
    }
    \label{f:SiIIext}
\end{figure}

Combining the information from \SiII*\ and H$_2,$ as well as other species sensitive to the density and UV radiation 
(e.g. \CI\ fine structure, whether CO molecules are present or not), can therefore provide constraints on the physical state of the gas and its distance to the central UV source.
In Fig.~\ref{f:SiIIext}, we compare the apparent optical depths of the \SiII$\lambda1808$ and \SiII*$\lambda$1264 lines,\footnote{We caution that \SiII*$\lambda$1264 is partly blended with \SiII*$\lambda$1265. However, the latter has an oscillator strength ten times smaller than the former, so the impact on the apparent optical depth remains small.} which are both optically thin. 
The ground state and excited profiles of singly-ionised silicon have similar velocity profiles with three main clumps, two of which coincide with the positions of H$_2$ absorption components. Yet, the strongest bulk component of \SiII* does not align with the H$_2$ components. In the top panel of Fig.~\ref{f:SiIIext}, we show the ratio of \SiII*/\SiII.  This ratio is not enhanced at the position of H$_2$ absorption, where instead it tends to be {consistently lower than average over several pixels}. Since molecular hydrogen traces the high-density regions in the absorbing gas, this may suggest that the high excitation of \SiII* is not due to high density but rather due to strong UV irradiation.

\subsubsection{Modelling warm and cold gas as successive slabs}

Since the quasar itself is likely the dominant source of photons striking the medium at distances $\lesssim$ few tens of kpc, the UV field received by 
each slab of gas (absorption component) actually depends on the transmitted field through the previous slab of gas 
on the line of sight. While understanding successive filtering slabs of gas makes the study of proximate systems significantly more complex than intervening ones (where the components can generally be treated independently), 
advantages are that the observed column densities correspond to the actual layers through which UV photons propagate, and we have prior knowledge about the shape of this field.\footnote{In turn, modelling of intervening DLAs is generally intended to reproduce the observed column densities, but the actual incident flux intensity and shape is not known (for example, whether it is attenuated or not before striking the cloud) and there is no reason for its direction to match that of the line of sight.}
However, the large number of components and their unknown respective physical locations along the line of sight does not allow for a full modelling of the system, which would have far too many parameters to vary.

We simplified the problem by considering two successive slabs of gas along the line of sight to get an idea of the physical conditions in the different phases. 
In this picture, one slab of 
warm gas, directly illuminated by the quasar, is likely responsible for most of the \SiII*, while the second slab, for which the incident field contains no more H-ionising photons, is in a cold state and contains the H$_2$. 

We used the photoionisation code Cloudy v17.1 \citep{Ferland2017}, running grids of constant density models with plane-parallel geometry illuminated on one side by the built-in cloudy AGN spectral energy distribution. The cosmic microwave and cosmic ray backgrounds were included. The gas phase abundances were set to 
match the observed ones, and we included dust using the grain mixture and size distribution described in Sect.~\ref{s:dust}. In the first runs, the abundance of dust was scaled to self-consistently (i.e. conserving mass) match the observed depletion factors of iron and silicon. This in turn implies depletion factors of about 50\% and 25\% for carbon and oxygen, respectively. 
We tested and found that the results concerning the first layer are actually little sensitive to the abundance of dust, except, naturally, for the calculated extinction, which tends to strongly exceed the observed one due to the large {\sl \emph{total}} column density in the model. However, as we see below, the first layer contains mostly ionised gas, as well as neutral gas with high temperatures ($>10^4$\,K), so it is most likely that grains, in particular small ones, are destroyed. We then assumed no grains in that phase --- but these are kept in the  second (cold) slab.

\begin{figure}
    \centering
    \includegraphics[angle=90,width=0.8\hsize]{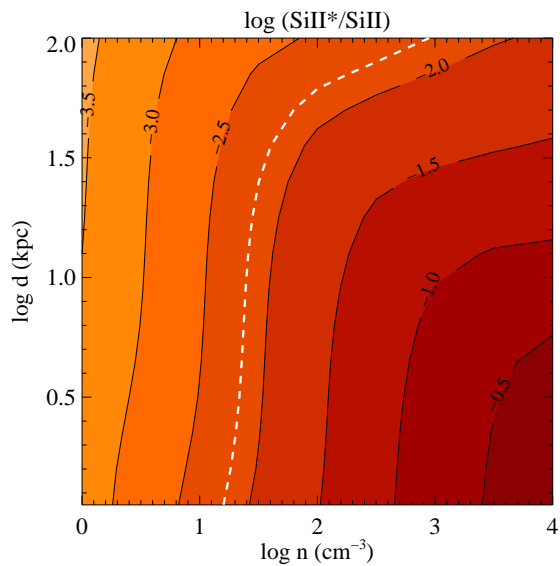}
    \caption{Excitation of ionised silicon fine-structure level as a function of density and distance to the quasar. The dashed white line corresponds to the observed {integrated ratio (i.e. $N($\SiII*)/$N($\SiII))}. 
    }
    \label{f:mod:SiIIs}
\end{figure}

The results concerning the excitation of silicon in the first slab are shown in Fig.~\ref{f:mod:SiIIs}, where the UV flux is converted into distance to the source using the observed quasar luminosity. 
These show that the observed excitation of \SiII*\ can be reproduced by gas with densities 
of the order of $\sim 10$~cm$^{-3}$ at any distance less than $\sim$50 kpc from the source. 
Exciting \SiII\ at the observed level, if located further, would require higher densities. However, this 
would also push the second slab further out, where the cold slab of gas containing H$_2$ would become too cold and little excited compared to observations, as we discuss later. 
We note that the two slabs  need not have the same velocities since 
the ionising cross-section is wide. 

\begin{figure}
    \centering
    \includegraphics[width=0.9\hsize]{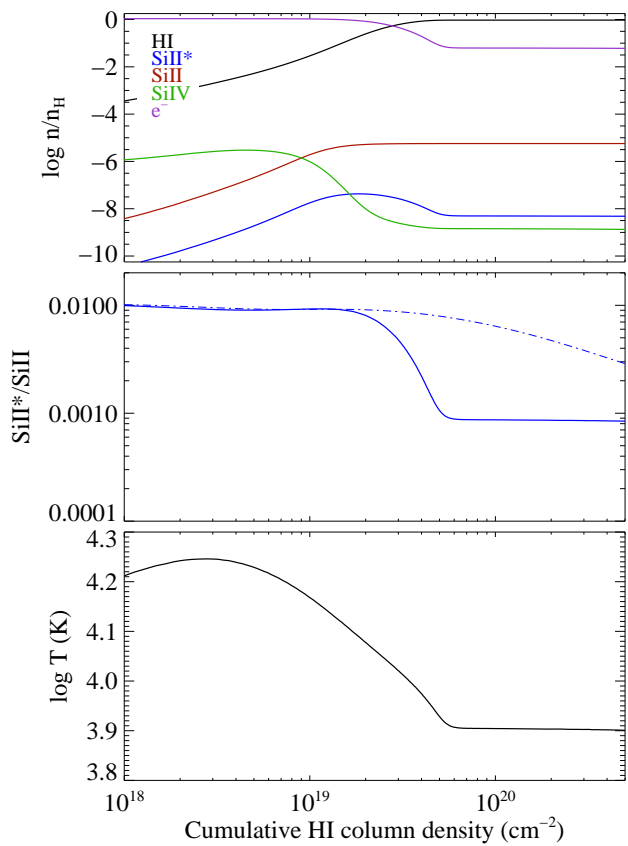}  
    \caption{Abundances of various species (top) and variation of \SiII*/\SiII\ ratio (middle) and temperature (bottom) as a function of cumulative \HI\ column density for a constant density model with $\log n/{\rm cm^{-3}} = 1$ at $d = 10$~kpc. The dashed-dotted line in the middle panel shows the cumulative $N(\SiII*)/N(\SiII)$ ratio.
    }
    \label{f:mod:SiIIsNHI}
\end{figure}

In Fig.~\ref{f:mod:SiIIsNHI}, we show the variation of number densities of neutral hydrogen, electrons,  
and singly- and thrice-ionised silicon, as well as the excitation of silicon and the electron temperature as a function of the cumulative \HI\ column density for an example model with $n=10$~cm$^{-3}$ and $d=10$~kpc. In the ionised gas, the excitation of 
\SiII\ is high owing to efficient collisions with electrons. 
The local \SiII*/\SiII\ ratio then drops by about a factor of ten after crossing the dissociation front, when the gas becomes neutral and both the electron density and the temperature decrease. The \SiII*/\SiII\ ratio then remains constant afterwards, as do $n_{e}$ and $T$. 
However, since singly-ionised silicon is not the main ionisation stage upstream, the overall {\sl \emph{column}} density ratio, $N(\SiII*)/N(\SiII)$ (which is what we observe), is actually $N(\HI)$-weighted, explaining the slow dependence 
on the stopping \HI\ column, as shown by the cumulative $N(\SiII*)/N(\SiII)$ ratio\footnote{{This is the value one would observe (i.e. $N(\SiII*)/N(\SiII)$) at the given depth into the cloud.}} (dashed-dotted curve) in the middle panel. 
    
Changing the UV field affects the depth of the dissociation front (i.e. the column density of upstream ionised gas), but the results remain almost unchanged with respect to the cumulative \HI\ column density.  This explains the low dependence of the $N(\SiII*)/N(\SiII)$ ratio on the distance to the UV source. It is interesting, however, to note that the range spread by \SiII*/\SiII\ at different depths (from about 0.01 in the ionised gas to 0.001 in the neutral gas) corresponds roughly to that seen in Fig.~\ref{f:SiIIs}, which is consistent with the fact that the absorption system also contains ionised gas. 
The main effect of increasing the volumic density is an upwards shift of the \SiII*/\SiII\ curve, as collisional excitation of \SiII\ increases both in the ionised gas (collisions with electrons) and in the neutral gas (collisions with \HI), so \SiII*/\SiII\ is roughly proportional to $n$ in the UV-independent regime. At much higher densities, the increasing dependence of \SiII*\ on the distance to the AGN is mostly due to the dependence of 
the gas temperature and ionisation fraction on the strength of the UV field.
Finally, we note that H$_2$ is not yet produced in the first slab, which is consistent with the non-detection of H$_2$ in the strongest \SiII*\ component.

{}

{}

\begin{figure*}
    \centering
    \begin{tabular}{c c c} 
          \includegraphics[angle=90,width=0.3\hsize]{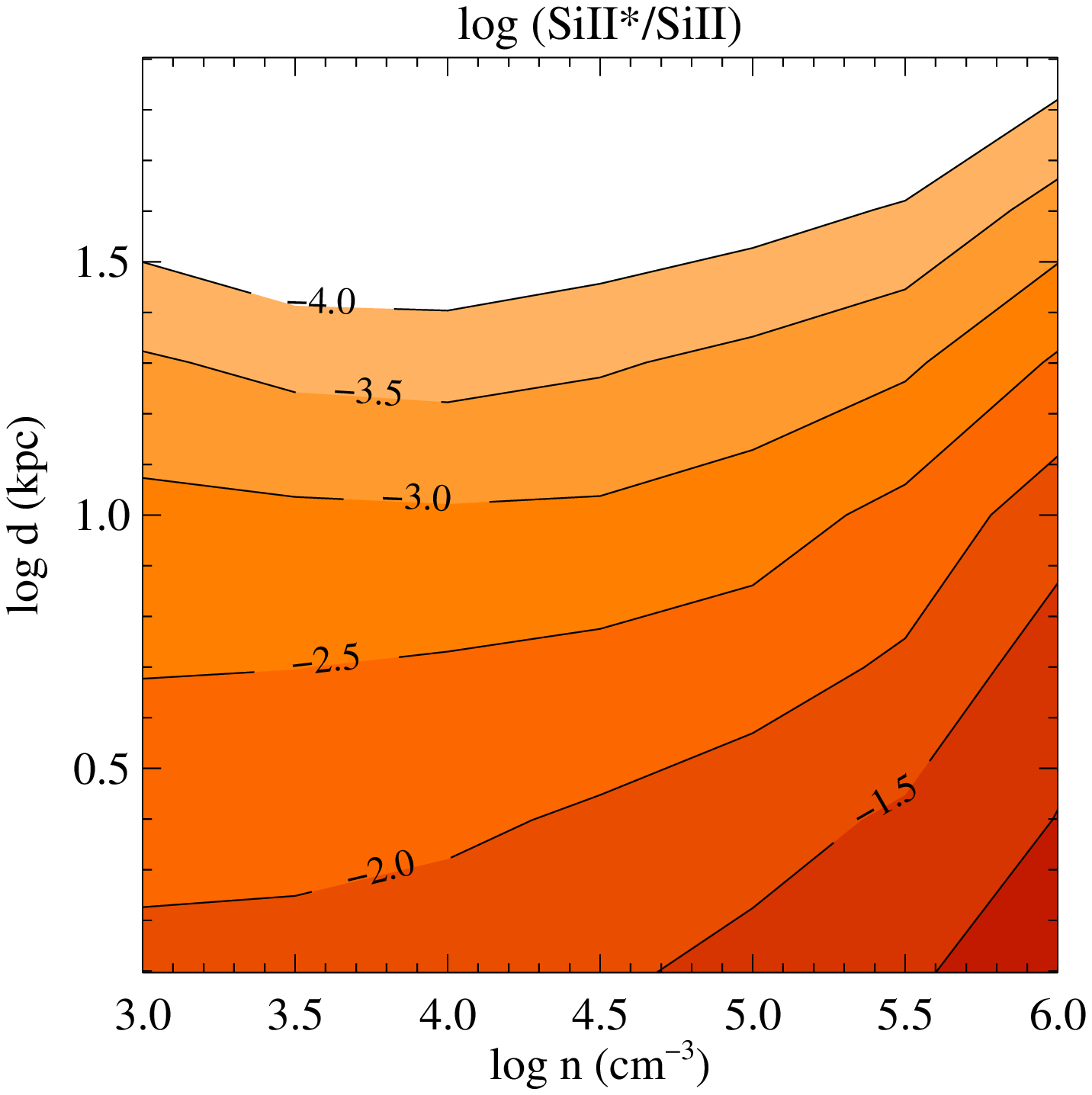}  &
          \includegraphics[angle=90,width=0.3\hsize]{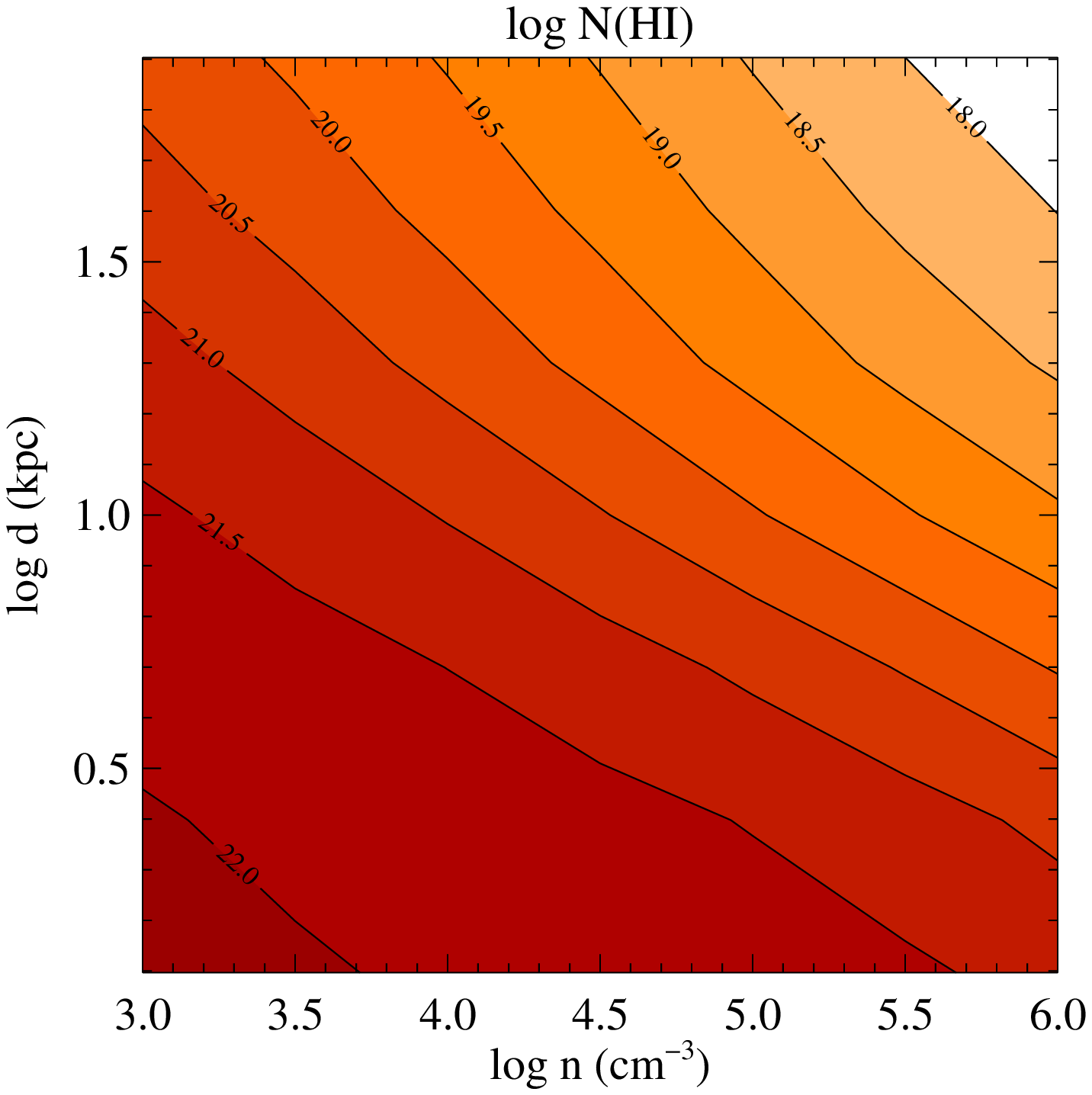}     &
         \includegraphics[angle=90,width=0.3\hsize]{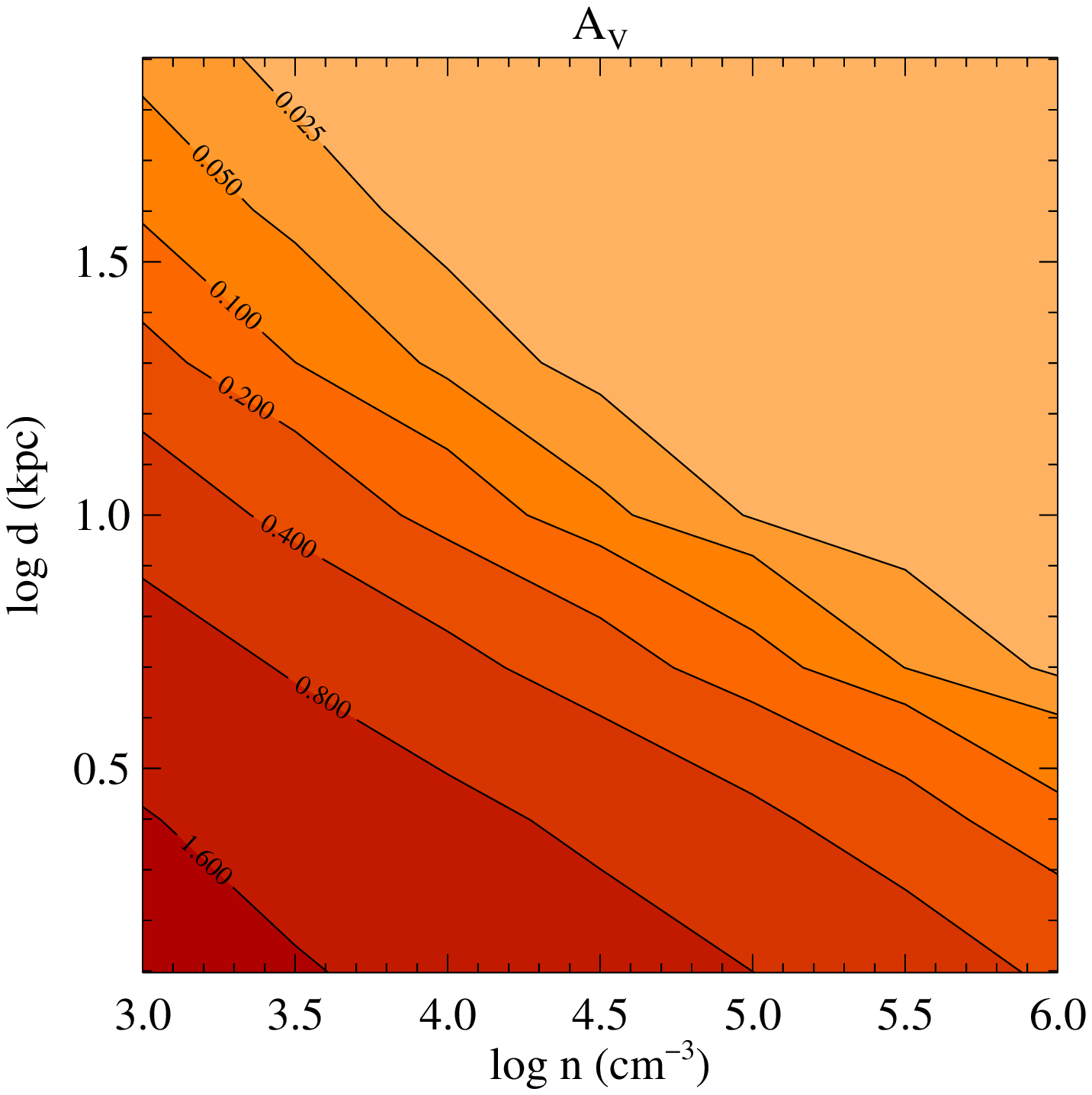}     \\
          \includegraphics[angle=90,width=0.3\hsize]{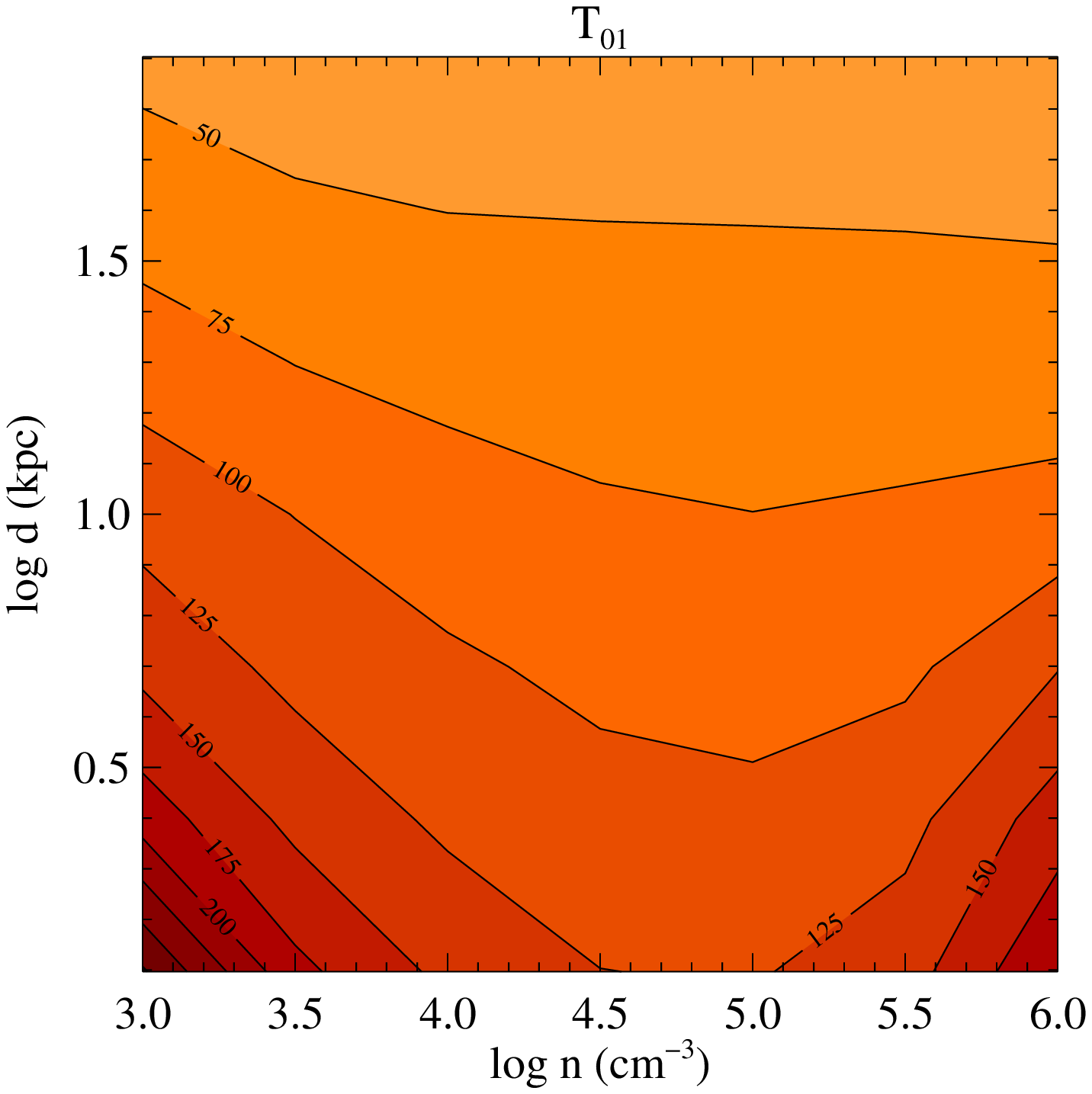}    &
          \includegraphics[angle=90,width=0.3\hsize]{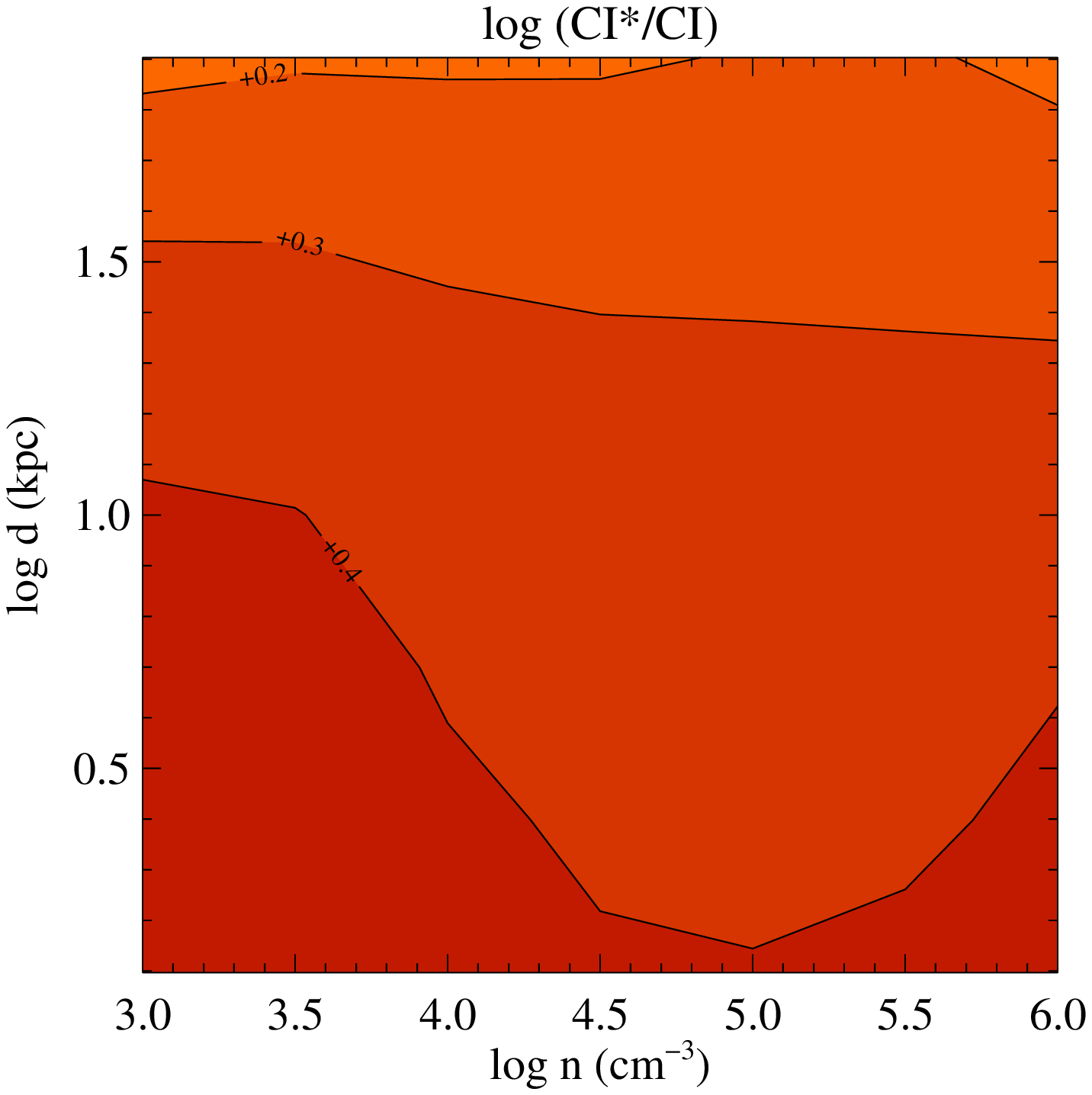}    &  
          \includegraphics[angle=90,width=0.3\hsize]{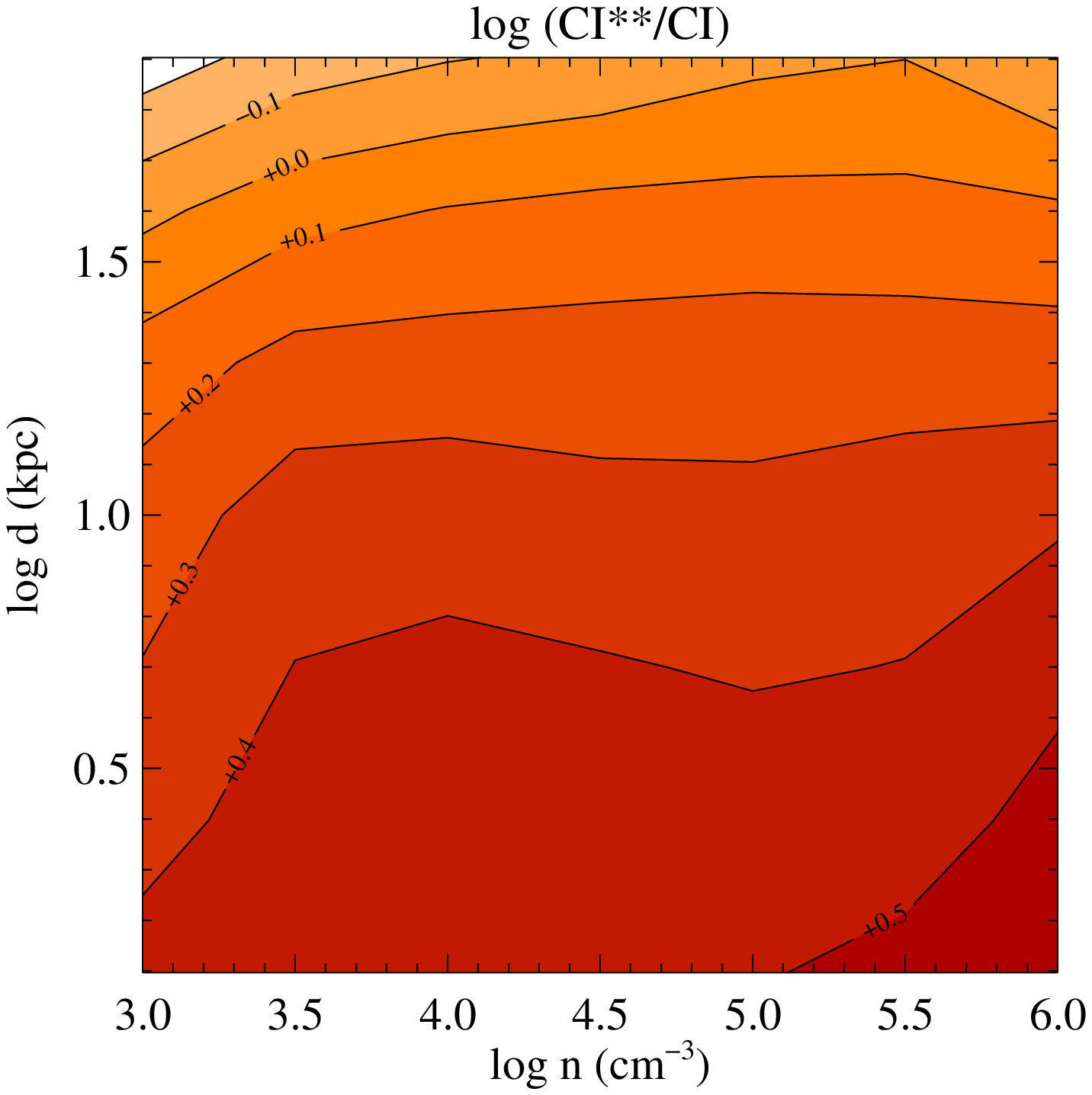}   \\  
          \includegraphics[angle=90,width=0.3\hsize]{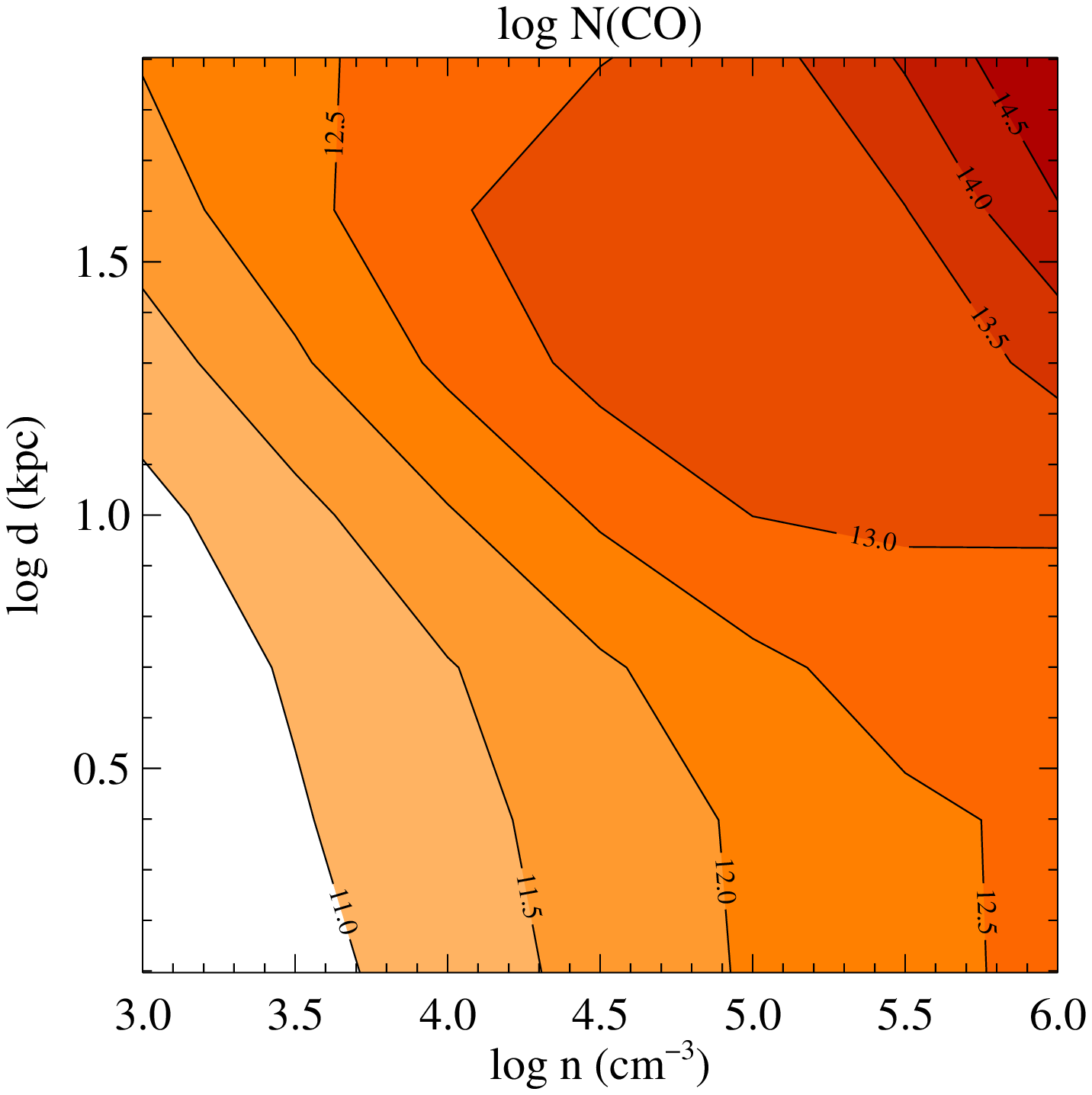}     &  
         \includegraphics[angle=90,width=0.3\hsize]{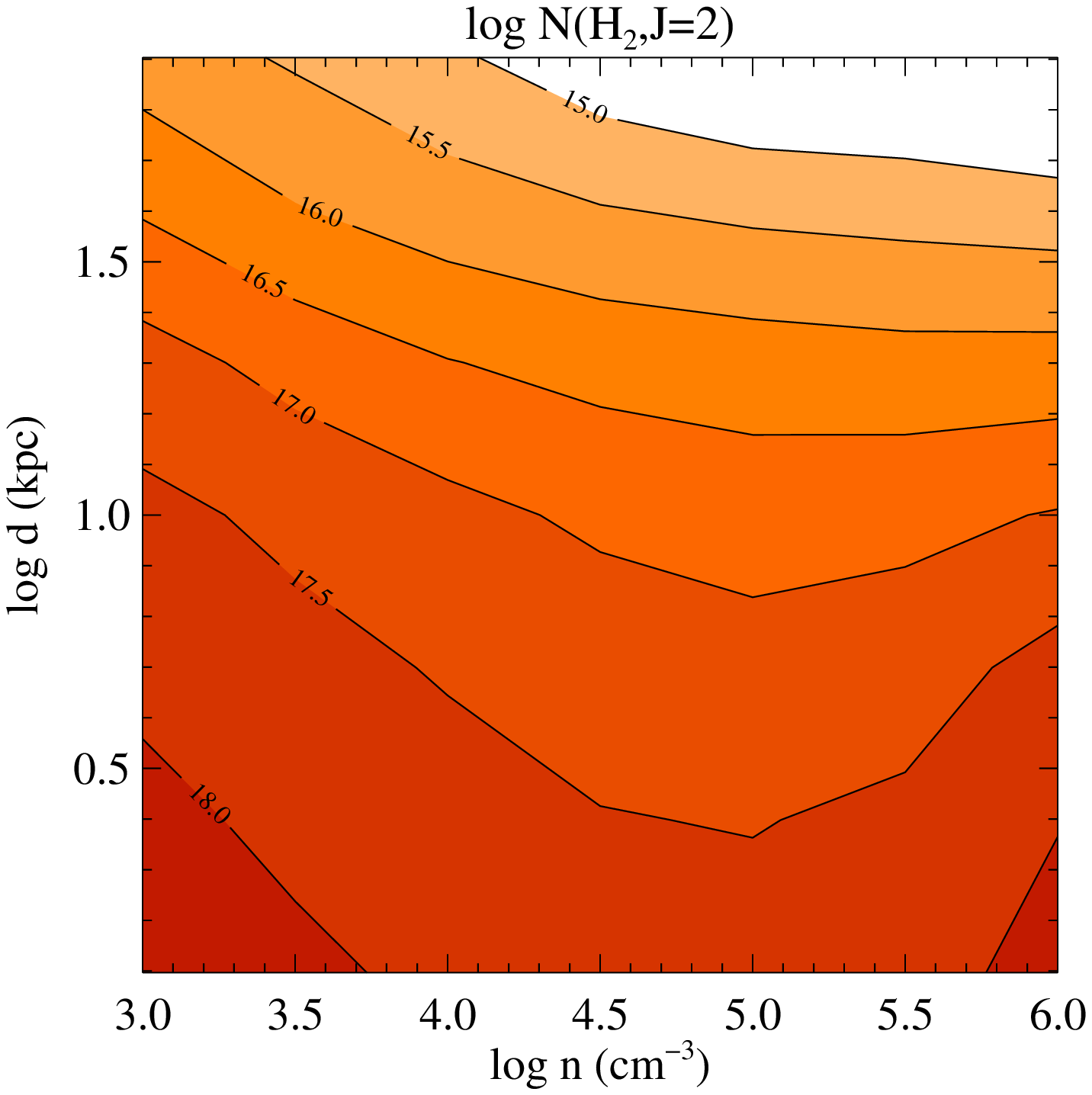}    &  
          \includegraphics[angle=90,width=0.3\hsize]{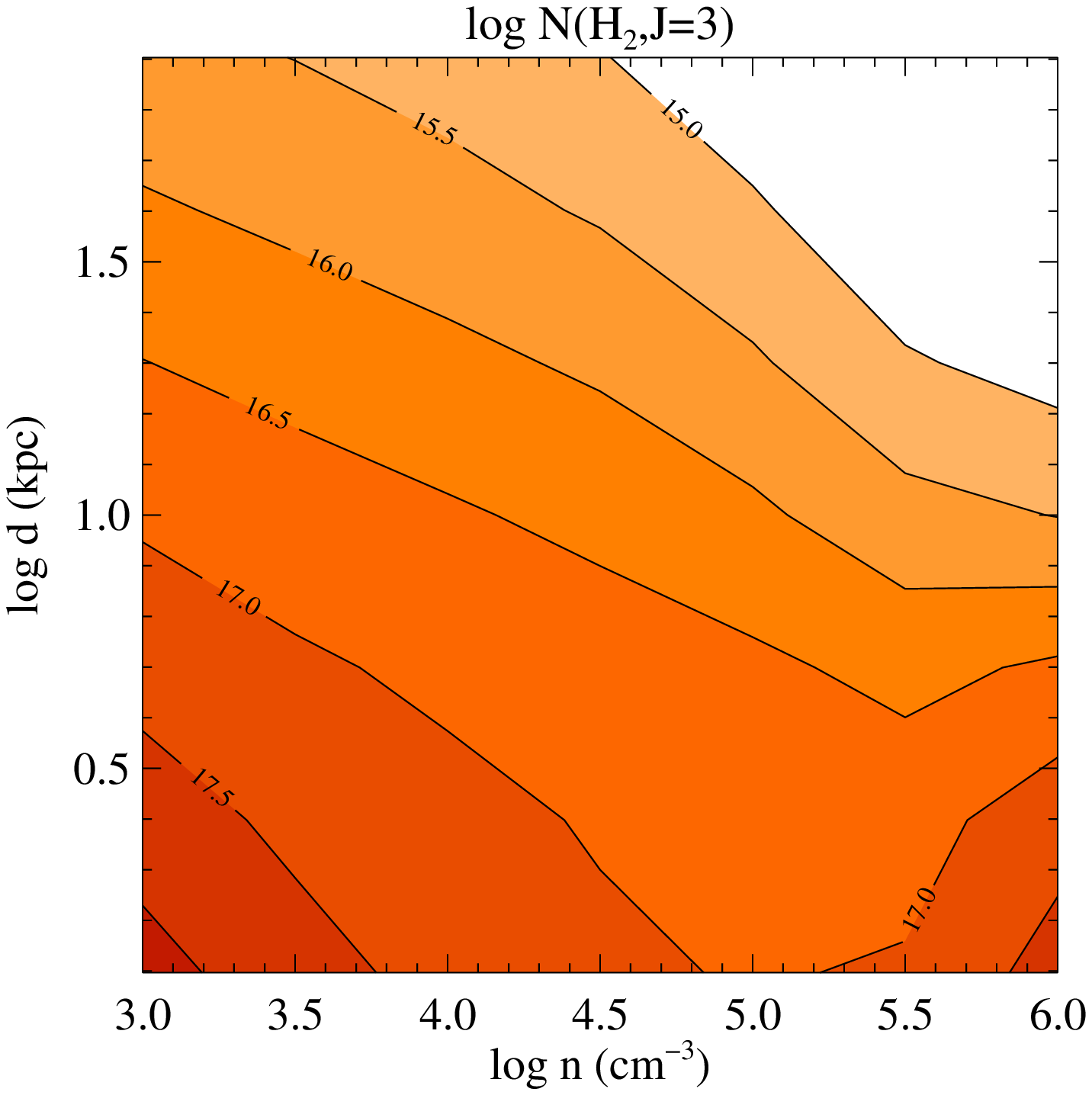}   \\  
    \end{tabular}
    \caption{Cloudy model predictions for the second slab of gas, with $\log N(\HH)$=19.3. }
    \label{f:mod:H2}
\end{figure*}

We modelled the second slab of gas again using constant density models at various distances from the quasar. However, the striking radiation field is now the transmitted field through the first slab. The main effect of the first slab on the incident field is a removal of hydrogen-ionising photons. This means that the assumed parameters for the first slab actually have little effect on the second slab. 
We stopped the calculations of this second slab when reaching $\log N($H$_2) = 19.3$, 
as a representative value for individual H$_2$ components. The results are shown in Fig.~\ref{f:mod:H2}. 

{}

The excitation of \SiII*\ is now significantly lower compared to the first slab for a given 
distance and density as the temperature decreases considerably. The \SiII*-to- \SiII\ ratio reaches the 
same level as in the first layer for distances of only a few kpc, for densities up to $\log n\sim5$, 
and it increases at higher densities.  
However, there is also indication that the excitation of 
silicon is less at the velocities of H$_2$ components, with log(\SiII*/\SiII) between roughly $-3$ and $-2.5$, than at other velocities (where, as we showed above, the high excitation is easily explained by the high pressure in the first layer).
In that case, the distance is slightly larger, between 5 and 12~kpc.  

The $T_{01}$ excitation temperature strongly depends on the UV flux, but more weakly on the density. The observed value of about 100~K corresponds to distances of the order of 5--15~kpc. 
At such distances, the high rotational levels are highly populated, even though they remain lower than what is observed for the total column in the model (equating that of component C). This is a common situation in the modelling of static slabs and depends much on the actual geometry and turbulence, so we do not consider the 0.5~dex as a strong mismatch between the data and model.

The excitation of neutral-carbon fine-structure levels has also proved to be a useful probe of the physical conditions in H$_2$-bearing DLAs \citep[e.g.][]{Silva2002}, in which it is usually detected \citep[e.g.][]{Srianand2005,Noterdaeme2018}. Here, the \CI*/\CI\ ratio is notably high in all three components, reaching the limit of a statistical weight ratio, and it appears to be mostly populated by UV pumping. The \CI*/\CI\ favours distances smaller than the above 10~kpc, it but remains consistent within the large uncertainties of the observations over most of the parameter space probed. 
For component C, where the constraints are slightly better, the excitation of \CI\ rejects distances larger than about 10~kpc at $\sim 2$\,$\sigma$. 

The strict upper limit on the \HI\ column density is consistent with these distances 
as long as the density is as high as $\log n \sim 4.5$ or more. The dust extinction is then also 
within the allowed range. Finally, the absence of CO at a detectable level ($\log N(\rm CO)<13.4$ (3\,$\sigma$) following the procedure presented by \citealt{Noterdaeme2018}) rejects the high density, large distance corner of the plot. 

Our model is of course very simplified, as the actual system presents numerous components at various velocities and unknown respective spatial location along the line-of-sight. The chemical properties, gas-phase metal abundances, and dust grain properties can also vary from one component to another. We also explored
a number of varying parameters, like turbulence, shape of incident radiation field, column density of the first layer, and the use of the simple removal of H-ionising photons instead 
of a full transfer calculation. All this typically affects the distances and densities further by a factor of two. We thus conclude that within a factor of a few, the distances of the clouds are of the order of {$\sim 10$~kpc}, and H$_2$ is produced in clouds with densities as high as $n\sim 10^{4.5}$\,cm$^{-3}$, located downstream, after H-ionising photons have been removed by a warm atomic layer. The latter is also responsible for most of the \SiII\ fine-structure excitation, {while \CI\ is only present in the denser molecular gas}.

\subsubsection{Modelling the excitation of H$_2$ components individually using Meudon PDR}

In order to further test our results on the molecular phase, we used the {\sc Meudon PDR} code \citep{LePetit2006} to model the population of all detected H$_2$ rotational levels in the different individual components separately. We ran grids of isobaric cloud models, with thermal pressure from $10^4$ to $10^8$\, $k_B$ K cm$^{-3}$ and illuminated by a UV field with intensity from 
1 to $3\times10^4$ times the Mathis field \citep{Mathis1983}, corresponding to distances of $\sim160$ to $1$ kpc from the central engine.\footnote{Assuming the UV field is dominated by the AGN. We note that the there is no option to set the external radiation field shape to AGN in the PDR Meudon code. However, in the UV range, the shape of the Mathis field is quite similar to that of typical AGN, so that uncertainties related to the shape of the field are negligible compared to those arising from the unknown geometry of the clouds.} 

\begin{figure}
    \centering
    \includegraphics[width=0.95\hsize]{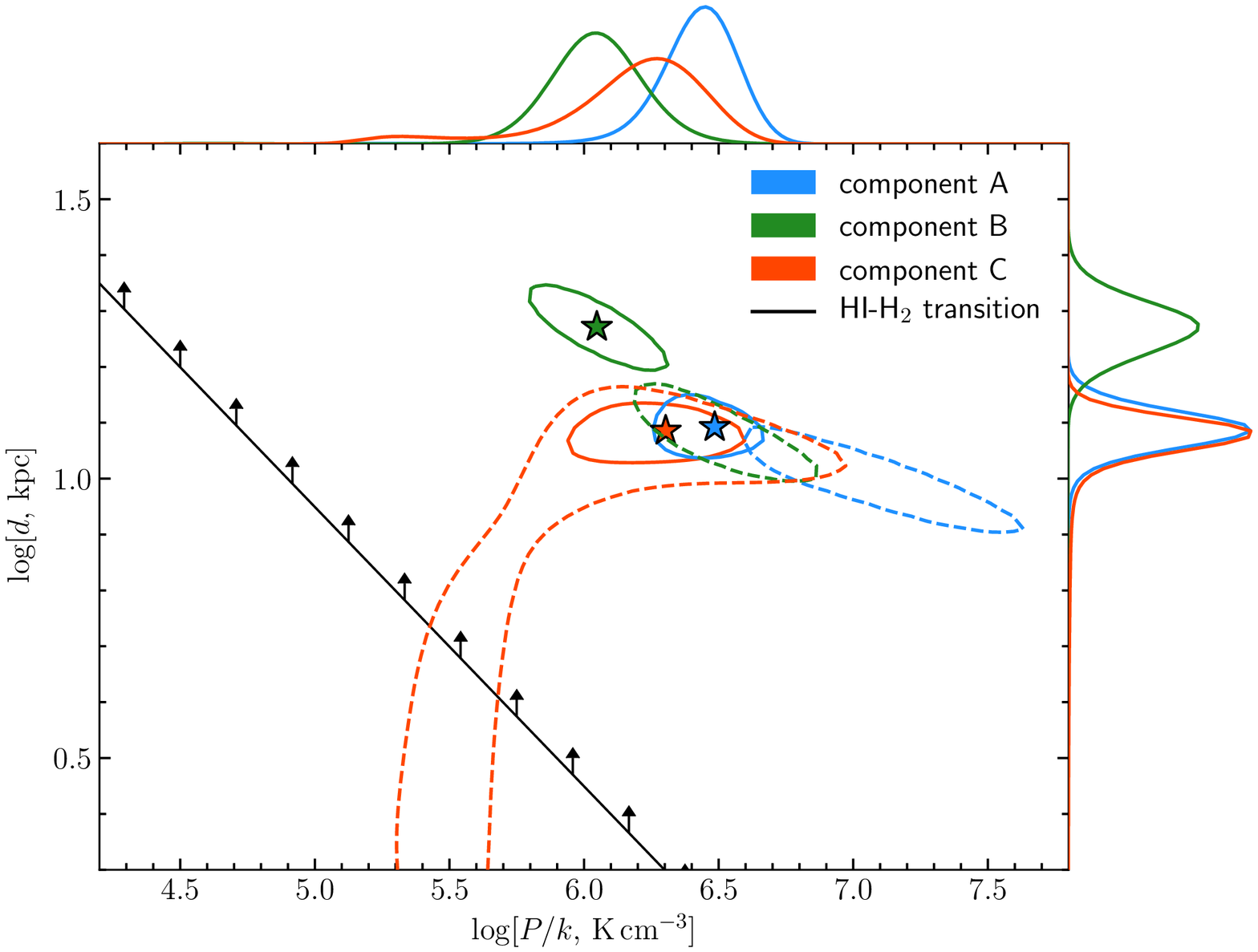}
    \caption{Constraints on thermal pressure, $P$, and distance from cloud to AGN, $d$, obtained by modelling the excitation of H$_2$ rotational levels with the {\sc PDR Meudon} code. The contours show formal regions that contain 0.683 probability distribution functions. The solid and dashed lines correspond to the estimates based on MCMC and vpfit result of H$_2$ fit (see Table~\ref{t:H2}), respectively. The 
    marginalised probability distribution functions (along right and top axes) and the best-fit values (stars) are shown for MCMC results only for the sake of clarity.
    The black line with upwards arrows show the constraints from the H$_2$-\HI\ transition, assuming the kinetic temperature is T$\sim 100$\,K, which is consistent with the observations.
    }
    \label{f:mod:H2_Meudon}
\end{figure}

\begin{figure}
    \centering
    \includegraphics[width=0.95\hsize]{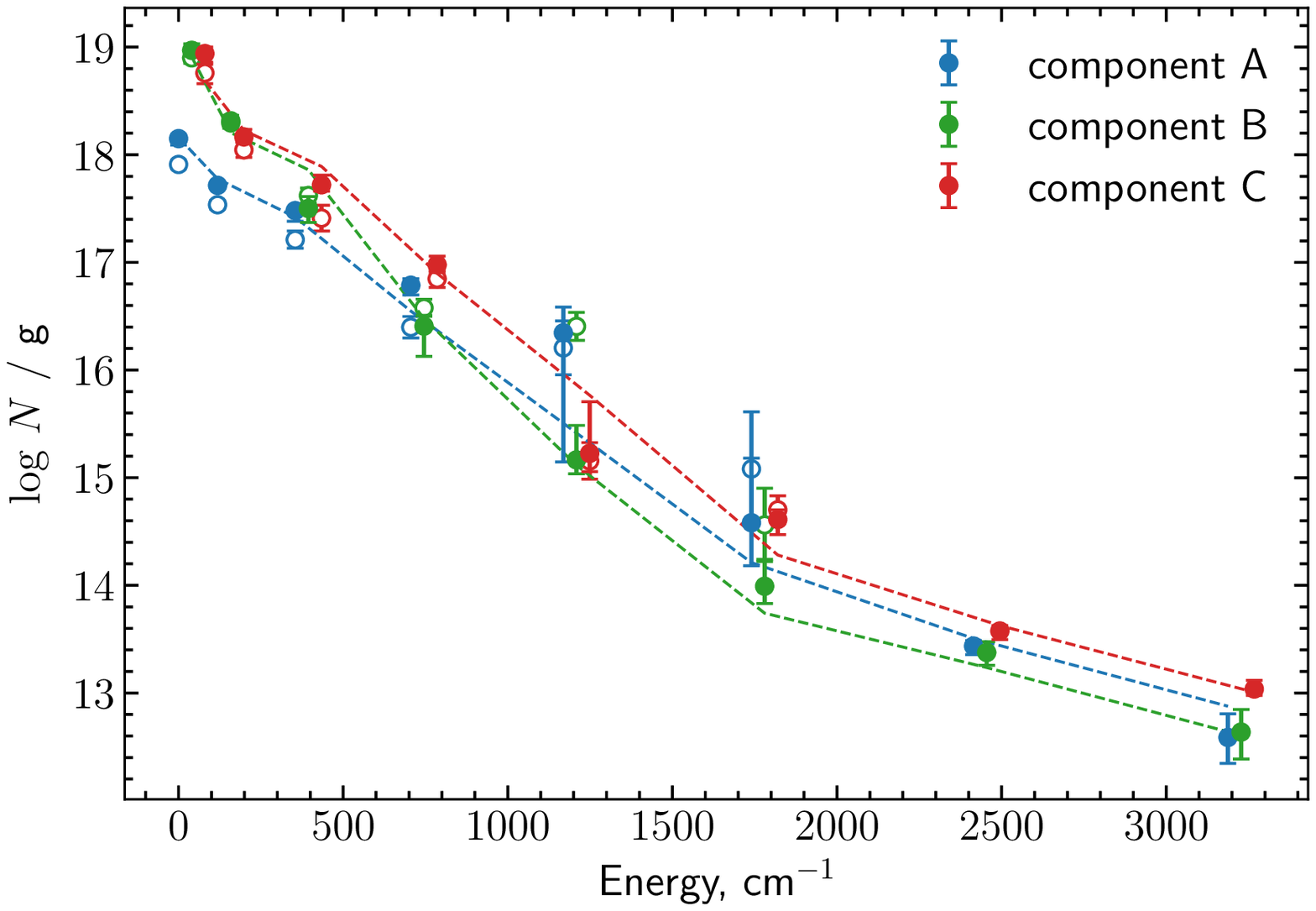}
    \caption{Comparison of model and observed excitation diagram of H$_2$ rotational levels. The filled and open circles represent the measurements using vpfit and a MCMC method (see Sect.~\ref{s:abs:h2}), respectively. The dashed lines show the best fit of the excitation diagram (corresponding parameters $d$ and $P$ marked by stars in Fig.~\ref{f:mod:H2_Meudon}). 
    The data points are artificially shifted on the x-axis to ease representation. }
    \label{f:mod:H2_Meudon_exc}
\end{figure}

We modelled each H$_2$ component independently, that is, we did not take into account shielding in H$_2$ UV lines
of one component by another component located closer to AGN.  
The column densities in the different rotational levels are all compared to the data using a simple least-square likelihood function. 
The constraints on the thermal pressure, $P$, and distance to AGN, $d$ for all three H$_2$ components are shown in Fig.~\ref{f:mod:H2_Meudon}. The constraints are similar for vpfit and MCMC H$_2$ fitting methods.
The most likely values correspond to 
high pressures $P / k_B\sim 10^6 - 10^{7}$\,K cm$^{-3}$ located at distances of 10--20 kpc for all three components. {The corresponding modelled excitation diagram is shown in Fig.~\ref{f:mod:H2_Meudon_exc}.}
The typical kinetic temperatures in the H$_2$ dominated cores of the clouds is then $\sim 200-300$\,K, which is slightly higher than the observed $T_{01}$. The number density in the H$_2$ -dominated medium is in the range  of $10^4$ to $10^5$ cm$^{-3}$, which is in agreement with the results using Cloudy. We also ran isochoric models and found that while isobaric models reproduce the observed H$_2$ excitation diagram slightly better, the constrained values of $d$, $P,$ and $n$ remain very similar. 
We note that the derived distances are also consistent with the limits predicted by the static H$_2$-\HI\ transition theory from \citet{Sternberg2014}, shown as a black line in Fig.~\ref{f:mod:H2_Meudon}: below these lines, the predicted \HI\ columns in the 
envelope of H$_2$ clouds would be higher than the total $N(\HI)$ observed.

While the distances are constrained to be 
of the order of 
10\,kpc, these do not 
allow us to determine which component is located closer to the AGN. In any case, since cold clumps can drop out from the bulk motion \citep[see e.g.][]{WU2020}, the information of the respective 
locations of H$_2$ clouds would not likely tell us whether the bulk outflow is decelerating or not. 

Interestingly, the high density derived in the H$_2$-bearing medium implies very small longitudinal sizes, of the order 
of $\sim 100$~a.u. only. This is actually comparable to the size of the accretion disc, assuming the relation from 
\citet{Morgan2010} and the black hole mass of $\log M_{\rm BH}/M_{\rm \odot}=8.6$ derived from width of the \CIV\ line (FWHM~$\sim 3000\,\kms$) and the rest-frame continuum luminosity at 1350~{\AA} using the calibration from \citet{Vestergaard2006}. However, we do not see any partial coverage of the background source by H$_2$. This could be due to the nuclear emission arising from a smaller area than expected, or the transverse extent of H$_2$-bearing gas being larger than the longitudinal one. The latter could be effectively reproduced by a large covering factor of H$_2$-bearing clumps around the component velocity. 
Given the typical (longitudinal) velocities of the components of $~\sim 400$\,\kms\ future observations on timescales of a few years could reveal a variability of the H$_2$ absorption lines, if the transverse velocities are of the same order.

{}

{}

\subsection{The leaking Ly$\alpha$ emission}

Sensitive integral field spectrographs, in particular MUSE at the VLT \citep{Bacon2010}, have revealed 
a near ubiquity of \lya\ nebulae around high-redshift quasars \citep[e.g.][]{Borisova2016,ArrigoniBattaia2018}. Extended \lya\ emission has 
been observed up to hundreds of kpc in some cases, and routinely over several tens of kpc. In turn, the strong glare 
from the quasar itself complicates the observations of the \lya\ emission at the small distances from the nucleus. 
In the case of \J, the presence of the DLA removes the nuclear emission, leaving us with free access to the 
\lya\ emission as soon as it is outside the BLR.  

To verify if the conjectured configuration of clumpy, gaseous outflows from
\J\ is a likely representation of the system (see cartoon in Fig.~\ref{f:cartoon}), we constructed a 3D model in an
adaptively refined mesh and carried out \lya\ radiative transfer (RT). 
The model is based upon our observational constraints, {as discussed below}; where we have no
constraints, we assumed plausible values {(in particular concerning the properties of the inter-cloud medium)}. In both cases, we then varied the
physical parameters in certain ranges to get a feeling for the importance and
degeneracies of these parameters.

The \lya\ RT was conducted using the Monte Carlo code {\sc MoCaLaTA}\footnote{
    \href{https://github.com/anisotropela/MoCaLaTA}
    {{\tt github.com/anisotropela/MoCaLaTA}}.
} \citep{Laursen2009a,Laursen2009b}. Given a grid of cells containing thephysical parameters of the gas (densities, temperature, and velocity field),
the code traces the paths of individual photons, producing a 3D (two spatial +
one frequential) cube that can be compared to our observation.
Given the large number of unknowns, rather than performing a stringent fit to
the observed spectrum (as has sometimes been done, see e.g.
\citealt{Verhamme2008,Gronke2017}), we aimed to construct a model that would at the same time be consistent with our various observational 
constraints, and --- most importantly --- roughly reproduce both the observed \lya\ spatial and spectral shapes.

The gas is assumed to have been expelled in a biconical outflow 
down through one of which we observe the quasar nucleus. 
Given the similar transverse and longitudinal length, we set the opening angle of the cone to $\theta\simeq45^\circ$. The number of low-ionisation absorption lines seen along the line of sight (covering factor $f_c \sim 10$) can 
then be reproduced distributing $N_\mathrm{cl}\simeq750\,000$ effective clouds in 
the outflows, each of radii $r_\mathrm{cl}\simeq40\,\mathrm{pc}$, $T_\mathrm{cl} = 10^4\,\mathrm{K}$ and $n_\mathrm{HI,cl} = 0.25\,\mathrm{cm}^{-3}$.
These are located between inner 
and outer radii of 16 and 20~kpc, respectively\footnote{These values are taken to be representative of the distances derived from the physical conditions and the fact that clouds are more easily ionised close to the quasar. As we aim at a global agreement between the model and actual data and still have many unknowns, we chose to fix them instead of considering these radii as 
free parameters.} and immersed into a mostly ionised inter-cloud medium with  $T_\mathrm{ICM} = 10^5\,\mathrm{K}$  and a residual neutral gas $n_\mathrm{HI,ICM} =
0.005\,\mathrm{cm}^{-3}$. Both phases have metallicity $Z = 0.5\,Z_\odot$. With
these numbers, the covering factor is $f_c = 10^{+3}_{-2}$ \HI\ column density is $\log N(\ion{H}{i})=20.63^{+0.09}_{-0.11}$.
The gas follows a velocity profile decreasing linearly from 250 km~s$^{-1}$ to
0 at the end of the jet. In addition to this, the clouds have a velocity
dispersion of $\sigma_{v,\mathrm{cl}} = 10\,\mathrm{km}\,\mathrm{s}^{-1}$.
Finally, photons are emitted from the centre with an intrinsic spectrum given
by the reconstructed unabsorbed emission (see Fig.~\ref{f:Lya}).

\begin{figure}
    \centering
    \includegraphics[width=\hsize]{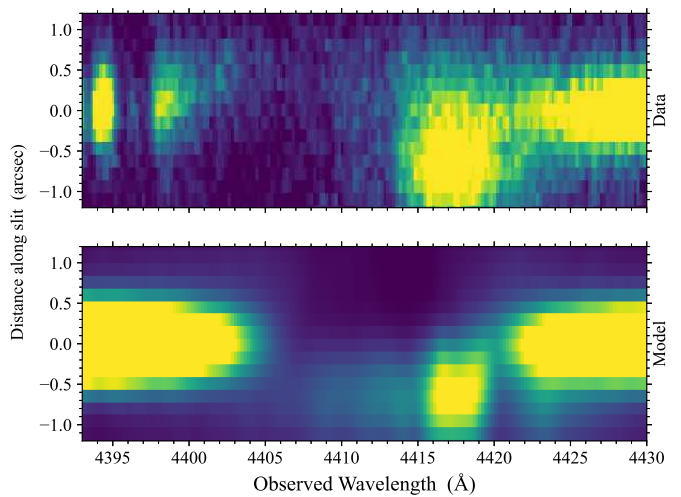}
    \caption{Comparison between simulated 2D spectrum of photons escaping in the particular direction of the
observer looking down the barrel of the jet (bottom) with the actual observed X-shooter 2D spectrum (top). The distance 
along slit is taken from the unresolved quasar continuum emission. 
We note that the observed spectrum is also affected by unrelated absorption, which we do not model here (e.g. at 4396$\AA$).
    \label{f:simspec}}
\end{figure}

\begin{figure}[!ht]
    \centering
    \includegraphics[width=\hsize]{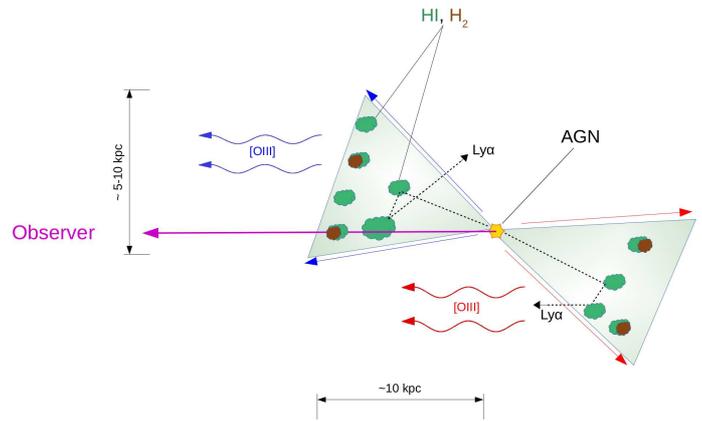}
    \caption{Cartoon of inferred configuration of the system. The emission properties 
    of the unresolved active nucleus (e.g. blueshifted broad [\OIII]) are not shown to avoid overcrowding the figure. \lya\ photons can be more easily scattered out from the cones towards the observer from the receding outflow and appear red peaked.}
    \label{f:cartoon}
\end{figure}

We generated a 2D spectrum of the photons escaping (in the particular direction of the
observer looking down the barrel of the jet) as would be seen with X-shooter, that is, with the 
same slit parameters, resolution, and seeing as observed. This is compared to the observed
spectrum in Fig~\ref{f:simspec}. While far from a perfect match, our model
reproduces $(i)$ the damped absorption line towards the nucleus, $(ii)$ the 
red-peaked leaking \lya\ emission, and $(iii)$ the spatial offset of this emission down the trace.
We note that \lya\ photons are also likely to be produced from recombination in the outflow 
as well as in the host galaxy, hence contributing to the overall observed profile.

\section{Conclusion} \label{s:con}

Active galactic nucleus feedback is a very complex issue that motivates significant efforts 
from the community in terms of theory, simulations, and observations. 
While progress in understanding galactic-scale outflows as the main observable evidence of feedback is being made quickly, \citet{Cicone2018} recently noted that these remain largely unconstrained as they imply various scales and different gas phases. Furthermore, the observation of outflows, especially at high redshift, is notoriously difficult.

Our VLT/X-shooter observations of the proximate molecular absorber towards \J\ suggest 
that spatially resolved emission of ionised gas and kinematically wide absorption in various phases towards the nucleus are different manifestations of a single outflowing process. From detailed investigation of the physical conditions in the cold phase of the absorbing gas, 
we derived a distance to the central engine of the order of 10~kpc. This, together 
with the similar kinematics and transverse extent of the ionised gas seen in emission, 
indicates a picture of a multi-phase quasar outflow, down the barrel, that is, intercepted by the line of sight to the bright nucleus. This configuration is corroborated by modelling of the resonant \Lya\ emission-absorption spectral and spatial profiles.

Of course, it may be reasonable to recall that this is likely an extremely simplified version of the actual geometry. It is, for example, well possible that the ISM in the quasar host is disrupted by the ionised outflow, giving rise to a much more complex structure of the various phases in the outflow. 
While the present observations open up a new way to study outflow properties, they do not yet tell us what the effect of these outflows on the host galaxy are. 
Understanding how much molecular outflows affect the host galaxy's reservoir, and hence whether outflows are effectively quenching star formation, could be addressed from spatially resolved observations of CO emission lines 
\citep[e.g.][]{Carniani2017}. 

Since the present analysis includes only a single object, 
it is legitimate to ask whether the prevalence of (multi-phase) outflowing gas is larger in quasars with proximate H$_2$ absorbers than in the overall population. By spreading out H$_2$ gas, outflows may contribute to increasing the H$_2$ absorption cross-section, and hence proximate H$_2$ absorbers could represent an effective way of pre-selecting these.  
If, as shown here, outflowing gas facilitates the scattering of \lya\ photons on galactic scales, then the presence of leaking \lya\ emission in a fraction of them could be an additional effective criterion. Finally, the presence of excited atomic species could indicate proximity of the gas to the central engine. Outflowing gas being responsible for a fraction of proximate H$_2$ absorbers could provide 
an explanation to the at least five-fold enhancement of the incidence rate of H$_2$-selected proximate DLAs compared to intervening statistics \citep{Noterdaeme2019} when \HI-selected proximate DLA show only marginal enhancement \citep{Prochaska2008} and likely missed systems with strong \lya\ leakage. Of course, a fraction of proximate H$_2$ absorbers could also originate from galaxies in the quasar group. 

We note that a correlation between a fraction of leaking \lya\ photons (not absorbed by the foreground cloud) and the strength of \SiII*\ in metal-selected proximate DLAs (without information of H$_2$) has been observed by \citet{Fathivavsari2020}.  The author interprets this as a decrease of the covering fraction of the background \lya\ emission by clouds with increasing compactness (i.e. of higher density for a given column density).
However, there may be no causality between the density inside individual (sub)-pc-scaled clouds and the fraction of leaking \lya\ photons (over kpc-scales) throughout the whole system. 
Instead, the prevalence of dense clouds and the 
\lya\ leaking fraction may both depend on the gas flow configuration, including the outflow viewing angle (see e.g. \citealt{Gupta2006} for a similar discussion on the detectability of H~{\sc i} 21-cm absorption) and quasar properties. 
Similarly,  \citet{denBrok2020} recently suggested that the morphologies of extended \lya\ nebulae can be used to understand the geometry of high-redshift AGN on circumnuclear scales. 
Nevertheless, the interpretation of single clouds partially covering the nucleus may still be valid for extreme cases where the \lya\ appears completely unabsorbed.

{}

\begin{acknowledgements}
We thank the anonymous referee for thorough reading of the paper and constructive comments. 
We are grateful to ESO and Paranal observatory staff for carrying out the observations in service mode.  
 We acknowledge support from the French {\sl Agence Nationale de la Recherche} under ANR grant 17-CE31-0011-01 / project ``HIH2'' (PI: Noterdaeme) as well as from the Russian-French collaborative programme (PRC). SB is supported by RSF grant 18-12-00301. PL acknowledges the warm and welcoming atmosphere of IAP during his visit.
 The Cosmic Dawn Center (DAWN) is funded by the Danish National Research Foundation under grant no. 140. JPUF acknowledges support from the Carlsberg foundation.
 We all thank our close relatives for facilitating good working conditions during the covid-19 confinement. 
 \end{acknowledgements}

\bibliographystyle{aa}
\bibliography{mybib.bib}
 
\end{document}